\documentclass{article}

\usepackage{graphicx}
\usepackage{amssymb}
\usepackage{amsmath, amsthm}
\usepackage[square,numbers]{natbib}
\usepackage[affil-it]{authblk}
\usepackage{color}
\usepackage{enumerate}
\usepackage{hyperref}
\usepackage{comment}
\usepackage{doi}

\newtheorem{thm}{Theorem}

\newtheorem{remark}[thm]{Remark}

\usepackage{caption}
\usepackage{subcaption}

\begin{document}

\title{Modeling wildfire dynamics through a physics-based  approach incorporating fuel moisture and landscape heterogeneity}

\author[1,$\ddagger$]{Adrián Navas-Montilla}
\author[2, 3]{Cordula Reisch}
\author[4]{Pablo Diaz}
\author[5]{Ilhan Özgen-Xian}

\affil[1]{Fluid Dynamics Technologies Group, Aragon Institute of Engineering Research (I3A), University of Zaragoza, Spain}
\affil[2]{Institute for Partial Differential Equations, Technische Universit\"at Braunschweig, Germany}
  \affil[3]{Department of Mathematics and Scientific Computing, Universität Graz, Austria }
\affil[4]{Department of Applied Mathematics,  University of Zaragoza, Spain}
\affil[5]{Institute of Geoecology, Technische Universit\"at Braunschweig, Germany}

\affil[$\ddagger$]{Corresponding author at Fluid Dynamics Technologies Group, Aragon Institute of Engineering Research (I3A), University of Zaragoza, Spain. E-mail: {\tt
    anavas@unizar.es}}

    \date{March 12, 2025}

\maketitle

\begin{abstract}

Anthropogenic climate change has increased the probability, severity, and duration of heat waves and droughts, subsequently escalating the risk of wildfires. Mathematical and computational models can enhance our understanding of wildfire propagation dynamics.  
In this work, we present a simplified Advection-Diffusion-Reaction (ADR) model that accounts for the effect of fuel moisture, and also considers wind, local radiation, natural convection and topography. The model explicitly represents fuel moisture effects by means of the apparent calorific capacity method, distinguishing between live and dead fuel moisture content. Using this model, we conduct exploratory simulations and present theoretical insights into various modeling decisions in the context of ADR-based models. We aim to shed light on the interplay between the different modeled mechanisms in wildfire propagation to identify key factors influencing fire spread and to estimate the model's predictive capacity.


\end{abstract}

\section{Introduction}

Over the next decades, hot and dry weather that creates favorable conditions for wildfires is expected to become more frequent \cite{aghakouchak2020climate}.
A recent report by the United Nations Environment Programme states that due to these hot and dry weather conditions, combined with expected land-use change, wildfires will become more frequent and intense, with a global increase in extreme fires \cite{UN:2022}. 
Additionally, anthropogenic climate change also increases the risk of co-occurring and cascading hazards, which lead to major societal impacts. The increase in frequency and severity of wildfires exacerbates the vulnerability of charred landscapes to flooding, landslides and debris flows \cite{aghakouchak2020climate}. Thus, there is a call for a radical change in wildfire management policies, shifting government investments from reaction and response to prevention, mitigation, and adaptation \cite{moreira2020wildfire}. 

Within the last 50 years, mathematical models for forest fire propagation have been developed with the aim of understanding and predicting the evolution of fire. These models range from physics-based models  \cite{asensio2023historical,burger_exploring_2020,ferragut2007numerical,grasso2020physics,  margerit2002modelling,san2023gpu, sero2002modelling, vogiatzoglou2024interpretable} to empirical models \cite{finney1998farsite, rothermel1972mathematical}, cellular automata \cite{alexandridis_cellular_2008, boters-pitarch_intelligent_2024,currie_pixel-level_2019}, as well as probabilistic models \cite{carmel_assessing_2009, tymstra_development_2010}. In what follows, we restrict our focus to physics-based models, which rely on balance laws and allow for gaining a better understanding of how the involved processes influence the fire spread.

Predicting the impact of wildfires on our ecosystems is challenging because wildfire propagation depends on multiple factors and processes across multiple scales, each addressing different physical and chemical processes involved in the combustion and fire propagation phenomena.  Knowledge of these scales is crucial for the formulation of a wildfire propagation model.  Depending on the purpose of the model, certain scales are more suitable than others.  When seeking a model with predictive capacity---i.e. one that operates faster than real time \cite{ferragut2007numerical,grasso2020physics,mandel2008wildland,san2023gpu, sero2008large,sullivan2009wildland, vogiatzoglou2024interpretable}---it is not feasible for the model to resolve all scales.  Instead, only the larger scales are explicitly represented in the model, while the processes occurring at the smaller scales must be approximated or modeled indirectly. Resolving these small-scale phenomena would require a very fine spatial discretization, resulting in prohibitively high computational costs.

Following  the criteria and nomenclature proposed by Sero et al. \cite{margerit2002modelling,sero2002modelling}, there are four relevant scales: the gigascopic scale, the macroscopic scale, the mesoscopic scale and the microscopic scale. The gigascopic scale encompasses fire spread over large areas, several hundred meters to kilometers. At this scale, the fire front can be represented as a one-dimensional curve advancing across the landscape. Looking more closely at the vegetation stratum, the macroscopic scale is defined to model the vegetation layer, the air above and the ground underneath. On this scale, the occupation density or surface coverage (SC) is defined and the width and height of the fire front are determined. A one-level finer scale is the mesoscopic scale, which allows us to characterize the fine structure of vegetation, such as the geometry of branches and leaves. At this scale, the vegetation stratum is regarded as a two-phase porous medium, consisting of a mixture of vegetation (solid) and air (gas). Finally, the smallest scale is the microscopic scale, which  enables the definition of individual components of vegetation, including solid, liquid, and gaseous phases. At this scale, pyrolysis and drying can be represented and the process by which wood is converted into by-products such as char and flammable gases is described.

In this work, we aim to resolve fire propagation at the macroscopic scale. The heterogeneity observed at the mesoscopic scale and below, and the related processes happening at those scales, will not be explicitly represented and will instead be modeled accordingly on the macroscopic scale. The smallest spatial discretization size will be of the order of magnitude of the mesoscopic scale, and therefore the vegetation stratum will be treated as a continuous porous medium with averaged properties.

One of the most significant controls on both wildfire risk and propagation in many regions of the world is the so-called \textit{fuel moisture content} \cite{Ruffault:2022b, Brown:2022}, which is defined as the weight of water in fuel expressed as a fraction of the weight of dry fuel.  Fuel moisture content is one of the primary factors determining how much of the fuel is available to burn, and how much fuel might be consumed during a wildfire, with an eventual impact on wildfire rate of spread (ROS) and other attributes such as flame dimensions and fuel consumption \cite{matthews2013dead,wagtendonk1977refined}.  Fuel moisture content is inversely related to the likeliness of ignition and fire intensity, as part of the available energy for ignition is absorbed as latent heat during evaporation before the fire starts \cite{dimitrakopoulos2001flammability}.  Knowledge of the fuel moisture content is thus required to predict fire behavior and it usually is an important input parameter in operational fire models \cite{asensio2023historical, marcozzi_fastfuels_2025,matthews2013dead,rothermel1986modeling}.

In the context of wildfires, fire scientists distinguish between live and dead fuel \citep{McGranahan:2021}.  Live fuels are associated with live vegetation and comprise plant compartments such as leaves and stems.  Dead fuels are associated with dead biomass, for example, litter and twigs.  
Live fuel moisture content is responsive to long-term climate and plant adaptations to drought \cite{yebra2013global}. On shorter time scales, depending on the plant functional type, it is affected by soil moisture (understorey shrubs) or seasonal plant dynamics (overstorey trees) \cite{Brown:2022}. In contrast, dead wood moisture content depends directly on air temperature,  relative humidity and precipitation. Dead wood exchanges water with the surrounding environment through a mechanism of vapor exchange and tends to reach equilibrium with the atmosphere \cite{matthews2013dead,Ross_2010} and sometimes the soil \cite{rakhmatulina2021soil}. It releases water when the surrounding air becomes dryer and absorbs water when the air is more humid. The rate at which dead wood exchanges moisture with the atmosphere depends on its size. 
Finer fuels like leaves, bark, and small twigs reach the equilibrium within minutes, while coarse fuels like larger twigs and logs level with their surroundings in hours or days \cite{matthews2013dead}.

Despite its significance, fuel moisture is not always explicitly accounted for in physics-based wildfire propagation models. This reduces the predictive capability of these models and hinders the study of fuel moisture effects on wildfire propagation.  To the best of our knowledge, in the context of  Advection-Diffusion-Reaction (ADR)-based models, only the PhyFire model \cite{asensio2023historical}, the model by Vogiatzoglou et al. \cite{vogiatzoglou2024interpretable} and the models by Séro-Guillaume and Margerit \cite{margerit2002modelling,sero2002modelling} explicitly represent the effect of fuel moisture. The model in \cite{asensio2023historical} uses a multi-valued operator to relate enthalpy and temperature that allows to represent the actual sensible and latent heating of water within the fuel. In the literature, this is referred to as the Stefan problem, originally introduced for the computation of phase change materials \cite{eyres1946calculation,swaminathan1993enthalpy,voller1987enthalpy}. This approach yields physically consistent results that have been successfully validated with observations \cite{ASENSIO2023105710}, showing a great performance in accurately simulating fire behavior and establishing the basis for an operational fire spread simulator \cite{prieto2017gis}. An alternative approach is proposed by Vogiatzoglou et al. \cite{vogiatzoglou2024interpretable}, where the endothermic phase of the reaction is explicitly modeled. Wood dehydration, disintegration and combustion are modeled
by two consecutive reactions using a first-order Arrhenius kinetics. They offer a higher level of detail in the reaction equations than average state-of-the-art ADR predictive models. Additionally, they provide a solid physical basis for all the terms in the model, enabling the accurate simulation of complex wildfire propagation patterns. Regarding the approaches developed by Margerit and Séro-Guillaume \cite{margerit2002modelling,sero2002modelling,sero2008large}, they offer a very exhaustive representation of the physical and chemical processes involved in the fire, including the effect of moisture. A hierarchy of different models, with increasing complexity, is obtained. To model the different sensible and latent water heating processes, different equations are used at the different stages. A Dirac-type distribution term is used in their ADR equation to model the effect of the latent heating of water at the evaporation temperature. 

In this work, we extend the ADR wildfire propagation model presented in \cite{reisch_analytical_2024} to explicitly represent fuel moisture effects, distinguishing between live and dead fuel moisture content when desired. The proposed   approach is based on the apparent calorific capacity method, which models the effect of thermal phase changes by considering an apparent (or effective) heat capacity, usually defined as a piecewise constant function of the temperature \cite{caggiano2018reviewing,swaminathan1993enthalpy}. Three different moisture models, representing the evaporation process and the constituents of fuel with a different level of detail, are presented.  The most comprehensive model captures the underlying physio-chemical mechanisms involved in the moisture balance of both live and dead fuel, allowing for an approximation of their specific heat independently.  

The objective of this paper is twofold. First, we aim to present an ADR model that accounts for the effect of live and dead fuel moisture, along with advection (wind), local radiation, natural convection and topography. We put special emphasis on the derivation of the equations of the model, based on the theory of two-phase porous flows. Second, we seek to understand the interplay between the different model mechanisms in wildfire propagation to identify key factors influencing fire spread and to estimate the model's predictive capacity. 
We conduct exploratory simulations and present theoretical insights into various modeling decisions.  
Furthermore, we investigate whether the model's behavior is consistent with laboratory experiments and field observations by carrying out some parametric analyses and qualitative comparisons. 

The paper is structured as follows: 
In Sec.~\ref{sec:model}, the variables of the model are defined, and the differential equations are derived from physical balance laws, with additional details provided in App. \ref{appendix_eqs}. We discuss modeling decisions on the wind, radiation, combustion, and, introduce different moisture models in Sec.~\ref{sec:moisture}.  The numerical schemes are presented in Sec.~\ref{sec:numerics}, accompanied by verification results in App.~\ref{appendix}. 
Sec.~\ref{sec:numericalresults} investigates the effects of the discussed mechanisms by numerical simulations. Two-dimensional simulations in Sec.~\ref{sec:2dim} lead the path to more realistic scenarios. 
Finally, Sec.~\ref{sec:conclusion} summarizes the highlights, critically discusses shortcomings and provides an outlook on future challenges. 

\section{Mathematical description of the model}\label{sec:model}

After discussing the representation of the vegetation layer, the advection diffusion reaction model is derived by physical balance laws. Variants of different terms included follow, and this section closes with fixing physical parameters for the simulations. In App. \ref{appendix_eqs}, a detailed derivation of the equations for the conservation of energy, which complements this section, can be found.

\subsection{Representation of the vegetation layer}\label{sec:mathmodel}

 The wildfire propagation model is defined in the spatial domain $\Omega=\Lambda \times [0,l]$, with $\Lambda\subset\mathbb{R}^2$ as the ground surface and $l$ [m] as the height.
Let $V=S\times [0,l] \subset \Omega$ be a control volume inside this domain, with $S \subset \Lambda$ being the ground layer of this control volume. Within $V$, a continuum approach is used to describe the biomass and air mixture, allowing to define bulk parameters to characterize the properties of this continuum.  Averaging all modeled quantities in $l$ over $S$ leads to two dimensional space-dependencies. Thus, all variables of interest will be defined on $S$,  with $\mathbf{x}=(x,y)\in \Lambda$ denoting the spatial location.  A mixture of two phases is assumed within~$V$: (i) the solid phase, composed of solid fuel (that is considered equal to the biomass), and (ii) the gaseous phase, which consists of air and combustion products \cite{margerit2002modelling,sero2002modelling,sero2008large}. The mixture can be modeled as a two-phase porous medium \cite{brennen2005fundamentals,vafai2015handbook}. The abstraction from the two-phase porous medium leads to the definition of bulk parameters for quantities within $V$ at the centroid $\mathbf{x}_c$ of $S$: solid fuel with volume $V_f$ [m$^3$], density $\bar{\rho}_f$ [kg/m$^3$], the specific heat $\bar{c}_{p,f}$ [kJ/(kg·K)] and (dry) mass $m_f=\bar{\rho}_f V_f$ [kg], as well as air and other combustion products with volume $V_a$ [m$^3$],  density $\bar{\rho}_{a}$ [kg/m$^3$], specific heat $\bar{c}_{p,a}$ [kJ/(kg·K)] and mass $m_a=\bar{\rho}_{a}V_a$ [kg], that is $m=m_f+m_a$.  The fuel density and specific heat may depend upon space, that is to say $\bar{\rho}_f=\bar{\rho}_f(\mathbf{x})$ and $\bar{c}_{p,f}=\bar{c}_{p,f}(\mathbf{x})$,  whereas the density and specific heat of air and combustion products, $\bar{\rho}_{a}$ and  $\bar{c}_{p,a}$, are assumed to be spatially constant.

The bulk density of the mixture inside $V$ is given by
\begin{equation}\label{eq:rhodef}
\rho(\mathbf{x}, t)=\frac{m_f(\mathbf{x}, t) + m_a(\mathbf{x}, t)}{V}=(\bar{\rho}_f(\mathbf{x})- \bar{\rho}_a) R_f(\mathbf{x}, t)+\bar{\rho}_a,
\end{equation}
where $R_f(\mathbf{x}, t)=\frac{V_f(\mathbf{x}, t)}{V}\in [0,1]$ is the solid fuel volume ratio inside the control volume. Note the relation $R_f=1-p$, with $p$ being the porosity. The density at the initial time is computed as 
\begin{equation}\label{eq:rhodef0}
\rho_0(\mathbf{x})=\rho(\mathbf{x}, 0)=(\bar{\rho}_f(\mathbf{x})- \bar{\rho}_a) R_{f,0}(\mathbf{x})+\bar{\rho}_a ,
\end{equation}
where $R_{f,0}(\mathbf{x})=R_f(\mathbf{x}, 0)$ is the solid fuel volume ratio at the initial time. The solid fuel volume ratio $R_{f,0}(\mathbf{x})$ is computed from the initial biomass distribution and coverage by
\begin{equation}
R_{f,0}(\mathbf{x})=\frac{W_{f,0}(\mathbf{x})}{\bar{\rho}_f(\mathbf{x}) l},
\end{equation}
where $W_{f,0}(\mathbf{x})=m_{f,0}(\mathbf{x})/\lvert S \rvert$ is the initial fuel load per surface unit, with $m_{f,0}$ the initial fuel mass distribution \cite{alonso2020estimating}. The initial fuel load $W_{f,0}$ [kg/m$^2$] usually depends upon the surface coverage, $SC\in [0,1]$, which can be obtained from remote sensing techniques.  The initial fuel load is expressed as 
\begin{equation}\label{eq:SC}
W_{f,0}(\mathbf{x}) = W_{f,\mathrm{max}}(\mathbf{x}) \cdot SC (\mathbf{x}),
\end{equation}
where $W_{f,\mathrm{max}}(\mathbf{x})$ is a theoretical maximum fuel load for a fuel type. 
Depending on the application, either the surface coverage $SC$ or the initial fuel load $W_{f,0}$ are suitable information given. 

The remaining fuel mass fraction relates to the quantities defined above as
\begin{equation}\label{eq:ydef}
Y(\mathbf{x}, t) = \frac{m_f(\mathbf{x},t)}{m_{f,0}(\mathbf{x})} = \frac{V_f(\mathbf{x},t)}{V_{f,0}(\mathbf{x})} = \frac{R_f(\mathbf{x},t)}{R_{f,0}(\mathbf{x})},
\end{equation}
where $m_{f,0}$ and $V_{f,0}$ are the initial mass and volume of the fuel. Note that $Y$ will hereafter be referred to as the remaining fuel fraction or the remaining biomass fraction for the sake of brevity.

In what follows, the dependency of the variables upon $\mathbf{x}$ and $t$ is dropped for the sake of simplicity. The expression to compute the bulk density from the remaining fuel  fraction is obtained from Eqs. \eqref{eq:rhodef} and \eqref{eq:ydef} and reads \cite{vafai2015handbook}
\begin{equation}\label{eq:rhodefY}
\rho=(\bar{\rho}_f- \bar{\rho}_a) R_{f,0} Y  + \bar{\rho}_a .
\end{equation}
Analogously, the bulk specific heat inside $V$ is computed as follows \cite{vafai2015handbook}
\begin{equation}\label{eq:cpdefY}
c_p=\frac{\left(\bar{c}_{p,f}\bar{\rho}_f- \bar{c}_{p,a} \bar{\rho}_a\right)R_{f,0}  Y  + \bar{c}_{p,a}\bar{\rho}_a }{\rho},
\end{equation}
where $\bar{c}_{p,f}$ is the specific heat of the fuel, that may also depend upon the temperature and will be addressed later.

When setting $Y=1$ in Eqs. \eqref{eq:rhodefY} and \eqref{eq:cpdefY} we recover the  bulk density and specific heat at the initial time
\begin{equation}\label{eq:rhocp0}
\rho_0=(\bar{\rho}_f- \bar{\rho}_a) R_{f,0}  + \bar{\rho}_a \,,\qquad c_{p0}=\frac{\left(\bar{c}_{p,f}\bar{\rho}_f- \bar{c}_{p,a} \bar{\rho}_a\right)R_{f,0}  + \bar{c}_{p,a}\bar{\rho}_a }{\rho_0}.
\end{equation}
The initial values for the parameters approximate the time-dependent bulk parameters sufficiently good, \cite{mandel2008wildland}.
A simple analysis of the order of magnitude of the terms in Eq.~\eqref{eq:rhocp0}, assuming that $\bar{\rho}_f\gg \bar{\rho}_a$ \cite{sero2008large}, reasons the approximations 
\begin{equation}\label{eq:rhocpApprx}
\rho_0\approx \bar{\rho}_f R_{f,0}=\frac{W_{f,0}}{l} \,,\qquad c_{p0} \approx \bar{c}_{p,f}.
\end{equation}

The thickness $l$ in the vertical direction can be seen as a relative height (with unit), herein assumed to be constant $l=1$ m for simplicity. In the vertically averaged system, the vertical component is not modeled directly, and in the following derivation of the model, the simplification from a three-dimensional to a two-dimensional model is highlighted. 

\subsection{Physical derivation of the ADR model}

Consider a fixed control volume defined as $V=S\times [0,l] \subset \Omega$, with $S \subset \Lambda$  the ground layer, and with $l$ the thickness in the vertical direction.   
The control volume  is open along the lateral and top boundaries.  In what follows, it is assumed that the fuel consumption rate is given by the following phenomenological equation \cite{asensio2023historical,sero2002modelling}
 \begin{equation}\label{eq:heatflux43} 
    \dot{Y} := \frac{\partial Y}{\partial  t}= -  \Psi(T)Y 
\end{equation}
which represents a first-order kinetic equation with the reaction rate $\Psi(T)$ being discussed in Sec.~\ref{subsec:modelling_combustion}. A highly simplified chemical reaction is considered,  where the fuel is assumed to be the only reactive specie, and the product is a gas with properties similar to air. It is assumed that all the fuel undergoes complete combustion and is fully converted into gas. Thus, the product production rate $\dot{m}_a^R$ [kg/s] is equal to the fuel consumption rate $\dot{m}_f={m}_{f,0}\dot{Y}$ [kg/s], so $\dot{m}_a^R=-\dot{m}_f$. 

To derive the ADR model, we apply the equations for the conservation of mass and energy within the selected control volume.  As mentioned before, all variables are homogenized in the vertical direction, $z$, thus the dependency of the variables on $z$ is dropped and gradients in $z$-direction are assumed zero.  All vector variables only have the $x$- and $y$-component and the spatial differential operator is defined as $\nabla = (\partial/\partial x, \partial/\partial y)^T$. In addition,  incompressibility conditions are assumed within the mixture. The equation for the conservation of mass inside $V$ can be expressed as
\begin{equation}\label{eq:masscons1} 
    \frac{d}{d t}\int_{V}\rho \,\mathrm{d}V + \int_{\partial V}\bar{\rho}_a (1-R_f) \mathbf{v}_a\cdot \mathbf{n} \,\mathrm{d}\Gamma =  -\dot{m}_\mathrm{out} \,,
\end{equation}
where $\rho  = \bar{\rho}_f R_{f} +  \bar{\rho}_a (1-R_{f})$ is the density of the mixture, $\mathbf{v}_a=(u_a,v_a)^T$ [m/s] is the actual (2D) velocity of the gas phase through the solid fuel  and  $\dot{m}_\mathrm{out} $ is the mass flow leaving the domain through the top layer, exiting into the atmosphere. Note that $\partial V$ represents the boundary surfaces of the control volume, excluding the top and bottom layers. Using the relation $\dot{R_f}=\dot{Y}R_{f,0}$ on the first term of Eq.~\eqref{eq:masscons1}, we  obtain
\begin{equation}\label{eq:masscons2} 
    \dot{m}_\mathrm{out} =   \dot{Y}R_{f,0}( \bar{\rho}_a- \bar{\rho}_f)V- \int_{\partial V}\bar{\rho}_a (1-R_f) \mathbf{v}_a\cdot \mathbf{n}\,\mathrm{d}\Gamma .
\end{equation}

Note that the mass flow leaving the domain is due to two contributions: (i) the first term on the right-hand side of Eq. (\eqref{eq:masscons2})  represents the excess mass of gas that cannot occupy the volume vacated by the fuel upon its consumption, it will be zero without combustion; (ii) the second term on the right-hand side of Eq.~\eqref{eq:masscons2}  represents the mass flow difference through the boundaries of the domain in 2D, due to variations in $R_f$ and a divergence of $\mathbf{v}_a$. Assuming that $R_f\lesssim 0.01$ \cite{vogiatzoglou2024interpretable}, the approximation $1-R_f\approx 1$ allows to rewrite Eq.~\eqref{eq:masscons2} as
\begin{equation}\label{eq:masscons21} 
    \dot{m}_\mathrm{out} \approx   \dot{Y}R_{f,0}( \bar{\rho}_a- \bar{\rho}_f)V- \bar{\rho}_a\int_{\partial V} \mathbf{v}_a\cdot \mathbf{n}\,\mathrm{d}\Gamma .
\end{equation}
Note that the second term will be zero when the incompressibility condition is satisfied in the 2D advection field, i.e. $\nabla \cdot \mathbf{v}_a = 0$, leading to
\begin{equation}\label{eq:masscons3} 
    \dot{m}_\mathrm{out} =   \dot{Y}R_{f,0}( \bar{\rho}_a- \bar{\rho}_f)V
\end{equation}

Now consider the equation for the conservation of energy. Assume that there is a local thermal equilibrium in the medium, that is, the temperature of the solid fuel, $T_f$ [K], and the temperature of the gases, $T_a$ [K], will coincide, $T_f=T_a=T$ \cite{sero2002modelling,sero2008large}. In addition, the effect of pressure variations and viscous dissipation is neglected, as shown in App.~\ref{appendix_eqs}. The conservation of energy inside $V$ can be expressed as (see App.~\ref{appendix_eqs}, Eq.~\eqref{eq:twophase1})
\begin{equation}\label{eq:energy1} 
    \frac{d}{d t}\int_{V}\rho h \,\mathrm{d}V + \int_{\partial V}\bar{\rho}_a h_a (1-R_f) \mathbf{v}_a\cdot \mathbf{n} \,\mathrm{d}\Gamma = \dot{{Q}}_\mathrm{cond}  + \dot{{Q}}_\mathrm{rad} - \dot{{Q}}_\mathrm{out} +  \dot{{Q}}_\mathrm{comb} \,,
\end{equation}
where $h$ [kJ/kg] is the specific enthalpy and $\rho$ [kg/m$^3$] is the density of the mixture, i.e. $\rho h = \bar{\rho}_fh_f R_{f} +  \bar{\rho}_a h_a (1-R_{f})$, with $h_f$ and $h_a$ being the specific enthalpy of the solid and gas phase, respectively.  $\dot{{Q}}_\mathrm{cond}$, $\dot{{Q}}_\mathrm{rad}$ and $\dot{{Q}}_\mathrm{out}$ [kJ/s] are the conduction, radiation and free cooling (atmospheric exchange) heat fluxes, respectively, and $ \dot{{Q}}_\mathrm{comb}$ [kJ/s]  is the reaction heat source due to combustion. These terms will be modelled following the approaches often used in the literature \cite{asensio2002wildland,burger_exploring_2020,grasso2020physics,mandel2008wildland,reisch_analytical_2024,vogiatzoglou2024interpretable}.

The advection term accounts for the advection of energy of the gaseous phase, $\bar{\rho}_a h_a$, and can be rewritten in terms of the so-called Darcy velocity $\mathbf{v}_D$ in the context of porous flows \cite{vafai2015handbook}
\begin{equation}\label{eq:advectTermMass0} 
\mathbf{v}_D=  (1-R_f)\mathbf{v}_a.
\end{equation}
 The enthalpy-weighed bulk advection velocity,  $\mathbf{v}=(u,v)^T$, allows to rewrite the advection term in Eq.~\eqref{eq:energy1} more compactly. The bulk velocity $\mathbf{v}$ is related to the actual flow velocity through the solid fuel,  $\mathbf{v}_a$, as  
\begin{equation}\label{eq:advectTerm} 
\mathbf{v}=  (1-R_f)\frac{\bar{\rho}_a \bar{c}_{p,a}}{\rho {c}_{p}}\mathbf{v}_a,
\end{equation}
with $\rho c_p=\bar{\rho}_f \bar{c}_{p,f} R_{f} +  \bar{\rho}_a \bar{c}_{p,a} (1-R_f)$ given by Eq.~\eqref{eq:rhocp0}, allowing to rewrite the advection term as 
\begin{equation}\label{eq:advectTerm2} 
     \int_{\partial V}\rho h \mathbf{v}\cdot \mathbf{n} \,\mathrm{d}\Gamma = \int_{\partial V}\rho h \beta\mathbf{v}_a\cdot \mathbf{n} \,\mathrm{d}\Gamma  
\end{equation}
where $\mathbf{n}$ is the surface normal vector and $\beta=(1-R_f)\frac{\bar{\rho}_a \bar{c}_{p,a}}{\rho {c}_{p}}$  can be regarded as a correction factor \cite{grasso2020physics,prieto2015sensitivity} and will be addressed in Sec. \ref{subsec:modelling_wind}.  Note that $\mathbf{v}_a$ will be a function of the surface wind velocity, $\mathbf{w}$ that needs to be calibrated depending on the vegetation type.  Given this and considering the level of simplification used in the model, the parameter $\beta$ will be retained as a calibration parameter, consistent with the studies referenced above.

Rewriting Eq.~\eqref{eq:energy1} with $H=\rho h$ the enthalpy per unit volume and $h=h(T)$ the specific enthalpy, which depends on the temperature, gives
\begin{equation}\label{eq:energy2} 
    \frac{d}{d t}\int_{V}H \,\mathrm{d}V + \int_{\partial V}H \mathbf{v}\cdot \mathbf{n} \,\mathrm{d}\Gamma =  \dot{{Q}}_\mathrm{cond}  + \dot{{Q}}_\mathrm{rad} - \dot{{Q}}_\mathrm{out} +  \dot{{Q}}_\mathrm{comb} .
\end{equation}

The conduction heat flux is given by Fourier's law as 
\begin{equation}\label{eq:heatflux1} 
    \dot{{Q}}_\mathrm{cond}= -\int_{\partial V} \dot{\mathbf{q}}\cdot \mathbf{n} \,\mathrm{d}\Gamma= \int_{\partial V} k_c\nabla T\cdot \mathbf{n} \,\mathrm{d}\Gamma \, ,
\end{equation}
where $\dot{\mathbf{q}}$ [kJ/(m$^2$·s)] is the heat flux vector and  $k_c$ [kJ/(m·K·s)] is the thermal conductivity of the bulk mixture. 

In this model, radiation is assumed to act only locally by assuming that the medium is optically thick. Following the Rosseland approximation, the radiation heat flux can be written in the form of a non-linear conduction heat flux as \cite{asensio2023historical,grasso2020physics}
\begin{equation}\label{eq:heatflux3} 
    \dot{{Q}}_\mathrm{rad}=  -\int_{\partial V} \dot{\mathbf{q}}_\mathrm{rad}\cdot \mathbf{n} \,\mathrm{d}\Gamma \approx \int_{\partial V} 4\sigma\delta\epsilon T^3 \nabla T\cdot \mathbf{n} \,\mathrm{d}\Gamma \, ,
\end{equation}
provided that the optical path length for radiation, $\delta$ [m], is smaller than the characteristic lengths of the modeled physical phenomena. Other parameters in the previous expression are the Stefan-Boltzmann constant, $\sigma$ [kJ/(m$^{2}$·K$^{4}$·s)], and the emissivity factor, $\epsilon$.

The free cooling heat flux can be expressed as
\begin{equation}\label{eq:cooling} 
   \dot{{Q}}_\mathrm{out} = \dot{{m}}_\mathrm{out}h_a+ \dot{{Q}}_\mathrm{out}' ,
\end{equation}
where $\dot{{m}}_\mathrm{out}h_a=\dot{{m}}_\mathrm{out}\bar{c}_{p,a} T$ represents the energy flux into the atmosphere due to the escape of mass from the control volume and $\dot{{Q}}_\mathrm{out}'$ models other heat transfer phenomena between the system and the atmosphere (natural convection, radiation, etc.). The term $\dot{{m}}_\mathrm{out}h_a$ can be expressed as a flux of the bulk enthalpy
\begin{equation}\label{eq:cooling2} 
   \dot{{m}}_\mathrm{out}h_a = H \left( \dot{Y}R_{f,0}V\left( \frac{\bar{\rho}_a- \bar{\rho}_f}{\rho}\right)- \int_{\partial V} \frac{\bar{c}_{p}}{\bar{c}_{p,a}}\mathbf{v}\cdot \mathbf{n}\,\mathrm{d}\Gamma \right).
\end{equation}
This term can be regarded as a generalization of the approach proposed in \cite{nieding_impact_2024}. For the sake of simplicity, let us assume
\begin{equation}\label{eq:heatflux2} 
    \dot{{Q}}_\mathrm{out} = \int_{ S} \hat{\alpha}(T-T_{\infty})\,\mathrm{d}\Gamma ,
\end{equation}
with $\hat{\alpha}$ [kJ/(m$^2$·K·s)] a global heat transfer coefficient coefficient, retained as a calibration parameter, and $T_{\infty}$ [K] the temperature of the air far above the control volume.

Finally, the reaction heat flux due to the combustion of fuel  is given by
\begin{equation}\label{eq:heatflux41} 
    \dot{{Q}}_\mathrm{comb}= - \int_{V} \bar{\rho}_f R_{f,0} \dot{Y}\mathcal{H}  \,\mathrm{d}V \, ,
\end{equation}
where $ \mathcal{H}$ [kJ/kg] is the combustion heat per unit mass of fuel and $\dot{Y}$ is the fuel consumption rate per unit volume, given by Eq.~\eqref{eq:heatflux43}. Inserting Eq.~\eqref{eq:heatflux43} into Eq.~\eqref{eq:heatflux41} gives
\begin{equation}\label{eq:heatflux44} 
    \dot{{Q}}_\mathrm{comb}= \int_{V} \bar{\rho}_f R_{f,0}  \mathcal{H} \Psi(T)Y \,\mathrm{d}V.
\end{equation}

Using the Gau{\ss} divergence theorem,  Eq.~\eqref{eq:energy2} can be rewritten as
\begin{equation}\label{eq:energy3} 
    \int_{V} \left(\frac{\partial H}{\partial t}  + \nabla \cdot (\mathbf{v}H) \right)\,\mathrm{d}V  = \int_{V} \nabla\cdot \left( k \nabla T \right)  \,\mathrm{d}V   -\int_{ V}  \alpha(T-T_{\infty})\,\mathrm{d}V + \int_{V} \bar{\rho}_f R_{f,0} \mathcal{H}\Psi(T)Y \,\mathrm{d}V \,,
\end{equation}
where $\alpha=\hat{\alpha}/l$   and 
\begin{equation}\label{eq:diffusion}
    k(T)= 4\sigma\epsilon\delta T^3 + k_c, 
\end{equation}
which gives the partial differential equation
\begin{equation}\label{eq:pde1} 
     \frac{\partial H}{\partial t}  + \nabla \cdot (\mathbf{v}H)   =  \nabla\cdot \left( k \nabla T \right)     - \alpha(T-T_{\infty}) +  \Psi(T)\bar{\rho}_f R_{f,0} \mathcal{H}Y \,.
\end{equation} 

The relation between $T$ and $H$ in Eq.~\eqref{eq:pde1} is given by a function $T=T(H)$ that may be multi-valued to model phase change due to the evaporation of fuel moisture \cite{ASENSIO2023105710,ferragut2007numerical}. In the literature, this is referred to as the Stefan problem, originally introduced for the computation of phase change materials in closed systems \cite{eyres1946calculation,swaminathan1993enthalpy,voller1987enthalpy}. Two aspects must be taken into account in this regard. First, the phase change  (fuel moisture evaporation) only occurs in one direction, since the water vapor then leaves the domain. Therefore, in the cooling-to-ambient-temperature process, the enthalpy-temperature relation $T=T(H)$ should be modified to prevent  physically unfeasible latent cooling (water vapor condensation), as resolved in \cite{asensio2023historical}. Second, since the advection term is written in conservative form, there may be physically unfeasible enthalpy accumulation in some spatial regions caused by a non-zero divergence of the velocity field. A correction based on Eq.~\eqref{eq:cooling} can be used. Recall that this non-zero divergence may be due to the dimensionality reduction of the problem, as well as to the representation of the effect of the topography proposed here (see Sec. \ref{subsec:modelling_wind}). Taking into account the degree of simplification used here and the previous observations, a definition of the problem in full conservation form, as in Eq.~\eqref{eq:pde1}, is not required for the purpose of this paper.

 We can find alternative approaches in the literature  to the multi-valued enthalpy operator method in Eq.~\eqref{eq:pde1} to include the effect of the phase change.  Most of these approaches were developed for other engineering applications. However, some of the difficulties mentioned above are likely to remain. The so-called \textit{heat source method} separates the enthalpy into two terms, one corresponding to the sensible heat and the other to the latent heat. The latter would be included as a source term in the classical heat equation \cite{caggiano2018reviewing,sero2008large,swaminathan1993enthalpy}. Another approach is called \textit{apparent calorific capacity method}, which models the effect of thermal phase changes by considering an apparent (or effective) heat capacity, usually defined as a piecewise constant function \cite{caggiano2018reviewing,swaminathan1993enthalpy}. This approach exploits the relation $\frac{\partial H}{\partial t}  = \frac{\partial H}{\partial T} \frac{\partial T}{\partial t}$, allowing to write 
\begin{equation} 
      \frac{\partial H}{\partial t} = \rho c_{p}  \frac{\partial T}{\partial t} \, ,
\end{equation}
where $c_{p}=c_{p}(T,Y)$ is the effective specific heat, which is a function of the temperature and may also depend upon biomass fraction. Further, the effective specific heat $c_p$ depends on the history of the burning process, modeling the different behavior when heating up (evaporation of moisture) and when cooling down after the burning. 
The effect of fuel moisture will be represented by a piecewise definition of this coefficient (see Section \ref{sec:moisture}). 

Motivated by the observations mentioned above, we propose to use a version of Eq.~\eqref{eq:pde1} in non-conservative form. The apparent calorific capacity approach gives an alternative formulation for the energy equation (see App.~\ref{appendix_eqs})
\begin{equation}\label{eq:pde1v2} 
    \rho c_{p}  \left( \frac{\partial T}{\partial t}  + \mathbf{v} \cdot \nabla T  \right)  =  \nabla\cdot \left( k \nabla T \right)     - \alpha(T-T_{\infty}) +  \Psi(T)\bar{\rho}_f R_{f,0} \mathcal{H}Y \,,
\end{equation}
where the term $\frac{\partial T}{\partial t}  + \mathbf{v} \cdot \nabla T$ represents the material derivative of the temperature and the product $\rho c_p=\bar{\rho}_f \bar{c}_{p,f} R_{f} +  \bar{\rho}_a \bar{c}_{p,a} (1-R_f) $ represents the bulk properties, see Eqs.~\eqref{eq:rhodefY} and \eqref{eq:cpdefY}. This version of the equation for the conservation of energy in non-conservative form is often found in the literature \cite{asensio2002wildland,mandel2008wildland,sero2002modelling,sero2008large,weber1991modelling,weber_toward_1991}. 

\begin{remark}
    Note that the velocity $\mathbf{v}$ in Eq.~\eqref{eq:pde1v2}  is not the actual flow velocity but rather a bulk velocity defined  as $\mathbf{v}=(1-R_f)\frac{\bar{\rho}_a \bar{c}_{p,a}}{\rho {c}_{p}}\mathbf{v}_a$ in Eq.~\eqref{eq:advectTerm} because  only the gaseous phase (a) is advected. This allows to express the left-hand side of Eq.~\eqref{eq:pde1v2} as a material derivative of the temperature, enabling the model to be defined as a conventional ADR equation. 
\end{remark}

Considering fuel properties at the initial time, i.e. $\rho = \rho_0 $ and $c_p=c_{p0}$ according to Eq.~\eqref{eq:rhocp0}, simplifies the equation further. When using the apparent capacity approach, the complete model reads
  \begin{align} 
   \left\{
  \begin{aligned}\label{eq:model1b}   
         \rho_0 c_{p0}  \left( \frac{\partial T}{\partial t}  + \mathbf{v} \cdot \nabla T  \right)  &=  \nabla\cdot \left( k \nabla T \right)     - \alpha(T-T_{\infty}) +   \Psi(T)\bar{\rho}_f R_{f,0}  \mathcal{H} Y ,  \\
 \frac{\partial Y}{\partial  t}&= -\Psi(T)Y ,        
     \end{aligned}
       \right. 
 \end{align} 
with $T=T(\mathbf{x},t)>0$ and $Y=Y(\mathbf{x},t)\in[0,1]$ being the problem variables, where $\mathbf{x}\in \Lambda \subset \mathbb{R}^2$ and $t\geq 0$. The specific heat, $c_{p0}$, is a function of the temperature and may also depend on the position. The following boundary and initial conditions
  \begin{align} 
   \left\{
  \begin{aligned}  \label{eq:ccb} 
\left(  k \nabla T    -  \rho_0 c_{p0}\mathbf{v} T  \right) \cdot \mathbf{n}=0 \quad \quad  & \mathbf{x}\in \partial\Lambda,\, t>0 , \\
T(\mathbf{x},0)=T_0(\mathbf{x}),\, Y(\mathbf{x},0)=Y_0(\mathbf{x}) \quad \quad  &  \mathbf{x}\in \Lambda ,
      \end{aligned}
       \right. 
 \end{align} 
 are used to complete Eq.~\eqref{eq:model1b}.
 Here, the zero flux boundary conditions are gained under the assumption of an incompressible, so a divergence free, advection velocity at the boundary. 
 Then, for any point $\mathbf{x} \in \partial \Lambda$ it yields $\nabla \cdot (\mathbf{v} T) = \mathbf{v} \cdot \nabla T$. 
 In the numerical simulations, the boundary conditions do not affect the overall system behavior because the initial ignition area is far away from the boundary and the computational time is small compared to the time needed for a fire front to reach the boundary.

It is worth highlighting that Eq.~\eqref{eq:pde1v2} can still be simplified using the assumptions in Eq.~\eqref{eq:rhocpApprx} and setting $l=1$, being expressed in terms of the initial fuel load per unit surface, $W_{f,0}$, and the  fuel heat capacity, as 
\begin{equation}\label{eq:pde1v2simple} 
    W_{f,0}\bar{c}_{p,f}  \left( \frac{\partial T}{\partial t}  + \mathbf{v} \cdot \nabla T  \right)  =  \nabla\cdot \left( \hat{k} \nabla T \right)     - \hat{\alpha}(T-T_{\infty}) +  \Psi(T) W_{f,0} \mathcal{H}Y \, .
\end{equation}

The model in Eq.~\eqref{eq:model1b}  can be recast in vector form---a notation that is more suitable for the description of the numerical scheme in the following sections---as 
\begin{equation}\label{eq:compactsys} 
    \frac{\partial \mathbf{U}}{\partial t}+\mathbf{A }\frac{\partial \mathbf{U}}{\partial x}  +\mathbf{B }\frac{\partial \mathbf{U}}{\partial y}  = \mathbf{C } \left(\frac{\partial \mathbf{F }(\mathbf{U})}{\partial x}+\frac{\partial \mathbf{G } (\mathbf{U})}{\partial y} + \mathbf{S} (\mathbf{U})  \right)  
\end{equation}
 with 
 \begin{equation}
\mathbf{A }=\begin{bmatrix}
u & 0 \\
0 & 0
\end{bmatrix}, \,  \mathbf{B }=\begin{bmatrix}
v & 0 \\
0 & 0
\end{bmatrix}, \, \mathbf{C }=\begin{bmatrix}
\frac{1}{\rho_0 c_{p0}} & 0 \\
0 & 1
\end{bmatrix}
\end{equation}
as the coefficient matrices and
\begin{equation}  \label{eq:matrixform} 
     \mathbf{U}=\left( \begin{array}{c}
       T  \\
       Y \\  
        \end{array}\right) ,\,\, \;
    \mathbf{F}=\left( \begin{array}{c}
       -k\frac{\partial T}{\partial x}  \\
       0 \\ 
    \end{array}\right)  ,\,\, \;
    \mathbf{G}=\left( \begin{array}{c}
       -k\frac{\partial T}{\partial y}   \\
       0 \\   
    \end{array}\right)  ,\,\, \;
    \mathbf{S}=\left( \begin{array}{c}
        - {\alpha}(T-T_{\infty})  +  {\bar{\rho}_f}\Psi(T) R_{f,0} H Y   \\
          -\Psi(T)Y  \\  
        \end{array}\right) 
  \end{equation}
being the vectors of variables, diffusive fluxes and source terms.

Considering $R_{f,0}=1$, $\bar{c}_{p,f}=\bar{c}_{p,a}$ and $\bar{\rho}_f=\bar{\rho}_a$ as constant values, the model in  \cite{reisch_analytical_2024} is recovered.

\subsection{Modeling wind and topography}\label{subsec:modelling_wind}

The advection velocity, $\mathbf{v}$, is modeled as a linear function of the wind velocity and the gradient of the topography as 
\begin{equation}\label{eq:wind-vel}
\mathbf{v}=\beta \mathbf{w} + {\gamma  \nabla Z} \, ,
\end{equation}
where $\mathbf{w}=\mathbf{w}(\mathbf{x},t)$ is the wind velocity, $Z=Z(\mathbf{x})$ is the topography and $\beta$ and $\gamma$ are calibration parameters, yet to be defined. Recall that only the energy within the fluid phase is advected and an expression for $\beta$ was derived in Eqs. \eqref{eq:advectTerm}--\eqref{eq:advectTerm2}
\begin{equation}\label{eq:beta}
\beta=\frac{(1-R_f)\bar{\rho}_a \bar{c}_{p,a}}{\bar{\rho}_f \bar{c}_{p,f} R_{f} +  \bar{\rho}_a \bar{c}_{p,a} (1-R_f)}
\end{equation}
with $\beta < 1$ in presence of fuel and $\beta = 1$ when only air is present ($R_f=0$). Therefore, this coefficient  can be regarded as a wind attenuation factor \cite{grasso2020physics}. 

In \cite{prieto2015sensitivity}, a similar expression for $\beta$ was derived without assuming thermal equilibrium between the gaseous and solid phases. This approach allowed for different values to be proposed inside and outside the flame. In the present work,  thermal equilibrium, i.e., $T_f=T_a=T$, is assumed and, as a result, $\beta$ does not depend on temperature. Furthermore, given the level of simplification in  the model, $\beta$ is treated as a constant calibration parameter. However, variations arising from the effects of moisture, through $\bar{c}_{p,f}$, and the temporal evolution of $R_f$ might not be negligible. These aspects are further discussed in Sec. \ref{sec:parametric}.

With the present formulation, a  velocity field with non-zero divergence can appear. If solving the  problem in conservative form, a correction is needed \cite{nieding_impact_2024}. Alternatively, when solving the non-conservative version as proposed no correction is required here.

Approximating the effect of topography by means of an additional advective effect is a simple approach that allows to retain in the model's solution the influence of topographic variations on wildfire propagation without introducing significant computational complexity  \cite{burger_exploring_2020,san2023gpu}. A different approach with a more meaningful physical basis would require modeling the effect of radiation globally, and not only locally through the Rosseland approximation \cite{asensio2023historical}.  Besides, there are other approaches in the literature which provide a higher level of detail in the relation between $\mathbf{w}$ and the actual wind velocity 10 m above the ground \cite{vogiatzoglou2024interpretable}.  

For further investigations, note that certain topographic influences are similar to an artificial wind velocity in this formulation. 
For example, a constant topographic slope transfers with the parameter $\gamma$ into a constant artificial wind velocity. 
If the fire is moving upslope, the artificial wind velocity is positive in this direction. If the fire moves downslope, the artificial wind velocity is negative.
Experiments indicate that the linear assumption is acceptable for topographic slopes up to $30^\circ$ \cite{sanchez-monroy_fire_2019}. Any further investigations that are carried out for a wind velocity can be transferred easily to a topography with a constant slope.

\subsection{Modeling local radiation}\label{subsec:modelling_diffusion}

In the model herein presented, only the local effect of radiation is considered, assuming that the medium is optically thick \cite{asensio2002wildland,grasso2020physics,weber_toward_1991}. The diffusion coefficient in Eq.~\eqref{eq:diffusion} includes some linear diffusion by conduction and a non-linear term due to radiation that results from the Rosseland approximation.
There is previous work where only the linear diffusion by conduction was considered \cite{mandel2008wildland,reisch_analytical_2024, weber1991modelling}. For calibrated parameters, the numerical simulations show a small influence of the non-linear diffusion by radiation, see \cite{burger_exploring_2020, weber_toward_1991}.

In this work, the influence of the radiation is investigated regarding two aspects: 
the overall influence of radiation and the dependency of the radiation term. 
The radiation depends on some fixed physical parameters, on the temperature $T$ and on the optical path length $\delta$.
The optical path length is a physical value that is difficult to estimate. The challenges are caused by the unknown structure of the vegetation and by the reduction of the spatial dimensions from three to two. Therefore,  $\delta$ is kept constant in the model following previous literature \cite{asensio2002wildland,grasso2020physics}.

The key aspect when considering local radiation resides in the strong non-linearity of the diffusion term. The dependency with the cubed temperature makes the diffusion mechanism very sensitive to this problem variable. The parametric analyses in the results section will show that this dependency is required to reproduce physically consistent results.

\subsection{Modeling combustion}\label{subsec:modelling_combustion}

A standard choice for modeling the combustion process is using the Arrhenius law \cite{weber_toward_1991}, leading to 
  \begin{equation}  \label{eq:psi_arrhenius} 
    \Psi_a(T)=s(T)A_a \exp{\left(-\frac{T_{ac}}{T}\right)},
 \end{equation} 
 with $T_{ac}$ the activation temperature, $A$ the pre-exponential factor and $s(T)$ an activation function
 \begin{equation}  \label{eq:s} 
    s(T)= \left\{ \begin{array}{ccc}
      0 & \hbox{if} & T<T_{pc} , \\
      1 & \hbox{if} & T\geq T_{pc}  ,
     \end{array}\right. 
 \end{equation} 
that triggers the combustion when the temperature is higher than the pyrolysis temperature $T_{pc}$. From this point onwards, it will be assumed that $T_{ac}=T_{pc}$. Assume also that pyrolysis and combustion occur sequentially, with gas oxidation beginning immediately after fuel devolatilization starts, thereby generating net exothermic heat. This allows $T_{pc}$ to be considered as the threshold for combustion.

Originally, molecular chemical processes motivated the Arrhenius law. In the wildfire model, the variables are acting on a much larger length scale, far away from the molecular scale. 
This motivates the approximation of the non-linear Arrhenius law by a constant function as \cite{asensio2023historical}
  \begin{equation}  \label{eq:psi_constant} 
    \Psi_c(T)=s(T) A_c.
 \end{equation} 
 The switch $s(T)$ in Eq.~\eqref{eq:s} for activating the combustion process above the ignition temperature is included for both approaches. 
Previous studies, see \cite{mitra_studying_2024}, show that the process of switching on is more important than the switching off. One can argue that the combustion processes continue even for temperature values below the ignition temperature once the fire was burning in this space.
On the other hand, the switching off after the burning process affects only the cooling process and the remaining biomass fraction.
It does not affect the maximal fire temperature nor the rate of spread of the fire. 
This paper focuses more on the rate of spread. Therefore, following the approach in \cite{reisch_analytical_2024}, the switch is modeled for the ignition and the cooling process.

Sec.~\ref{sec:simulation_combustion} compares the two approaches for the combustion function $\Psi$  by means of numerical simulations.

\subsection{Modeling fuel moisture }\label{sec:moisture}

The  bulk specific heat $\bar{c}_{p0}$ in Eq.~\eqref{eq:rhocp0} depends on the specific heat of the fuel $\bar{c}_{p,f}$. In general, the specific heat of the fuel is a function of the temperature. However, it can also be interesting to consider its dependency on the biomass fraction in order to model the effect of fuel moisture evaporation and degradation of the biomass, i.e. $\bar{c}_{p,f}=\bar{c}_{p,f}(T,Y)$. 

In this section, the apparent calorific capacity method \cite{swaminathan1993enthalpy} is used to model the effect of fuel moisture. This method models the effect of thermal phase changes by considering an apparent (or effective) heat capacity, usually defined as a piecewise constant function. Thus, the specific heat $\bar{c}_{p,f}$ will be defined as a piecewise constant function featuring an apparent specific heat to account for the effect of fuel moisture. 

Four different approaches for the definition of $\bar{c}_{p,f}$ are introduced. The first approach is referred to as NM (neglects moisture) and considers the specific heat of dry wood. The second approach is called the simple two-stage moisture model (S2M) and features a piecewise definition of the apparent specific heat with two stages in the heating process. The third approach is the three-stage moisture model (S3M) and considers three stages, being able to represent evaporation at a quasi-constant temperature. The fourth approach is called the complete two-stage moisture model (C2M), which is an extension of the S2M approach to distinguish between green and dead wood as the fuel of the fire.

\subsubsection{Neglecting moisture (NM)}

Assuming zero fuel moisture,  the specific heat for the fuel can be considered constant, i.e. identical to the specific heat of the dry fuel,
\begin{equation}
    \bar{c}_{p,f}=c_{p,f_0} \, .
\end{equation}
This simplification reduces the system in Eqs.~\eqref{eq:compactsys}--\eqref{eq:matrixform} to the temperature--biomass ADR model commonly found in the literature, see, e.g. \cite{reisch_analytical_2024}.

\subsubsection{Simple two-stage moisture model (S2M)} \label{sec:S2M}

As a first approach to model fuel moisture, the moisture content $M$ is defined as the ratio between the mass of water, $m_w$, and the mass of dry wood, $m_0$, as
\begin{equation}
    M=\frac{m_w}{m_0}.
\end{equation}

For consistency with Eq.~\eqref{eq:model1b} and to agree with literature standards \cite{alonso2020estimating,Ross_2010}, a dry basis is assumed. The increment of specific enthalpy when heating from the ambient temperature, $T_\infty$, to the evaporation temperature of water, $T_w$, before evaporation (sensible heating) is defined as 
\begin{equation}\label{eq:dh_01}\Delta h_{\mathrm{sen},1}:=\frac{\Delta H_{\mathrm{sen},1}}{\rho_{f_0}}=c_{p,f_0}(T_w-T_\infty)+{M}c_w(T_w-T_\infty) ,
\end{equation}
where $\rho_{f_0}$ is the density of the dry fuel, $c_{p,f_0}$ is the specific heat of the dry fuel and $c_w$ is the specific heat of water. For simplicity, assume $\rho_{f_0}\approx \bar{\rho}_f$, which was defined in Sec.~\ref{sec:mathmodel} as an input parameter. Accordingly, the increment of specific enthalpy  during evaporation (latent heating) is defined as
\begin{equation}\label{eq:dh_02}\Delta h_\mathrm{lat}:=\frac{\Delta H_\mathrm{lat}}{\rho_{f_0}}= M L_w \, ,
\end{equation}
where $L_w$ is the specific latent heat of water. In the same way, the increment of specific enthalpy when heating from $T_w$ to $T_{pc}$ after evaporation (sensible heating) is  
\begin{equation}\label{eq:dh_03}\Delta h_{\mathrm{sen},2}:=\frac{\Delta H_{\mathrm{sen},2}}{\rho_{f_0}}=c_{p,f_0}(T_{pc}-T_w)
\end{equation}
with the pyrolysis temperature $T_{pc}$.

In the complete heating process, i.e. when heating from $T_\infty$ to $T_{pc}$ (sensible and latent heating), the increment of specific enthalpy   is defined as 
\begin{align}\label{eq:dhsimple}
\Delta h &:=\frac{\Delta H}{\rho_{f_0}}=\Delta h_{\mathrm{sen},1} + \Delta h_{\mathrm{sen},2} +\Delta h_\mathrm{lat} \\
&=c_{p,f_0}(T_{pc}-T_\infty)+{M}c_w(T_w-T_\infty) + ML_w . 
\end{align}

The effective specific heat  for the complete heating process from $T_\infty$ to $T_{pc}$ is then
\begin{equation}\label{cpeffsimple}
  c_{p, \mathrm{eff}}:= \frac{\Delta H}{\rho_{f_0}(T_{pc}-T_\infty)}=\frac{\Delta h}{T_{pc}-T_\infty}, 
\end{equation}
leading to
\begin{equation}\label{cpeffsimple2}
  c_{p, \mathrm{eff}}= c_{p,f_0}  + M\left[\frac{c_w(T_w-T_\infty) + L_w }{(T_{pc}-T_\infty)}\right] \, .
\end{equation}
Recall that $c_{p, \mathrm{eff}}$ in Eq.~\eqref{cpeffsimple2} is defined in a dry basis, for consistency with Eq.~\eqref{eq:model1b}.

The actual fuel-specific heat is then defined as a piecewise constant function of the temperature by
 \begin{equation}  \label{eq:cpfsimple} 
  \bar{c}_{p,f}= \left\{ \begin{array}{ccccl}
      c_{p, \mathrm{eff}} & \hbox{if} & T<T_{pc} & \hbox{ and }  & Y=1 ,\\
      c_{p,f_0} & \hbox{if} & T\geq T_{pc}  &  \hbox{ or } & Y<1 . 
     \end{array}\right. 
 \end{equation}

\begin{figure}
    \centering
    \begin{subfigure}{0.49\textwidth}
    \includegraphics[width=\linewidth]{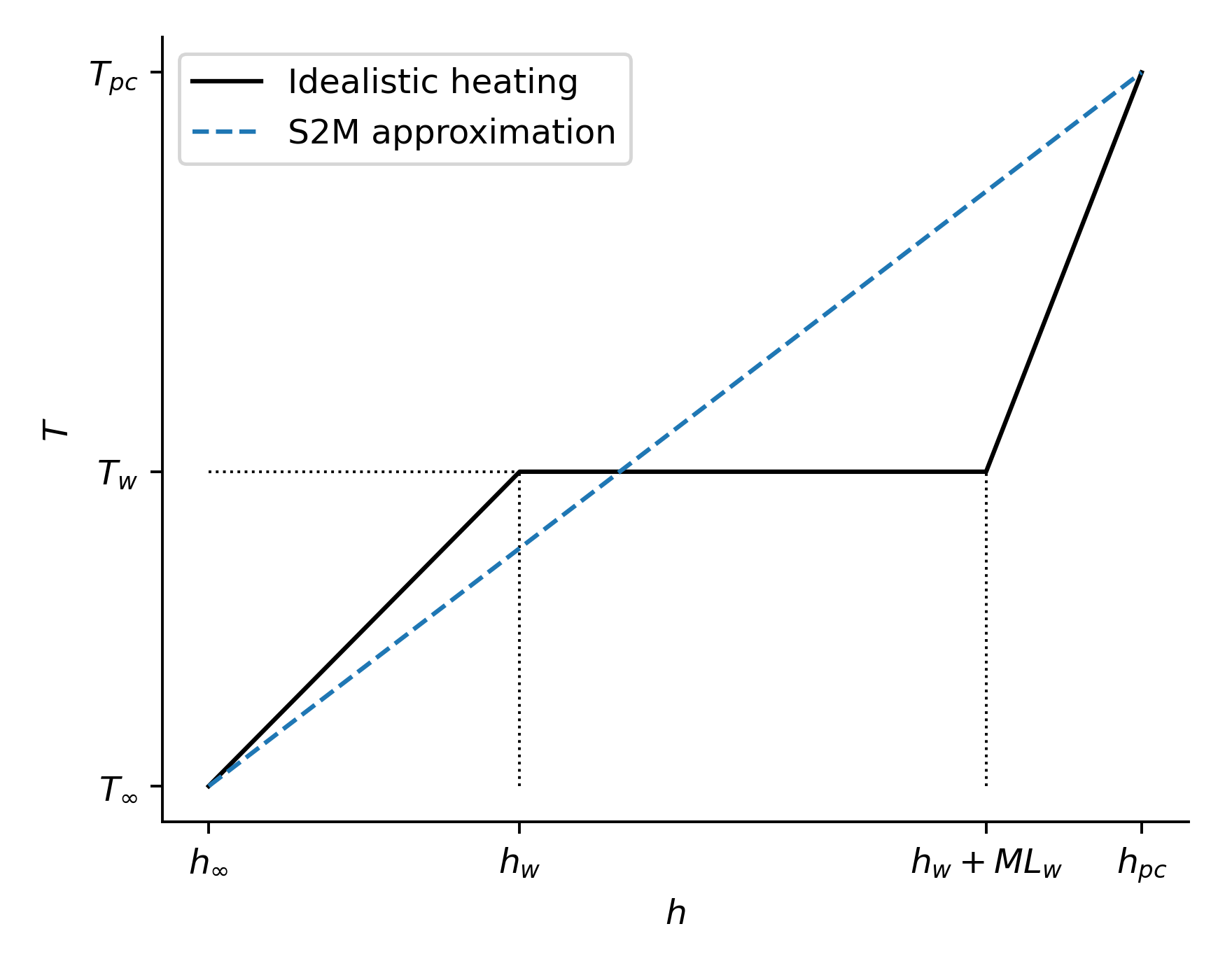} 
    \caption{Idealized and S2M-approximated $h$-$T$-curve.\label{fig:heatingS2M}}
    \end{subfigure}
    \begin{subfigure}{0.49\textwidth}
    \includegraphics[width=\linewidth]{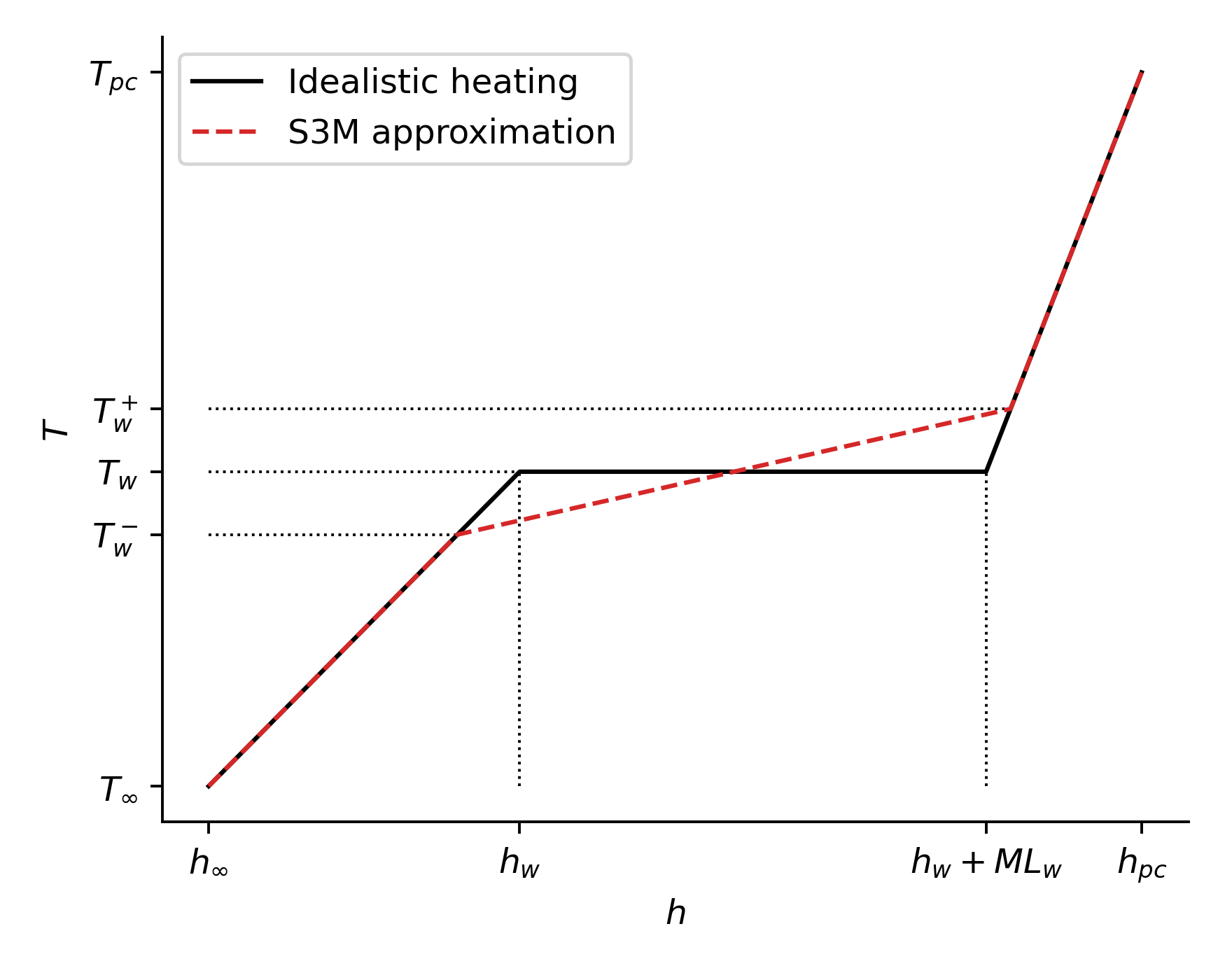}        
    \caption{Idealized and S3M-approximated $h$-$T$-curve.\label{fig:heatingS3M}}
    \end{subfigure}
    \caption{ 
    Plot of idealized, S2M-approximated, and S3M-approximated $h$-$T$-curves. Note the relations $\Delta h_{\mathrm{sen},1}=h_w-h_{\infty}$ and $\Delta h_{\mathrm{sen},2}=h_{pc}-(h_w+ML_w)$.}

\end{figure}
 
\subsubsection{Simple three-stage moisture model (S3M)}

Using the ideas in \cite{caggiano2018reviewing,voller1990fixed}, we consider an artificial finite-width evaporation region in the $h$-$T$ diagram, where the initial and final evaporation temperatures are
\begin{equation}\label{eq:deltas1}
T_{w}^-=T_w-0.5\Delta T_w\,,\qquad T_{w}^+=T_w+0.5\Delta T_w \, ,
\end{equation}
with $\Delta T_w$ the width of the evaporation interval.

An approximate enthalpy change for the sensible heating process is then defined by
\begin{equation}\label{eq:dhw1}
\Delta \tilde{h}_{\mathrm{sen},1}=c_{p,f_0}(T_{w}^--T_\infty)+{M}c_w(T_{w}^- -T_\infty),
\end{equation}
as well as an approximate enthalpy change for the latent heating process
\begin{equation}\label{eq:dhw2}
\Delta \tilde{h}_\mathrm{lat}=c_{p,f_0}(T_{w}^+ - T_{w}^-)+{M}\left(c_w(T_{w} -T_{w}^-)  + L \right) .
\end{equation}
The apparent  specific heat for the sensible heating process is 
\begin{equation}\label{eq:cpeff21}
  c_{\mathrm{sen},1}= \frac{\Delta \tilde{h}_{\mathrm{sen},1}}{T_{w}^--T_\infty}, 
\end{equation}
whereas the apparent specific heat for the latent heating process is 
\begin{equation}\label{eq:cpeff22}
  c_{\mathrm{lat}}= \frac{\Delta \tilde{h}_\mathrm{lat}}{T_{w}^+ - T_{w}^-}.
\end{equation}

The actual fuel-specific heat used in the model is expressed as a piecewise constant function that leads to a different apparent specific heat for the sensible heating, latent heating and sensible heating after evaporation processes, as 
 \begin{equation}  \label{eq:cpeff3} 
  \bar{c}_{p,f}= \left\{ \begin{array}{ccccl}
      c_{\mathrm{sen},1} & \hbox{if} & T\leq T_{w}^- &\hbox{ and } &Y=1, \\
      c_\mathrm{lat} & \hbox{if} &  T_{w}^-  < T \leq T_{w}^+ &\hbox{ and } &Y=1,\\
      c_{p,f_0} & \hbox{if} & T\geq T_{w}^+ & \hbox{ or } &Y<1 .
     \end{array}\right. 
 \end{equation}

\subsubsection{Complete two-stage moisture model (C2M)}\label{sec:C2M}

The amount of water in the wood affects the process of heating the fuel from ambient temperature ($T_\infty$) to the pyrolysis temperature ($T_{pc}$). In the S2M model,  the fuel accumulates water homogeneously. Although this simplification can be useful in certain situations, in a more realistic approach the moisture content depends on the type of fuel which, for simplicity, will be referred here as wood.  In this regard, two types of wood are distinguished: dead and live (or green) wood. The second type refers to the trees or bushes which are still alive. 

The reason for making this classification is that their interaction with water, and therefore their moisture content, differs completely. Dead wood moisture content depends directly on temperature and relative humidity, whereas live wood moisture content depends essentially on the tree species and is independent of the environmental conditions.  
In the C2M model, the fuel will be modeled as a mixture of green and dead wood. Therefore, fuel mass will be defined as follows
\begin{equation}
    m_f=r m_g+(1-r) m_d,
\end{equation}
where $m_g$ and $m_d$ are the masses of green and dead wood, respectively, and $r\in [0,1]$ is the ratio of green wood to the wood mixture. The effective moisture content of the mixture is 
\begin{equation}
    M=r M_g+(1-r) M_d,
\end{equation}
where $M_g$ and $M_d$ are the moisture content of the green wood and dead wood, respectively. 

The moisture content of dead wood is a function of both, relative humidity and temperature
of the surrounding air. In wood, water is accumulated in two ways: in the lumens (as free water) or at the cell walls, forming a bound state as it interacts with the polysaccharides of the membrane of the cells. Water is accumulated in the lumens only after the fiber saturation point is reached, which is the threshold at which no more water can be held at the cell walls. The fiber saturation point is similar for all types of wood, and occurs when the moisture content is around 0.3 \cite{Ross_2010}. Green wood is generally above the fiber saturation point. Thus, the study of the behavior of wood below the fiber saturation point implicitly refers to dead wood. 

Another important concept for dead wood is the equilibrium moisture content (EMC), which is defined as the moisture content at which the wood is neither gaining nor losing water. If the environmental conditions change, wood does not reach immediately the EMC. In fact, the rate  at which dead wood exchanges moisture with the atmosphere to reach EMC depends crucially on the size of the sample. Dead fuel is separated into fine fuels –leaves, bark and twigs with a diameter lower than 6 mm– and coarse fuels---larger twigs and logs \cite{matthews2013dead}. The US National Fire Danger Rating System \cite{burgan19881988} classifies fuel sizes according to the time it takes to reach equilibrium with the atmosphere. Fine fuels are within the 1-hour fuel class, whereas coarse fuels are within the 10-, 100- and 1000-hour classes. However, in this study no variation in time of the relative humidity is assumed and thus $M_d$ can be approximated by the EMC; for a more complete approach representing time variations, see for example \cite{perello2024adaptable}. Tab.~\ref{tab:my_label} gives values of EMC, hereafter $M_d$, at $T=300K$. 

\begin{table}
    \centering
       \caption{The moisture content $M_d$ of dead wood depending on the relative humidity (RH) of the environment for fixed temperature ($T=300K$). 
    }
      \label{tab:my_label}
       \begin{tabular}{c|c c c c c c c c c}
        RH (\%) &10 &20&30&40&50&60&70&80&90 \\
        \hline
        $M_d$ &0.024 &0.044&0.061&0.076&0.091&0.108&0.129&0.157&0.202\\
    \end{tabular}
\end{table}

In general, values of $M_d$ are independent of the wood species and can be approximated with the empiric formula \cite{Ross_2010}
\begin{equation}\label{eq:ms}
    M_d=\frac{18}{W}\bigg(\frac{K\phi}{1-K\phi}+\frac{K_1K\phi+2K_1K_2K^2\phi^2}{1+K_1K\phi+K_1K_2K^2\phi^2}\bigg),
\end{equation}
where $\phi\in [0,1]$ is the relative humidity of the air. For temperature in Celsius, the constants in Eq.~\eqref{eq:ms} are
\begin{equation}
\left\{
\begin{aligned}
    W&=349+1.29\,T+0.0135\,T^2 , \\ 
    K&=0.805+0.000736\,T-0.00000273\,T^2, \\ 
    K_1&=6.27-0.00938\,T-0.000303\,T^2, \\ 
    K_2&=1.91+0.04007\,T-0.000293\,T^2.
\end{aligned}
\right. 
\end{equation}

As in Eq.~\eqref{eq:dhsimple}, the enthalpy balance is
\begin{equation}\label{deltahp}
\Delta h=c_{p}(T_w-T_\infty) + M L_w+ c_p'(T_{pc}-T_w ),
\end{equation}
where $c_{p}$ is the specific heat of the mixture of moist fuels.
Then, the effective specific heat adapted to the initial conditions of environment and wood is
\begin{equation}\label{cpeff}
  c_{p, \mathrm{eff}}= \frac{\Delta h}{T_{pc}-T_\infty}, 
\end{equation}
and the approach
\begin{equation}
    \Delta h=c_{p, \mathrm{eff}} (T-T_\infty),
\end{equation}
is displayed as a straight blue dashed line in Fig. \ref{fig:heatingS2M}.

The heat capacity of wood depends on the temperature and moisture content of the wood, but it is practically independent of density or species. In order to compute $c_p$ in Eq.~\eqref{deltahp}, referred to the mixture of wood from the value of the moisture content $M_{g}$ and $M_d$, the expressions 
\begin{equation}\label{eq:cpmoistures}
\left\{
\begin{aligned}
  c_p&=r\, c_{p,g}+(1-r)\,c_{p,d}, \\ 
c_{p,d}&=c_{p,f_0}+M_{d}c_w+s'(M_d) A_d, \\
c_{p,g}&=c_{p,f_0}+M_{g}c_w+s'(M_g) A_g 
\end{aligned}
\right.
\end{equation}
are used, where $A_i$ is a correction factor required below fiber saturation which accounts for the water accumulated at the cell walls of the wood where, as said above, it forms a bound state. An empirical formula for the value of $A_i$ depending on the temperature and moisture content is  \cite{Ross_2010}
\begin{equation}
    A_i=M_i(b_1+b_2T+b_3M_i)(1+M_i),
\end{equation}
with $b_1 = -0.06191, b_2 = 2.36 \cdot 10^{-4}$, and $b_3 = -1.33\cdot 10^{-4}$, 
with temperature given in Kelvin and with $i=d$ for dead wood and $i=g$ for green wood. The function $s'(M)$ in Eq.~\eqref{eq:cpmoistures} is a switch 
 \begin{equation}  \label{eq:sprime} 
    s'(M_i)= \left\{ \begin{array}{ccc}
      1 & \hbox{if} & M_i<0.3 , \\
      0 & \hbox{if} & M_i\geq 0.3  ,
     \end{array}\right. 
 \end{equation} 
to be activated below fiber saturation ($M_i<0.3$) where the correction factor $A_i$ needs to be taken into account. Above fiber saturation ($M_i\geq0.3$) a mixture rule is used to estimate the specific heat. For green wood, it is known that is $M_g>0.3$ \cite{Ross_2010}, therefore the switch is zero, $s'(M_g)=0$.

Finally, the specific heat of the fuel is computed as 
 \begin{equation}  \label{eq:cpf} 
  \bar{c}_{p,f}= \left\{ \begin{array}{ccc}
      c_{p, \mathrm{eff}} & \hbox{if} & T<T_{pc} , \\
      c'_p & \hbox{if} & T\geq T_{pc}  ,
     \end{array}\right. 
 \end{equation} 
and will be used in Eq.~\eqref{eq:cpdefY} for the C2M model.

\subsection{Model parameters}

This section summarizes the parameters required in the model. Tab.~\ref{table:general-parameters1} includes environmental (ambient forcing) and fuel-dependent parameters. Although these parameters may exhibit spatial heterogeneity, the table presents their default values. Tab.~\ref{table:general-parameters2} lists additional model parameters which, despite having a physical basis, are treated as calibration parameters and should not vary with spatial position. Among these are the optical path length, wind and slope correction factors, and the ambient cooling coefficient. Note that the latter does not correspond to a physically realistic value, as it also accounts for the energy loss owing to the mass escaping through the top of the domain, as well as for heat dissipation by radiation in the vertical direction, which is not explicitly represented in the model. Finally, Tab.~\ref{table:general-parameters3} contains other parameters and physical constants, which are fixed.

The default parameter values in the tables are based on previous studies and preliminary calibration. Rounded values are used, consistent with the order of magnitude found in the reference sources provided in the tables.

\begin{table}
\centering
\caption{Spatially dependent input data for the model, including the ambient forcing (top), basic fuel parameters (middle) and complete fuel moisture parameters (bottom). $^\dagger$ In a dry basis.  $^\mathsection$ For S2M and S3M models. }
\label{table:general-parameters1}
\begin{tabular}{lcccc}
Parameter & Symbol  & Default value & Units & Ref. \\
\hline
Ground elevation  & $Z$ & - &\hbox{m} & - \\
Wind velocity  & $\mathbf{w}$ & - &\hbox{m·s$^{-1}$} & -\\
\hline
Surface coverage & $SC$ & 1.0 &\hbox{-}  & -\\
Initial fuel volume ratio & $R_{f,0}$  & 0.01 &\hbox{-}  & \cite{sero2008large,margerit2002modelling}\\
Fuel density$^\dagger$  & $\bar{\rho}_{f}$ & 400 &\hbox{kg·m$^{-3}$}  & \cite{mell2007physics}  \\
Initial fuel load$^\dagger$  & $W_{f,0}$ & 4 &\hbox{kg·m$^{-2}$}  & \cite{margerit2002modelling} \\
Specific heat of dry fuel  & $c_{p,f0}$ & 1.0 &\hbox{kJ·kg$^{-1}$·K$^{-1}$} & \cite{mell2007physics}\\
Fuel pyrolysis temperature  & $T_{pc}$ & 600 &\hbox{K} & \cite{grasso2020physics} \\
Fuel combustion heat  & $\mathcal{H}$ & 4000 &\hbox{kJ·kg$^{-1}$} & -\\
Reaction rate  & $A_c$ & 0.01 &\hbox{s$^{-1}$} & -\\
Pre-exponential factor & $A_a$ & 0.0173 &\hbox{s$^{-1}$} & -\\
Fuel moisture content$^\mathsection$ & $M$ & 0.1 &\hbox{-} & \cite{Ross_2010}\\
\hline
Live (green) fuel moisture & $M_g$ & 0.9 &\hbox{-} & \cite{Ross_2010}\\
Live to dead fuel ratio & $r$ & 1.0 &\hbox{-} & -\\
Relative humidity & $\phi$ & 0.3 &\hbox{-} & -\\
\hline
\end{tabular}
\end{table}

\begin{table}
\centering
\caption{Calibration parameters. }
\label{table:general-parameters2}
\begin{tabular}{lcccc}
Parameter & Symbol   & Default value & Units& Ref.\\
\hline
Thermal conductivity  & $k_c$ & 0.0001&\hbox{kW·m$^{-1}$· K$^{-1}$} & \cite{grasso2020physics}\\
Optical path length & $\delta$ & 1.0 & \hbox{m} & \cite{grasso2020physics,margerit2002modelling}\\
Ambient cooling coefficient  & $\hat{\alpha} $ & 0.05&\hbox{kW·m$^{-2}$·K$^{-1}$}  & \cite{margerit2002modelling}\\
Wind correction coefficient  & $\beta$ & 0.02 &\hbox{-}  & - \\
Slope correction coefficient  & $\gamma$ & 0.04 &\hbox{-} &- 
\end{tabular}
\end{table}

\begin{table}
\centering
\caption{ Other parameters and physical constants.}
\label{table:general-parameters3}
\begin{tabular}{lccc}
Parameter & Symbol   & Value & Units \\
\hline
Ambient temperature & $T_{\infty}$ & 300 &\hbox{K} \\
Homogenisation height & $l$ & 1 &\hbox{m} \\
Evaporation temperature of water & $T_{w}$ & 373&\hbox{K} \\
Latent heat of water & $L_{w}$ & 2257&\hbox{kJ·kg$^{-1}$} \\
Specific heat of water  & $c_{w}$ & 4.19&\hbox{kJ·kg$^{-1}$·K$^{-1}$} \\
Air density  & $\rho_a$ & 1.0 & \hbox{kg·m$^{-3}$} \\
Stephan Boltzmann constant   & $\sigma$ & 5.670E-11 & \hbox{kW·m$^{-2}$·K$^{-4}$} \\
Emissivity factor (vegetation)   & $\epsilon$ & 0.9 & \hbox{-} \\
\end{tabular}
\end{table}

\section{Numerical model}\label{sec:numerics}

The numerical model is based on a Finite Volume Scheme in space and two variants of time stepping. 
We introduce the used methods here, and give numerical convergence results in App.~\ref{appendix}.

\subsection{Finite Volume spatial discretization}\label{subsec:fvm}

The initial--boundary value problem in Eqs.~\eqref{eq:model1b}--\eqref{eq:ccb} is defined in the domain $ [0,t_f] \times \Lambda$, where $\Lambda= [x_1,x_2] \times [y_1,y_2] $ is the spatial domain.  Using the finite volume approach, the spatial domain is discretized in $N_x \times N_y $  cells, defined as
\begin{equation}
	\Lambda_{ij}=\left[x_{i-\frac{1}{2}},x_{i+\frac{1}{2}}\right] \times \left[y_{j-\frac{1}{2}},y_{j+\frac{1}{2}}\right] \,, 
\end{equation}
with $i=1,...,N_x,\, j=1,...,N_y$. We consider a Cartesian grid, with a grid spacing $\Delta x$ and $ \Delta y$ in each of the Cartesian directions.  At time $t^n$,  the conserved quantities are defined as cell averages as
\begin{equation}  \label{cellaverage1} 
{\mathbf{{U}}_{ij}^n}=\frac{1}{\Delta x \Delta y }\int_{\Lambda_{ij}}\mathbf{{U}}(\mathbf{x},t^n)\, \mathrm{d}x \, \mathrm{d}y   \,.
\end{equation} 
The semi-discrete form of Eq.~\eqref{eq:compactsys} is written as
\begin{equation}\label{eq:semi}
	  \frac{\partial \mathbf{{U}}_{ij}}{\partial t}   = \mathcal{L}(\mathbf{{U}}_{ij}) ,
\end{equation}
where $\mathcal{L}(\mathbf{{U}}_{ij})$ is the following discrete operator
\begin{equation}\label{semi2}
 \mathcal{L}(\mathbf{{U}}_{ij})   = -   \frac{ (\mathbf{A}\Delta_x\mathbf{{U}})_{i,j} }{\Delta x}  -   \frac{ (\mathbf{B}\Delta_y\mathbf{{U}})_{i,j} }{\Delta y} +    \mathbf{C}_{ij}\left(\frac{\mathbf{{F}}_{i+1/2,j}  - \mathbf{{F}}_{i-1/2,j} }{\Delta x}  + \frac{\mathbf{{G}}_{i,j+1/2}  - \mathbf{{G}}_{i,j-1/2}}{\Delta y} +  \bar{\mathbf{S}}_{ij} \right)\, ,
\end{equation}
where $(\mathbf{A}\Delta_x\mathbf{{U}})_{i,j}$ and  $(\mathbf{B}\Delta_y\mathbf{{U}})_{i,j}$ are the non-conservative products, $ \mathbf{{F}}_{i\mp1/2,j}$ and  $\mathbf{{G}}_{i,j\mp 1/2}$ are the diffusive fluxes at cell interfaces and
\begin{equation}\label{sourcetermdef}
  \bar{\textbf{S}}_{ij} \approx  \frac{1}{\Delta x \Delta y }\int_{\Lambda_{ij}} {\textbf{S}}(\mathbf{{U}}) \, \mathrm{d}x \,\mathrm{d}y    
\end{equation}
is the approximation of the spatial integral of the source terms inside cells, yet to be defined. Note that $\mathbf{C}_{ij}$ is the evaluation of $\mathbf{C}$ using cell-averaged data at $(i,j)$.

The non-conservative products  are only non-zero for the first equation and are computed using an upwind formulation based on a high-order (7-th order) WENO reconstruction \cite{shu1998,reisch_analytical_2024}. The diffusive fluxes are only non-zero for the first equation, that is
 \begin{equation}  \label{eq:numfluxes} 
     \mathbf{{F}}_{i\mp1/2,j}=\left( \begin{array}{c}
       {{F}}_{i\mp1/2,j} \\
       0\\  
        \end{array}\right) ,\,\, \;
    \mathbf{{G}}_{i,j\mp 1/2}=\left( \begin{array}{c}
      {{G}}_{i,j\mp 1/2}  \\
       0 \\ 
    \end{array}\right)  \,\, \; .
  \end{equation}
These fluxes are approximated by means of second order central differences. For instance, let us consider the approximation of the  flux in the $x$ direction. In what follows, the subscript $j$ will be omitted for the sake of simplicity. At the interface $i+1/2$, the diffusive flux ${{F}}_{i+1/2}$ is  computed as 
 \begin{equation}  \label{eq:numfluxes2} 
       {{F}}_{i+1/2}= - \widetilde{k}\left(\frac{T_{i+1}-T_{i}}{\Delta x} \right)
  \end{equation}
where $\widetilde{k}=0.5( k_{i+1}+k_{i})$, so it is the mean of the non-linear diffusion in Eq.~\eqref{eq:diffusion}.

The source terms are approximated as 
 \begin{equation}  \label{eq:sources1} 
      \bar{\textbf{S}}_{ij}=\left( \begin{array}{c}
        - \alpha(T_{ij}-T_{\infty})  +  \Psi(T_{ij})\rho_{0,ij} H Y_{ij}  \\
         - \Psi(T_{ij}) Y_{ij} \\  
        \end{array}\right) .
  \end{equation}

\subsection{Time stepping}

The Strong Stability Preserving Runge–Kutta 3 (SSP-RK3) scheme \cite{ghosh2016well,sanKara2015} for stepping in time the semi-discrete Eq.~\eqref{eq:semi} is used. It is expressed as follows

\begin{align}\label{eq:nummeth_RK3}
	\begin{aligned}
		 \mathbf{{U}}_{ij}^{(1)} &=  \mathbf{{U}}_{ij}^n + \Delta t\mathcal{L} ( \mathbf{{U}}_{ij}^n), \\
		\\		
		 \mathbf{{U}}_{ij}^{(2)} &= \frac{3}{4} \mathbf{{U}}_{ij}^n + \frac{1}{4} \mathbf{{U}}_{ij}^{(1)} + \frac{1}{4}\Delta t\mathcal{L} ( \mathbf{{U}}_{ij}^{(1)}), \\
		\\
		 \mathbf{{U}}_{ij}^{n+1} &= \frac{1}{3} \mathbf{{U}}_{ij}^n + \frac{2}{3} \mathbf{{U}}_{ij}^{(2)} + \frac{2}{3}\Delta t\mathcal{L} ( \mathbf{{U}}_{ij}^{(2)}),
	\end{aligned}
\end{align}
 where $\Delta t$ is the time step, which is computed dynamically according to the Courant--Friedrichs--Lewy ($\mathsf{CFL}$) stability condition to preserve the stability of the solution \cite{courant1967partial}.

\section{Numerical results}\label{sec:numericalresults}

One-dimensional (1D) cases are first considered to assess the influence of different parameters. 
For the 1D simulations presented herein, the computational domain is $[0,500]$ m and the initial conditions are
\begin{equation}\label{eq:ic}
   T(x,0) = 
		\left\{
		\begin{array}{lll}
      	670 \mbox{ K}  &\mbox{if} &  245< x <255,  \\
        300 \mbox{ K} & \multicolumn{2}{l}{\mbox{otherwise}}  \\  
		\end{array}
		\right. \text{and} \qquad Y(x,0)=1.0.
\end{equation}
For all cases, the computational mesh is composed of $N_x=1600$ cells and $\mathsf{CFL}=0.1$. Default parameters from Tabs.~\ref{table:general-parameters1}--\ref{table:general-parameters3} are used, unless other configuration is specified. 

\subsection{Comparison of combustion functions}\label{sec:simulation_combustion}

\begin{figure}[hb]
\begin{subfigure}{0.53\textwidth}
        \includegraphics[width=\linewidth]{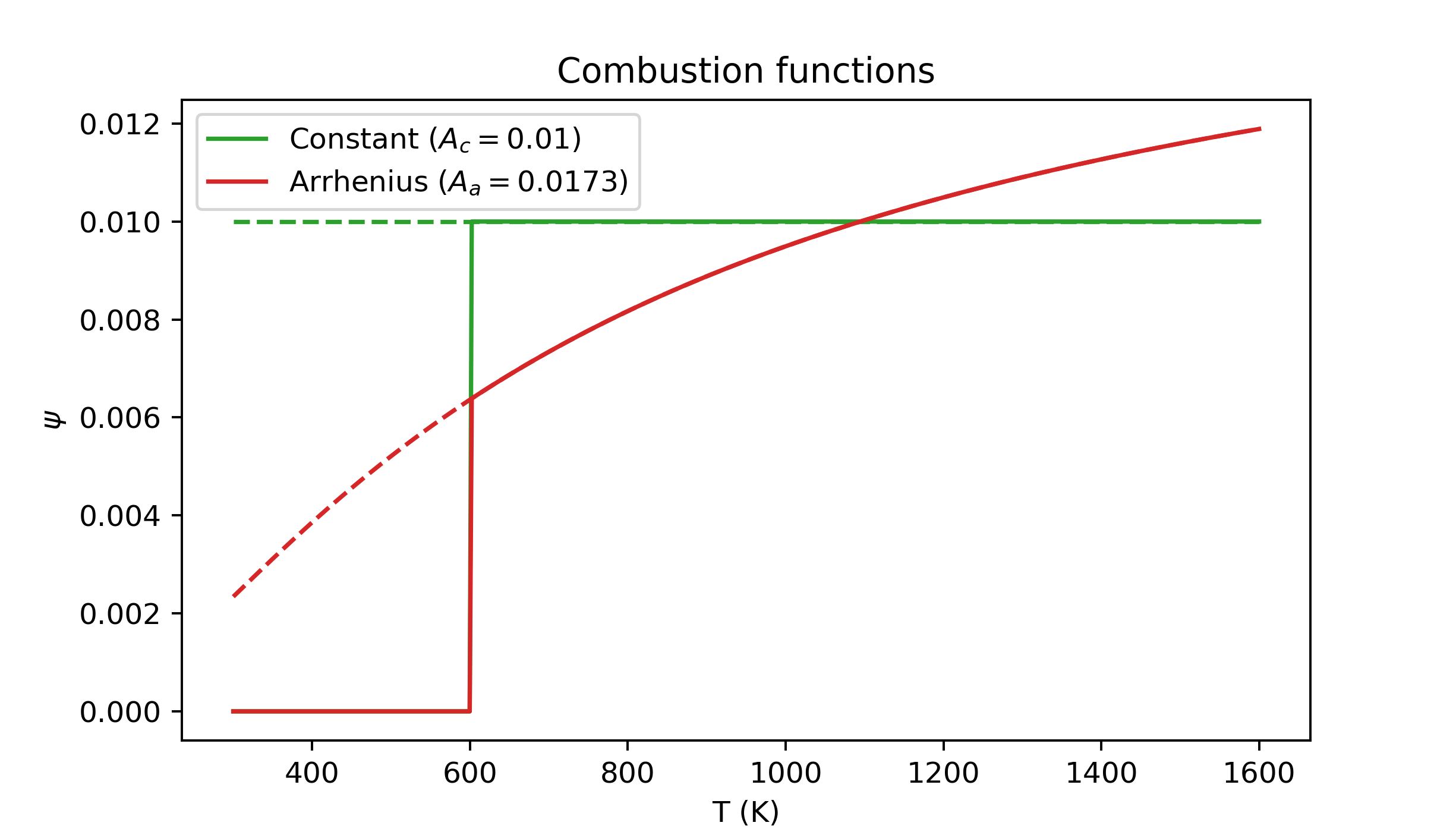}
        \caption{Combustion functions}
            \label{fig:combustion_funct_comparison}
\end{subfigure}
\begin{subfigure}{0.455\textwidth}
    \includegraphics[width=\linewidth]{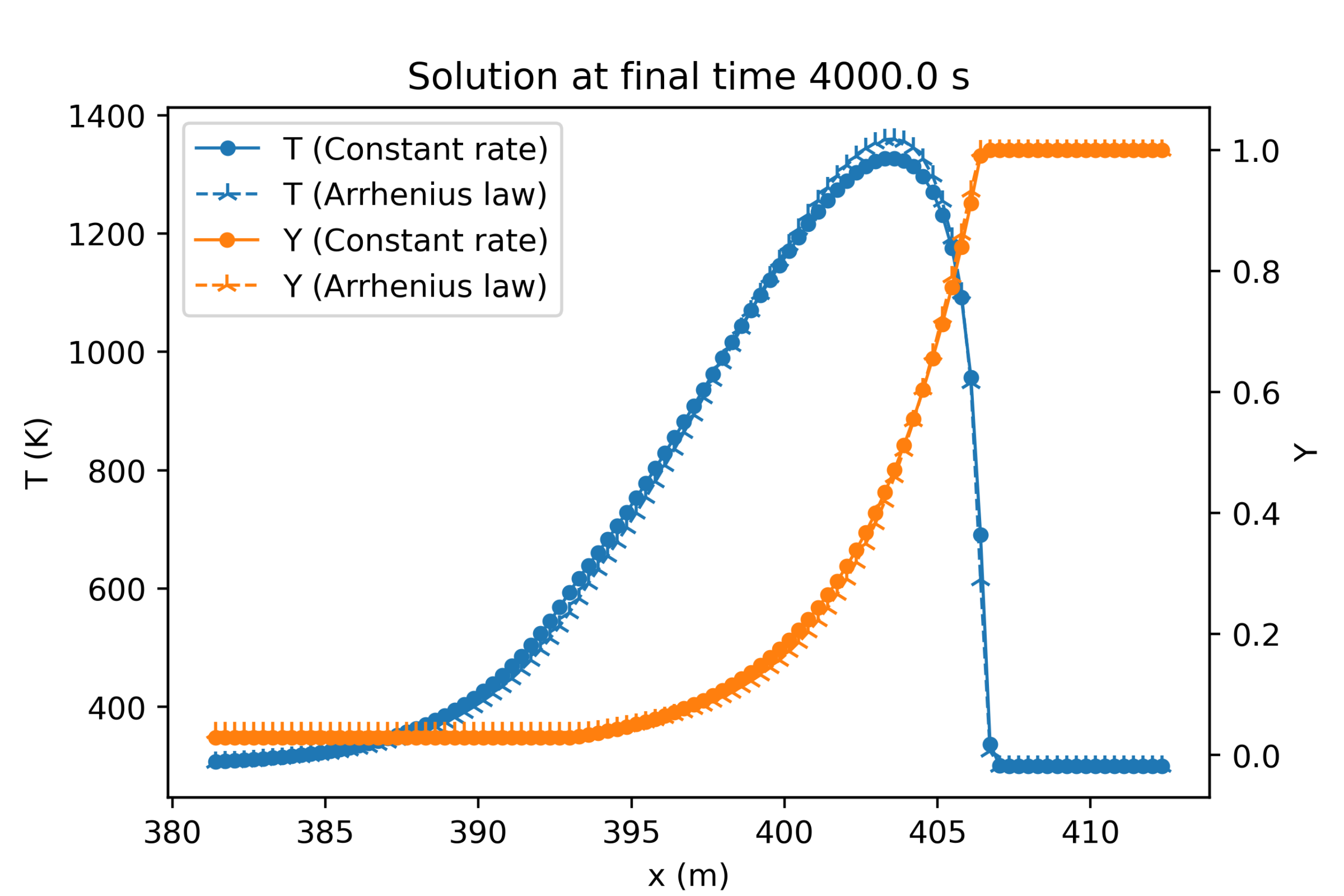}
    \caption{Solutions for $M=0.1$, $w=0$}
        \label{fig:sol_combustion}
    \end{subfigure}
    \caption{Comparison of the simulations using the two combustion functions. (a) Comparison of the combustion function using the Arrhenius law in Eq.~\eqref{eq:psi_arrhenius} and with a constant term in Eq.~\eqref{eq:psi_constant}. (b) Solutions of the ADR model~\eqref{eq:model1b} using the two combustion functions. The graphs widely overlap.}
    \label{fig:comparision_combustion}
\end{figure}

In Sec.~\ref{subsec:modelling_combustion},  two approaches for the combustion function were presented. 
Using the fixed parameters in Tab.~\ref{table:general-parameters3}, the parameter $A_c$ of the point-wise constant combustion function in Eq.~\eqref{eq:psi_constant} is determined such that it approximates the non-linear Arrhenius law Eq.~\eqref{eq:psi_arrhenius} in the relevant temperature domain. 
Fig.~\ref{fig:combustion_funct_comparison} shows this comparison.
The relevant temperature domain is $[T_{pc}, T_\mathrm{up}]$, where $T_\mathrm{up}$ approximates an upper bound for temperatures in a fire. 

For comparing the influence of the combustion function, the advection term is set zero ($w=0$) and  the solutions of the ADR-model are computed for a moisture level $M=0.1$. 
The direct comparison of the solutions in Fig.~\ref{fig:sol_combustion} shows that both, the traveling wave profile and the traveling wave speed are almost identical for the two combustion functions.
The maximal temperature is slightly higher for the combustion function using the Arrhenius law. This difference has its origin in the approximation in Fig.~\ref{fig:combustion_funct_comparison}, where the constant combustion function has a lower value for large temperatures than the Arrhenius combustion function. 
The differences of the solutions can be reduced even further by applying calibration techniques for the parameter $A_c$. Because of the good approximation, the assumption of a constant reaction rate instead of the Arrhenius rate is convincing. 

\begin{remark}
    Compared to other simplifications in the modeling process, the assumption of the Arrhenius reaction rate that is based on molecular reactions is a change of precision. 
    The model assumption of a constant reaction rate~\eqref{eq:psi_constant} is therefore a valid assumption for a macroscopic wildfire model. 
\end{remark}
In the further investigations, the piecewise constant reaction rate in Eq.~\eqref{eq:psi_constant} is used instead of the non-linear Arrhenius law. 

\subsection{Influence of radiation models}

The radiation leads to a non-linear diffusion term, compare Eqs.~\eqref{eq:heatflux3} and \eqref{eq:diffusion}. 
When studying the influence of various mechanisms on the system's behavior in \cite{reisch_analytical_2024}, the non-linear diffusion by radiation was not considered, and the diffusion was modeled only by conduction leading to a linear diffusion term with a modified larger constant $k_c$. Here, the differences in solutions caused by the simplification of a purely linear diffusion compared to the combined effect of radiation and conduction are demonstrated first. 

\begin{figure}[hbt]
    \centering
    \includegraphics[width=0.6\linewidth]{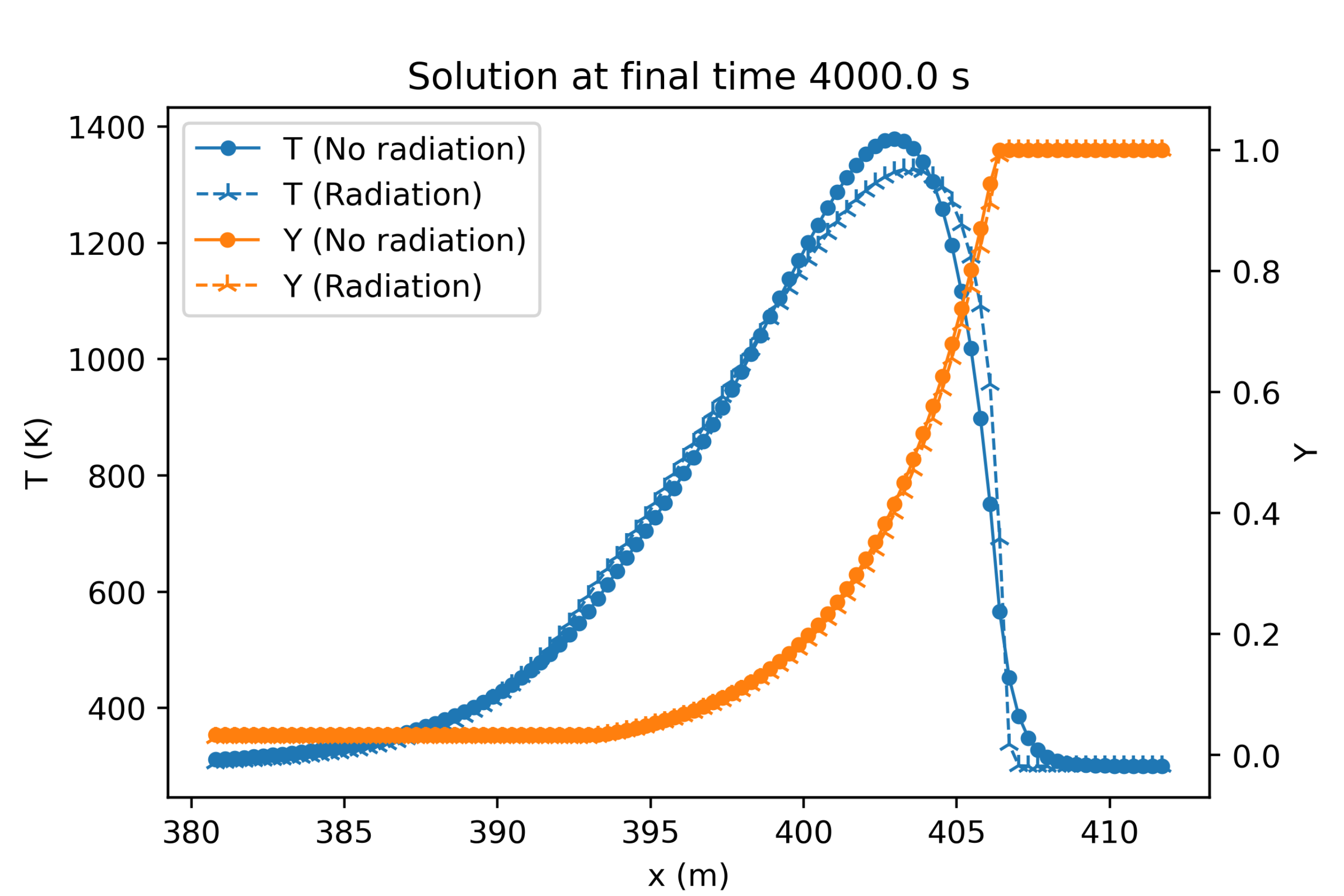}
    \caption{Comparison of the solution profiles for the temperature $T$ and the fuel $Y$ at a fixed time with and without radiation. The parameter $k_c$ differs in the two cases for compensating the lack of radiation. }
    \label{fig:profile_radiation}
\end{figure}

Fig.~\ref{fig:profile_radiation} shows the comparison of the solutions at a time $t$ for both cases, with and without radiation. 
In the case without radiation, the conduction parameter $k_c$ was modified for providing a better fit to the model with radiation, namely $k_c=0.1662~\text{kW\,m\textsuperscript{-1}K\textsuperscript{-1}}$. 
The temperature progression in the cooling process shows small differences between the models, but the steep traveling wave front and the rate of spread are almost identical. 
Fig.~\ref{fig:profile_radiation} could give the impression that it is not relevant to include the radiation. 
Further investigations in Sec.~\ref{sec:simulation_moisture} show the relevance of this term besides the physical derivation of the model:
The non-linear radiation term depends on $T^3$, leading to large values for high temperatures. Large diffusive effects are connected to a faster traveling wave speed, compare the analytical approximations in \cite{reisch_analytical_2024}. 
A higher traveling wave speed leads to a lower biomass consumption during the shorter combustion time. Further simulations will show that the different mechanisms level each other in a meaningful way. Besides, some parametric analyses in the next sections  will evidence that the non-linear dependency on the temperature is required to yield physically consistent results.

\subsection{Influence of fuel moisture models}

The different models for the fuel moisture in Sec.~\ref{sec:moisture} are now compared for some specific scenarios.

\subsubsection{Comparison of three moisture models}

First, consider a green wood to total wood ratio $r=1$, meaning that all wood is green wood, and a moisture content of $M=0.1$ under no wind conditions. The rest of the parameters are set as default.
In this case, the only difference between the simple and the complete two-stage moisture models S2M and C2M is that the latter considers a correction factor, $A_i$, in the specific heat approximation Eq.~\eqref{eq:cpmoistures} below wood fibre saturation (moisture content below  0.3). Note that the value of $M=0.1$ used in this test case would be unrealistic for green wood, but it has been chosen as an example for comparison of the models.

Tab.~\ref{table:case-moist-ros} gives the rate of spread for the different moisture models and compares with the model without any moisture effects. 
The model without moisture has a much higher rate of spread compared to all moisture models. 
The difference between the two-stage models and the three-stage model is rather small.

\begin{table}
\centering
\caption{Rate of spread of the traveling wave for different moisture models. The green wood to dead wood ratio is $r=1$, the moisture level is $M=0.1$, and the wind speed is $w=0$. }
\begin{tabular}{l|cccc}
Moisture model & NM  & S2M  & C2M & S3M \\
\hline
ROS (m/s)  & 0.0565 & 0.0382 & 0.0386 & 0.0390 \\
\end{tabular}

\label{table:case-moist-ros}
\end{table}

\begin{figure}
    \centering
        \includegraphics[width=0.6\linewidth]{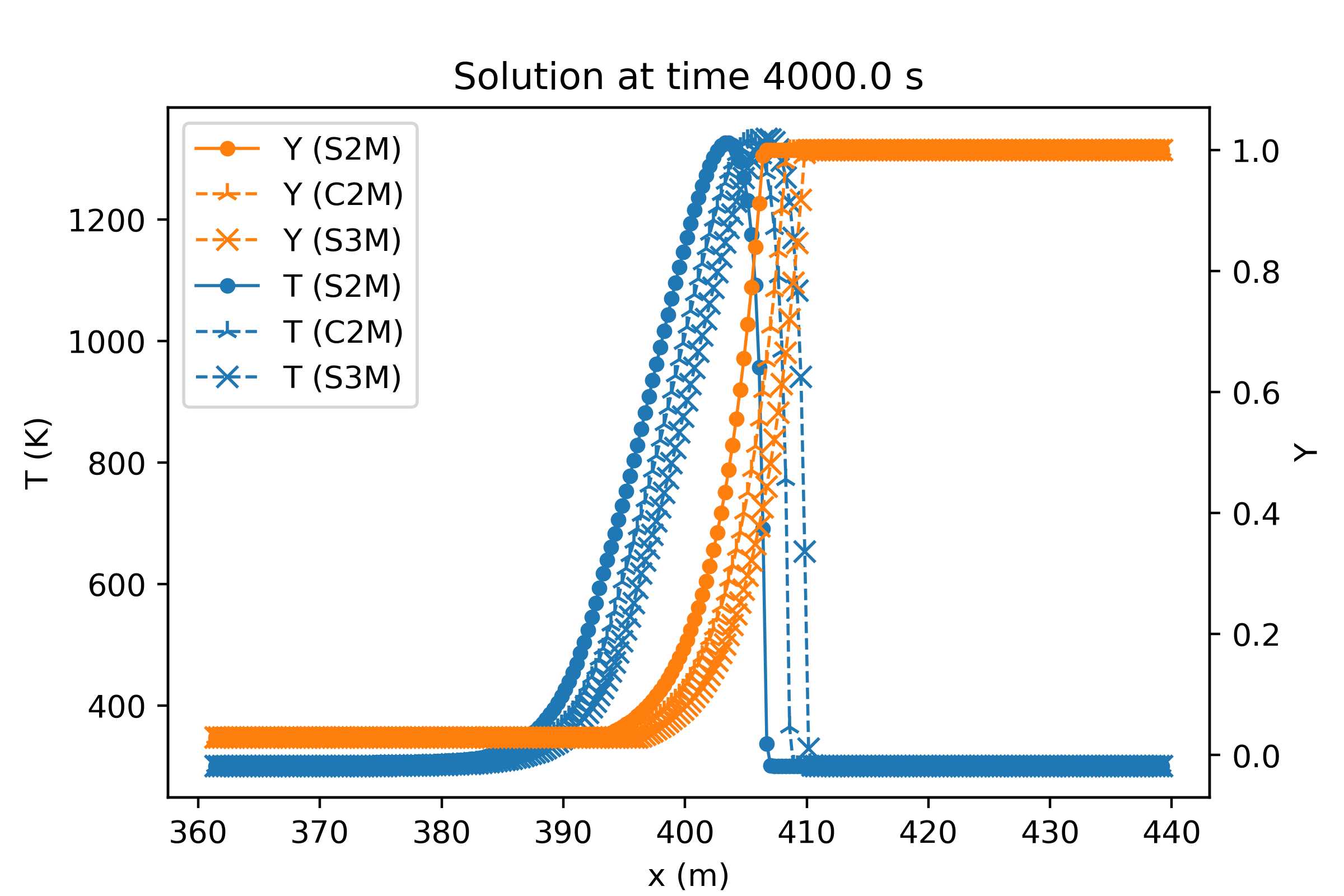}
    \caption{Comparison of the traveling wave profiles for the temperature $T$ and the fuel $Y$ at a fixed time for two-stage moisture models (S2M, C2M) with the three-stage moisture model S3M. The moisture content is $M=0.1$, the wind speed is $w=0$, and the live/dead ratio is $r=1$. 
    }
    \label{fig:profiles_moistmodels}
\end{figure}

Fig.~\ref{fig:profiles_moistmodels} shows the profiles of the traveling waves for the two- and three-stage models at the same time. 
The traveling wave front of solutions for the three-stage model S3M is slightly in front of the fronts of the two-stage models S2M and C2M. 
This higher rate of spread is in accordance with Table~\ref{table:case-moist-ros}.
Apart from the different rate of spread, the profiles have a similar shape. 
The two-stage models are in this setting very similar due to $r=1$, meaning that the correction factor $A_i$ has a  small impact on the approximation of the specific heat. 

\begin{figure}
    \centering
        \includegraphics[width=0.99\linewidth]{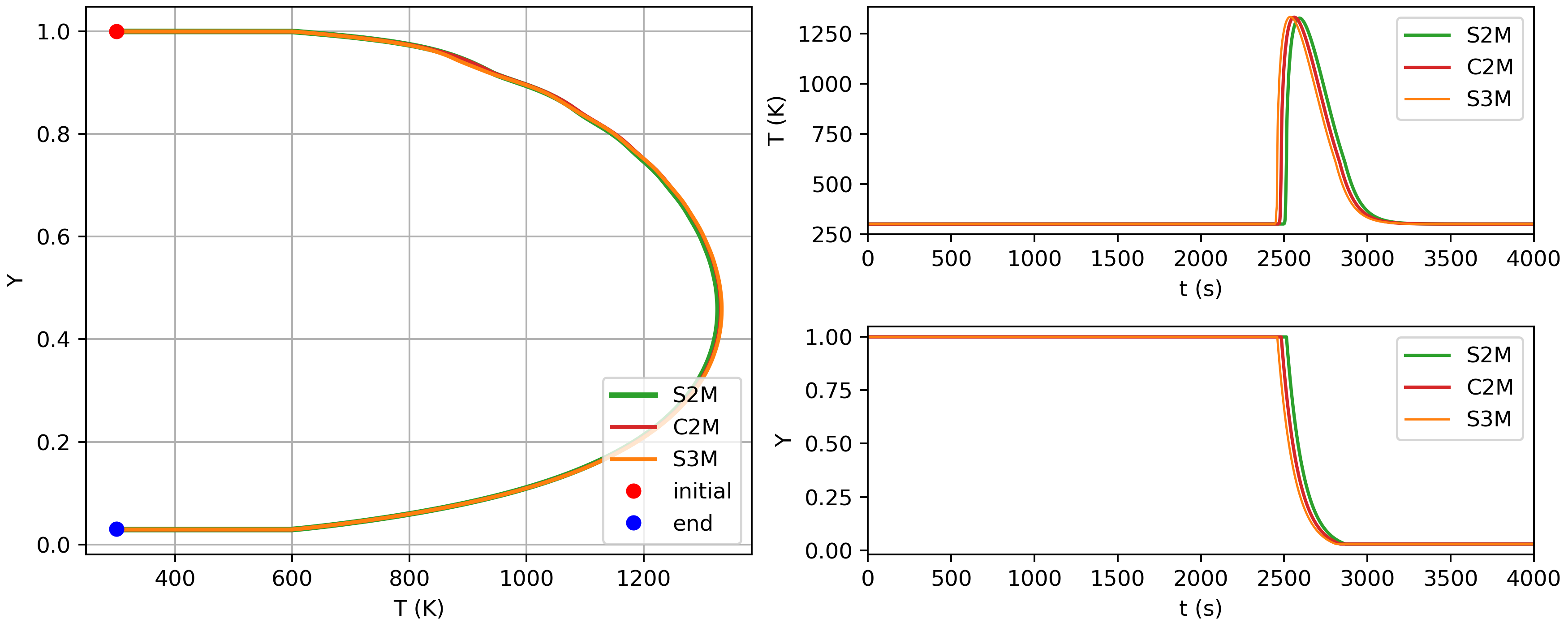}
    \caption{Comparison of  traveling wave profiles for the temperature $T$ and the fuel $Y$ for two-stage and three-stage moisture models with fixed $x=350$ m, and parameters $M=0.1$, $r=1$, $w=0$. Left: Phase portrait of the solutions at a fixed $x$; right: time-dependent solutions showing the traveling wave profiles. 
   }
    \label{fig:phasespace_moistmodels}
\end{figure}

 The phase portraits of the solutions provide further insight into the traveling wave profiles.
They highlight differences in the solution behavior very well, compare the use in \cite{reisch_analytical_2024}. 
To represent the phase portraits, we fix a certain point $x$ where the traveling wave will pass after some time. 
This point $x$ should not be too close to the maximum of the initial data for ensuring that the traveling wave front is already in its stable shape. 
For this point $x$, the solution values $T(x,t)$ and $Y(x,t)$ are plotted over each other, leading to the phase plot in Fig.~\ref{fig:phasespace_moistmodels}(left). 
The phase plot shows that the traveling wave front is following nearly the same $T$-$Y$-curve, independent of whether a two-stage or a three-stage model was used. Additionally, the time-dependent solutions for this fixed point $x=350$ m is shown, giving the profiles in Fig.~\ref{fig:phasespace_moistmodels}(right). This figure shows again the slightly slower rate of spread of the two-stage models.
Note that Fig.~\ref{fig:phasespace_moistmodels}(right) shows the time dependence; note that a faster traveling wave reaches the fixed spatial location in a shorter time.

\begin{figure}
    \centering
        \includegraphics[width=0.7\linewidth]{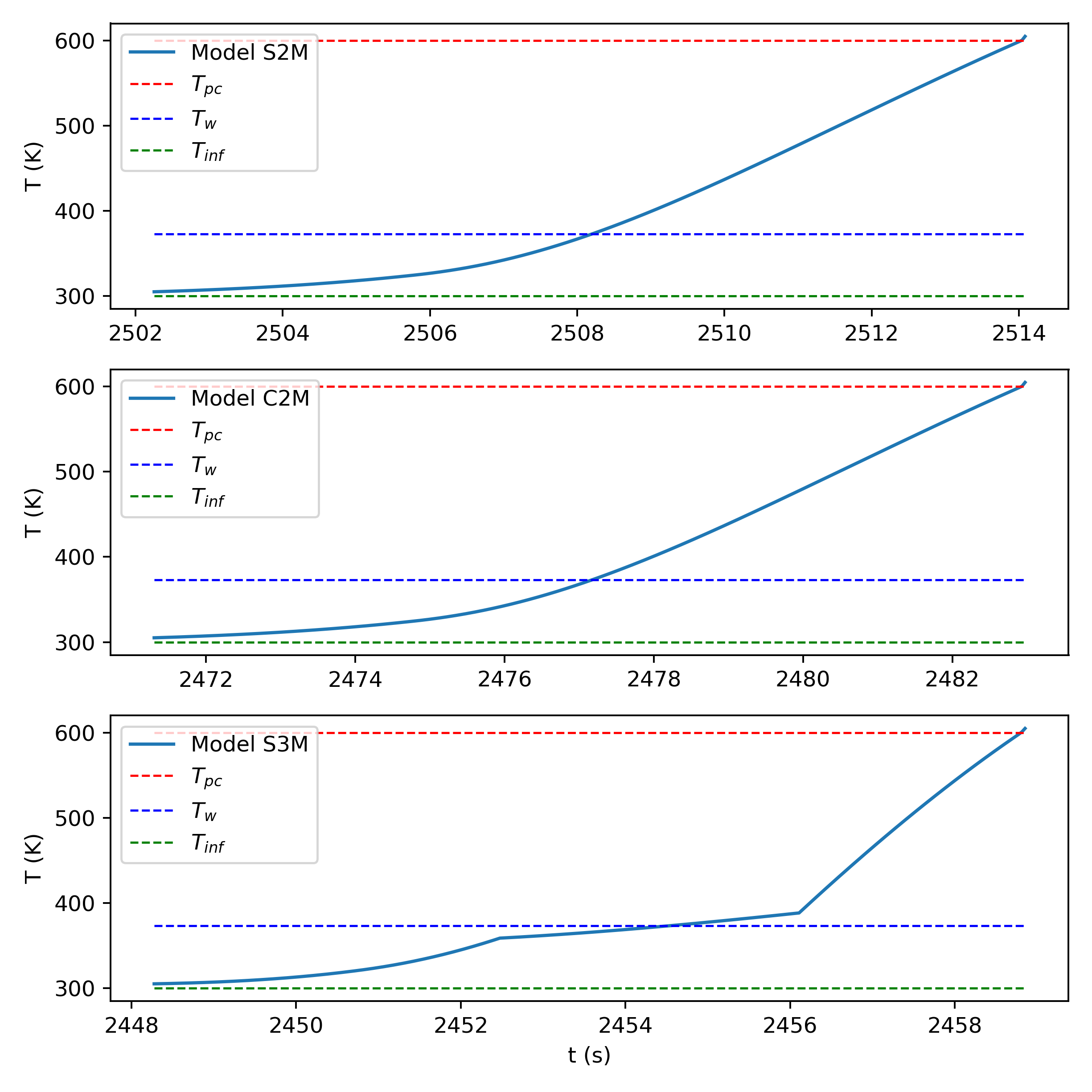}
    \caption{Time dependent heating process for the two-stage models S2M, C2M, and the three-stage model S3M. 
    The evaporation plateau leads to a slowed heating process for temperatures around the evaporation temperature $T_w$.
    Fixed parameters are  $x=350$ m, $r=1$, $M=0.1$ and $w=0$.
   }
    \label{fig:heating_moistmodels}
\end{figure}

The next step is to better visualize the differences between the 2- and 3-stage fuel moisture models. The 3-stage model explicitly represents the latent heating process into account, resulting in some plateau in Fig.~\ref{fig:heatingS3M}. This difference is not visible in the phase portrait in Fig.~\ref{fig:phasespace_moistmodels}, because it takes place while $Y=1$, following a straight line with increasing temperature.  In contrast, Fig.~\ref{fig:heating_moistmodels} shows very well the differences of the time-dependent temperature curves at a fixed $x=350$ m during the heating processes. The  parameters are fixed as in the previous figures. The solution using the S3M model shows the latent phase with a slightly increasing plateau, due to the width of the evaporation interval in Eq.~\eqref{eq:deltas1}. This evaporation plateau leads to a slowed heating process for temperatures around the evaporation temperature $T_w$. Contrarily, the S2M and C2M models show a monotonically increasing heating process without the evaporation plateau. In these models, the effect of latent heating is treated as a sensible heating and therefore evaporation cannot be explicitly represented.

In total, the comparison of the two-stage models S2M and C2M with the three-stage model S3M shows some differences in the preheating phase, so for temperatures between the environment temperature, $T_\infty$, and the ignition temperature, $T_{pc}$. The small differences observed in the ROS may have a numerical origin, since the effective specific heat in the preheating process should be the same in both cases.

\begin{remark}
The S3M model does not improve the solution in terms of ROS and burned zone, but involves a higher complexity. Therefore, from a practical point of view, we recommend using the S2M rather than the S3M model. Regarding the preference between the S2M and C2M model,  we recommend to use the C2M model when detailed information about the fuel characteristics and structure is available.
\end{remark}

In the next step, the S2M model is fixed and the effect of varying moisture levels is studied.

\subsubsection{Influence of moisture content}\label{sec:simulation_moisture}

The following investigations use the simple two-stage moisture model S2M in Sec.~\ref{sec:S2M}. Further, advection effects are not regarded, so $w=0$. In comparison to the last section, the moisture content $M$ is now varied.

\begin{figure}[bht]
    \centering
        \includegraphics[width=0.6\linewidth]{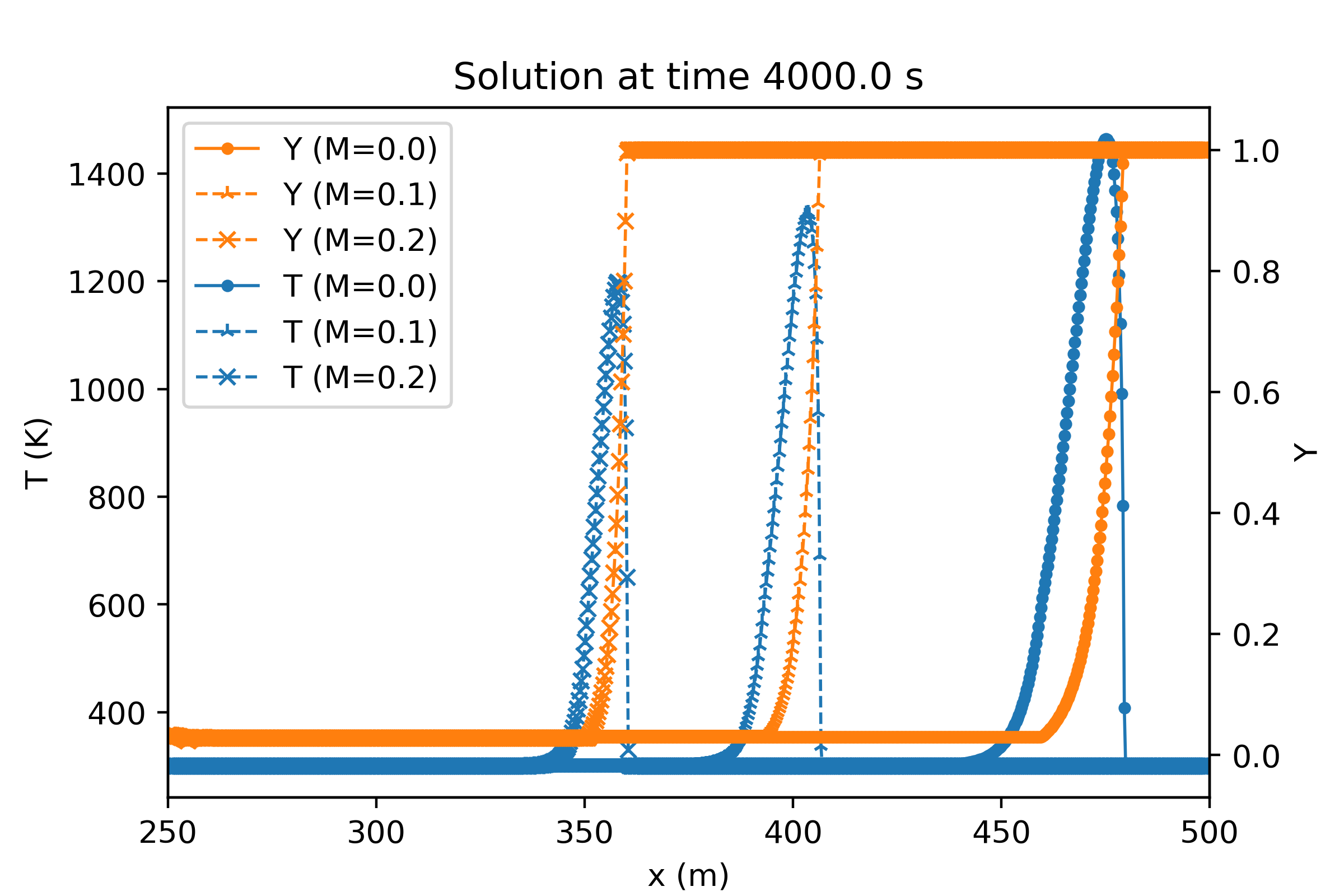}
    \caption{Traveling wave solution profiles at a fixed time $t$ for the two-stage moisture model S2M for different moisture levels. Fixed values are $r=1$ and $w=0$. High moisture levels lead to a slower rate of spread and a lower maximal temperature.}
    \label{fig:profiles_difM}
\end{figure}

Fig.~\ref{fig:profiles_difM}  shows the solution at $t=4000$ s for moisture contents $M=0$, $M=0.1$ and $M=0.2$.  These moisture content values are representative of dead fuels with relative humidity levels below 90\%. The traveling wave profiles in Fig.~\ref{fig:profiles_difM} show a decrease in the maximal temperature and in the traveling wave speed with increasing moisture content $M$. The surviving biomass fraction $Y$ is identical for all three values of $M$ even though the maximal temperature is different in all cases. 
In the complex system behavior, the larger maximal temperature values are compensated by the faster traveling wave speeds. Note that this inverse dependency between maximal temperature and traveling wave speed is due to the non-linear radiation term. In previous investigations \cite{nieding_impact_2024}, it was shown that a higher rate of spread is connected to a larger remaining biomass fraction $Y$. 
In this case now, the higher maximal temperature would lead to a smaller remaining biomass fraction, but the faster rate of spread compensates this effect. 

The slower traveling wave speed for higher moisture levels is connected to a slower preheating process. 
Tab.~\ref{table:case-preheating-times} gives the preheating time, so the time needed for heating to the pyrolysis temperature $T_{pc}$, for different moisture levels.

\begin{table}[b]
\centering
\caption{Time to reach $T_{pc}$ (preheating time) for different moisture levels, using the S2M model, $w=0$, $r=1$. }
\begin{tabular}{r|ccccc}
$M$ & 0.0  & 0.1  & 0.2 & 0.3 & 0.4  \\
\hline
preheating time (s)  & 8.5 & 11.8 & 17.5 & 26.8 & no propagation \\
\end{tabular}
\label{table:case-preheating-times}
\end{table}

The preheating time is non-linear related to the moisture level. 
In particular, for this parameter setting, a moisture level of $M\geq 0.4$ stops the traveling wave behavior. 
The longer the preheating time is, the slower are the traveling waves because it needs longer to start the combustion process when the traveling wave front pushes forward. 

\begin{figure}
    \centering
        \includegraphics[width=0.99\linewidth]{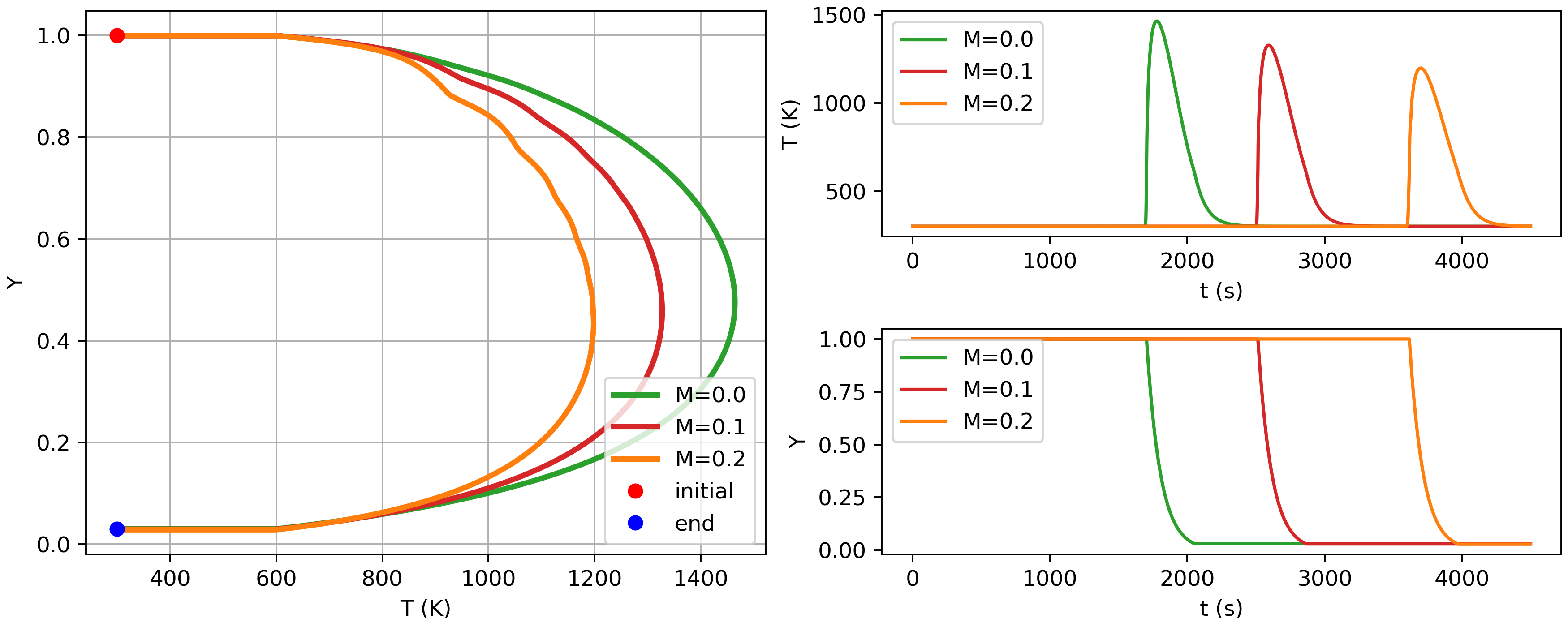}
    \caption{Comparison of different moisture levels in the S2M model with fixed $x=350$ m, $w=0$, $r=1$. Left: phase portrait of the solutions at a fixed $x$; right: time-depending solution profiles showing the traveling wave front. Higher moisture level lead to a slower ROS and a lower maximal temperature. 
    }
    \label{fig:phase_difM}
\end{figure}

Besides the influences on the traveling wave speed and the maximal temperature, the moisture content changes the dynamical behavior for large moisture levels with still propagating fronts. 
Fig.~\ref{fig:phase_difM} shows the phase portrait for the simple two-stage moisture model S2M and the three moisture levels from Fig.~\ref{fig:profiles_difM}.
The trajectory for the highest moisture level, $M=0.2$, shows the most visible bumps.

\begin{figure}[h]
    \centering
        \includegraphics[width=0.8\linewidth]{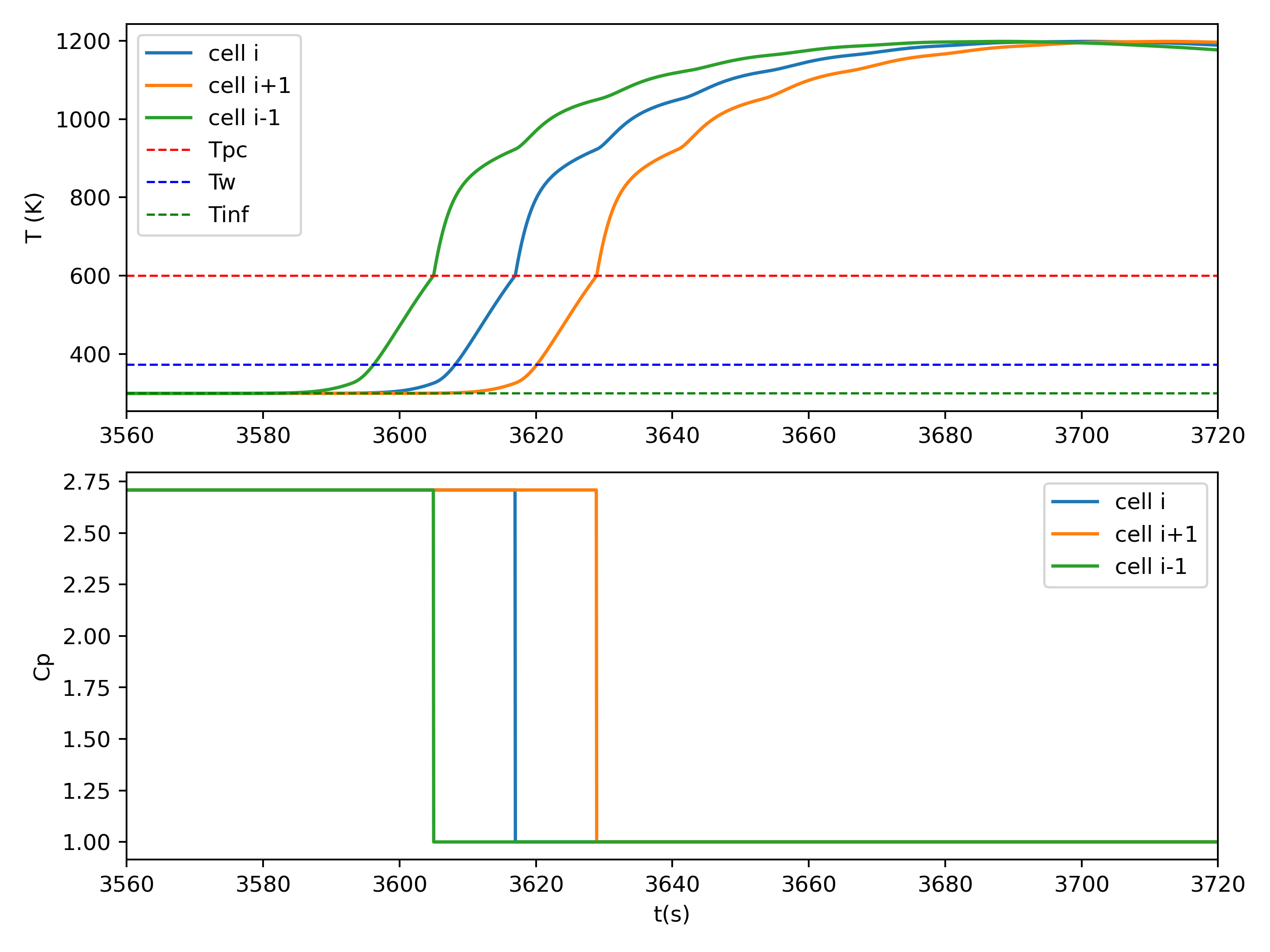}
            \caption{ Time-dependent temperature curves for three neighboring discretization cells (upper), and specific heat values in the three cells (lower). The moisture model is S2M with $M=0.2$, $w=0$, $r=1$ and $x=350$~m. 
             The curves show a similar behavior with some time delay. The discontinuous values of $c_p$ lead to bumps in the temperature curves.
            }
    \label{fig:discrete_difM}
\end{figure}

The bumps in the phase portrait are a consequence of spatial discretization in the numerical simulations. 
Fig.~\ref{fig:discrete_difM} shows in the upper figure the time-dependent temperature curves for three neighboring discretization cells. 
The three curves show a time-delayed identical behavior, including the bumps. This is because the fire front takes the form of a self-similar solution.
In the lower plot of Fig.~\ref{fig:discrete_difM}, the specific heat value of the three discretization cells is given over time. 
The discrete nature of the simulation requires a jump of the specific heat value once the system exceeds the pyrolysis temperature. 
This jump occurs at different times for the three cells. 
Due to the interaction by advection and diffusion, each cell is influenced as well by the jumps in the neighboring cells, resulting in a changed behavior visible as bumps in the temperature-time curve or the phase plot.

The occurrence of bumps is independent of the discretization parameters and desired for describing the temperature dependent evaporation process. They do not change the traveling wave behavior, see Fig.~\ref{fig:phase_difM}. Therefore, they are noticed as a feature of the model, not distracting any other observations. 

\subsection{Investigation of the parameter-dependency of ROS}\label{sec:parametric}

The S2M model is used for some further investigations on the dependency of the rate of spread on the moisture content, the wind velocity, the relative humidity and the surface coverage.
These investigations lead the path towards more realistic scenarios and deepen the understanding of the influence of moisture models. 

\paragraph*{Effect of wind and moisture}

The effect of wind on the solution of an ADR model without moisture-influence was studied in \cite{reisch_analytical_2024}. There, it was shown that the wind strongly influences the evolution of traveling wave fronts in only one or in two directions. Further, the maximal temperature of the traveling wave in wind direction is larger than the maximal temperature of a model without wind.
A higher rate of spread due to wind leads to less burned biomass, compare \cite{nieding_impact_2024}.

\begin{figure}[htb]
    \centering
    \begin{subfigure}{0.49\textwidth}
        \includegraphics[width=\linewidth]{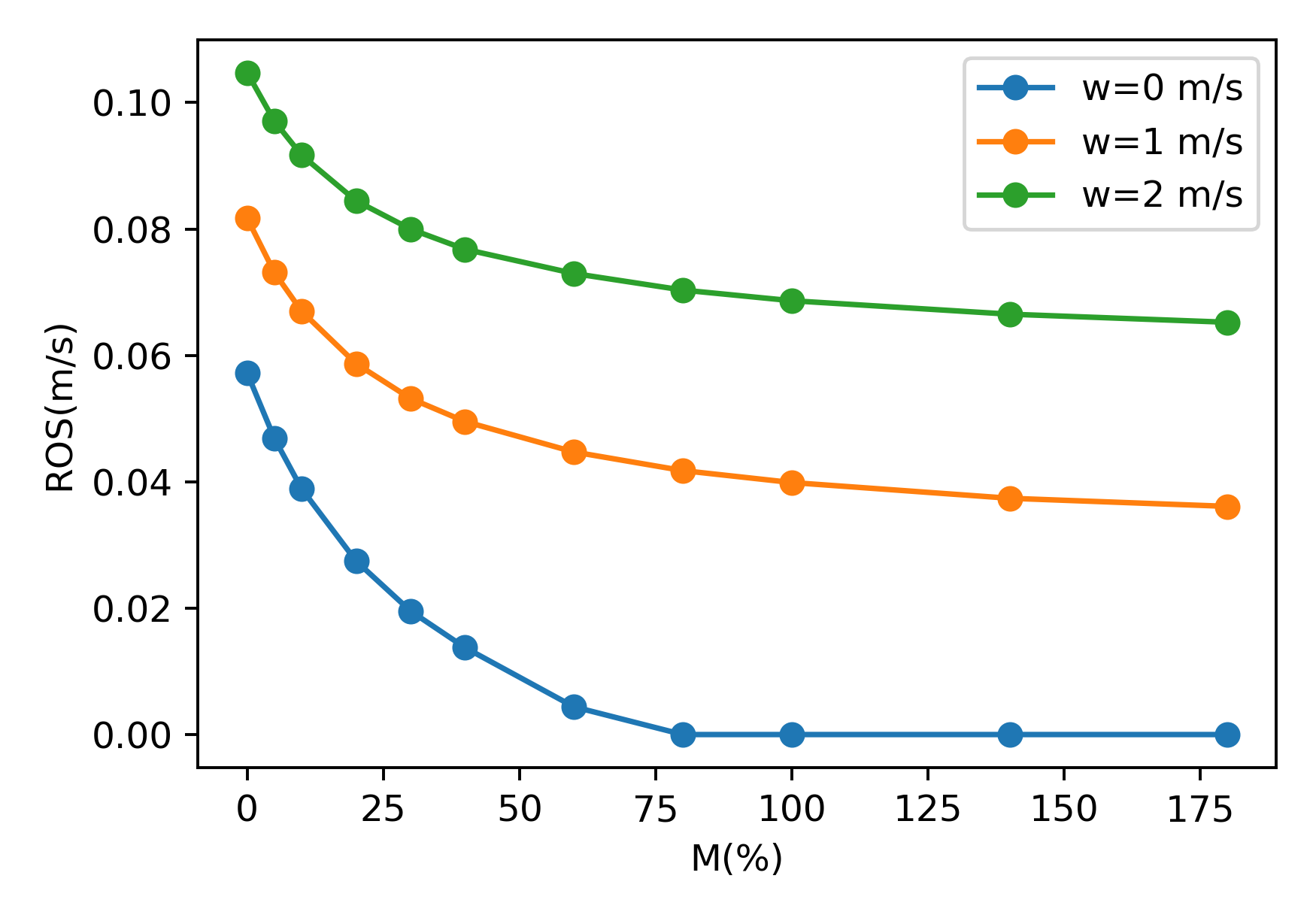}
        \caption{Varying moisture content}
        \label{fig:ROS_moisture_a}
        \end{subfigure}
        \hfill
    \begin{subfigure}{0.49\textwidth}
            \includegraphics[width=\linewidth]{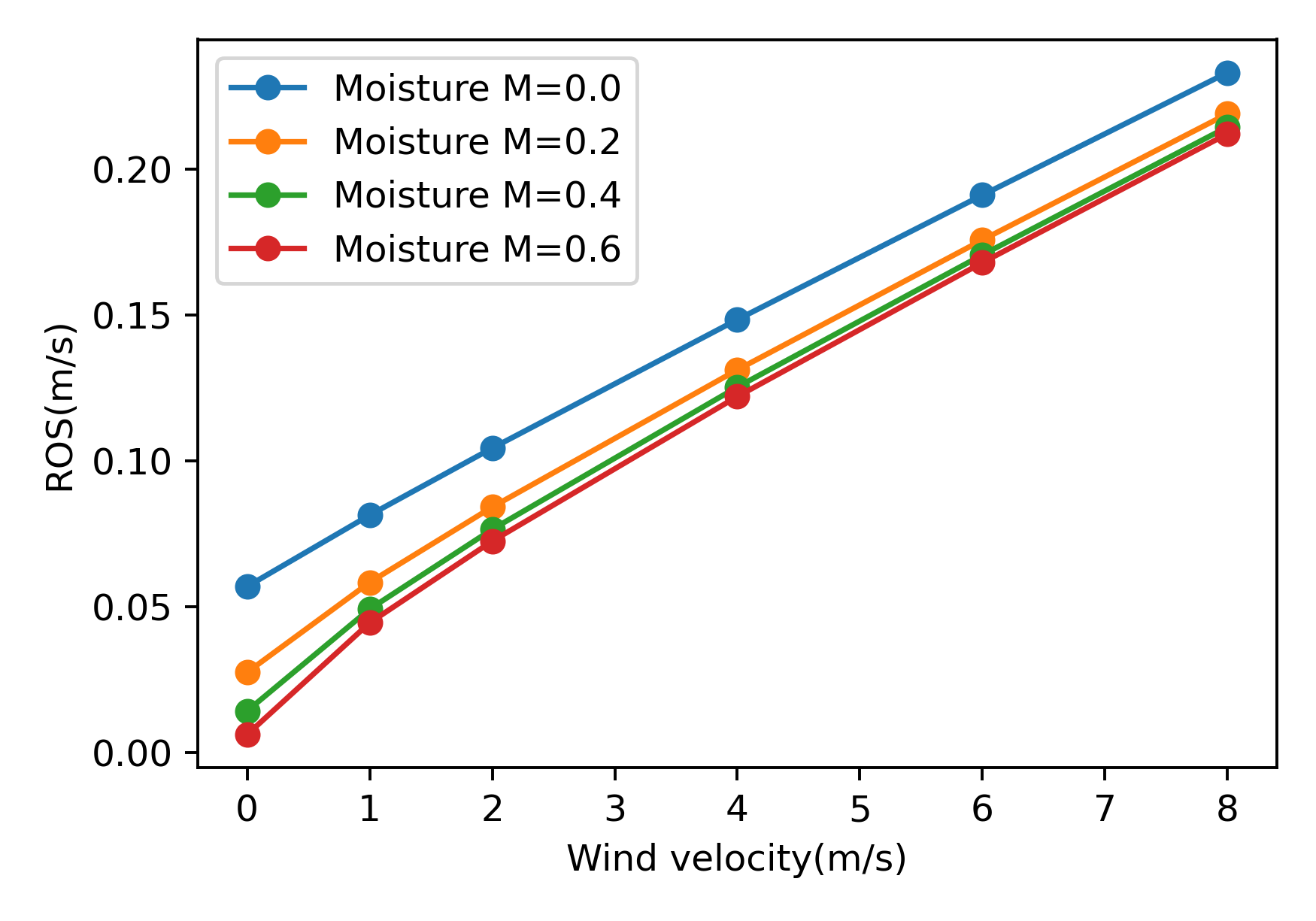}
            \caption{Varying wind velocity}
             \label{fig:ROS_moisture_b}
    \end{subfigure}
    \caption{Dependency of the rate of spread (ROS) on the moisture content $M$ and the wind velocity $w$. The S2M model is used for modeling the moisture. Higher moisture levels lead to smaller rate of spread while a higher wind velocity increases the rate of spread. If the wind is zero and the moisture level is high, the traveling wave stops.}
    \label{fig:ROS_moisture}
\end{figure}

Fig.~\ref{fig:ROS_moisture_a} shows the rate of spread (ROS) of the model in Eq.~\eqref{eq:model1b} with respect to the total moisture content, $M$, computed using the model S2M in Eq.~\eqref{eq:cpeff3}. In the case without wind, $w=0$, the traveling wave stops for larger moisture contents. The fire front stops and the fire is not spreading any further. 
This solution's behavior is different for situations with a small wind, $w=1$ or $w=2$ m/s. In these cases, the moisture content slows down the rate of spread but does not lead to a stopping of the fire front. 

\begin{remark}
    The observation that the traveling wave solution with some wind $w>0$ does not stop is in accordance with practical observations, although the underlying mechanisms differ  in reality and in this model.   In reality, a strong wind may lead to a heating of the fuel far ahead of the traveling wave front and therefore to an evaporation of the fuel moisture before the fire front arrives.   This long-distance effect, however, is not represented in the current mathematical model. Here, the model only considers local radiative effects through the Rosseland approximation, which acts as a diffusion mechanism. As a result, heat is transferred between adjacent areas in a manner similar to conduction. To properly model this effect,  the non-local effect of radiation should be considered as proposed in \cite{ferragut2007numerical,sero2008large}. This implies the resolution of additional  equations for the radiation, thus increasing the complexity of the model.

    In the mathematical model, the advection term dominates the effects of a varying specific heat. Note that the specific heat is multiplying both terms (local and convective derivatives) on the left-hand side of Eq.~\eqref{eq:pde1v2}. This has a physical meaning, as the advection term is modeling a transport of temperature of the gaseous phase across the cell boundaries and does not depend on the bulk properties. Besides, the rise of temperature due to wind, as observed in \cite{reisch_analytical_2024}, enlarges the temperature gradient being relevant for all heat transport mechanisms. 

    Consequently, the effect of advection dominates any effects of moisture in the model.     This results in the non-zero rate of spread in Fig.~\ref{fig:ROS_moisture} for non-zero wind and any moisture content.     Nevertheless, the effect of moisture decreases with wind up to a point where this effect may be underestimated. At this point, the consideration of non-local radiation effects could improve our results.
    
\end{remark}

The stopping rate of spread in Fig.~\ref{fig:ROS_moisture_a} for $w=0$ is in accordance with an empirical model for laboratory and field experimental data in \cite{rossa2016laboratory}. 
There, the rate of spread shows a qualitatively similar dependency on the moisture content. The result in Fig.~\ref{fig:ROS_moisture_a} for non-zero wind speed reproduce the results in \cite{ASENSIO2023105710} qualitatively.
In \cite{ASENSIO2023105710}, an advection-diffusion-reaction model for temperature and fuel was fitted to various fuel types in the literature, including the experiments in \cite{rossa2016laboratory}. 
The rate of spread in \cite{ASENSIO2023105710} relates to the fuel moisture content with an exponential decay function. This exponential decay is visible as well in Fig.~\ref{fig:ROS_moisture_a}, which validates our modeling assumptions.

\begin{remark}
 The impact of increasing fuel moisture on the ROS would be higher if, instead of considering a constant factor $\beta$, the expression in Eq.~\eqref{eq:beta} is used, introducing a dependency $\beta=\beta (M)$. As $M$ increases, $\bar{c}_{p,f}$ also increases, leading to a decrease in $\beta$, which results in a larger reduction of the ROS for high moisture and wind values. Nevertheless, given the level of simplification in our model,  we limit ourselves to a constant value for $\beta$ and defer this option for future research. This approach could improve the results for strong wind situations, where the effect of moisture is underestimated.
\end{remark}

Fig.~\ref{fig:ROS_moisture_b} shows the dependency of the rate of spread on the wind velocity for three different moisture levels. 
The different moisture levels become almost irrelevant for high wind velocities but play a more important role for lower wind velocities. 
Fig.~\ref{fig:ROS_moisture_b} shows a tendency of convergence for large wind velocities towards a common rate of spread, at least in the case that the moisture percentage is non-zero. 

In Sec.~\ref{subsec:modelling_wind}, the connection between wind velocity and topographies with constant slopes was drawn. 
Recall the key idea: Even though the investigations in Fig.~\ref{fig:ROS_moisture} were carried out for a wind velocity, they are valid as well for a non-flat topography with a constant slope. Further effects of the topography on the spread of fire will be studied with simulation in two spatial dimensions in Sec.~\ref{sec:2dim}.

\paragraph{Effect of relative humidity}

In the next step, the relative humidity and the live to dead ratio $r$ of the fuel are varied. 
The live to dead ratio is relevant only for the complete two-stage moisture model C2M, which will be used for the following investigations.

\begin{figure}
    \centering
        \begin{subfigure}{0.49\textwidth}
             \includegraphics[width=\linewidth]{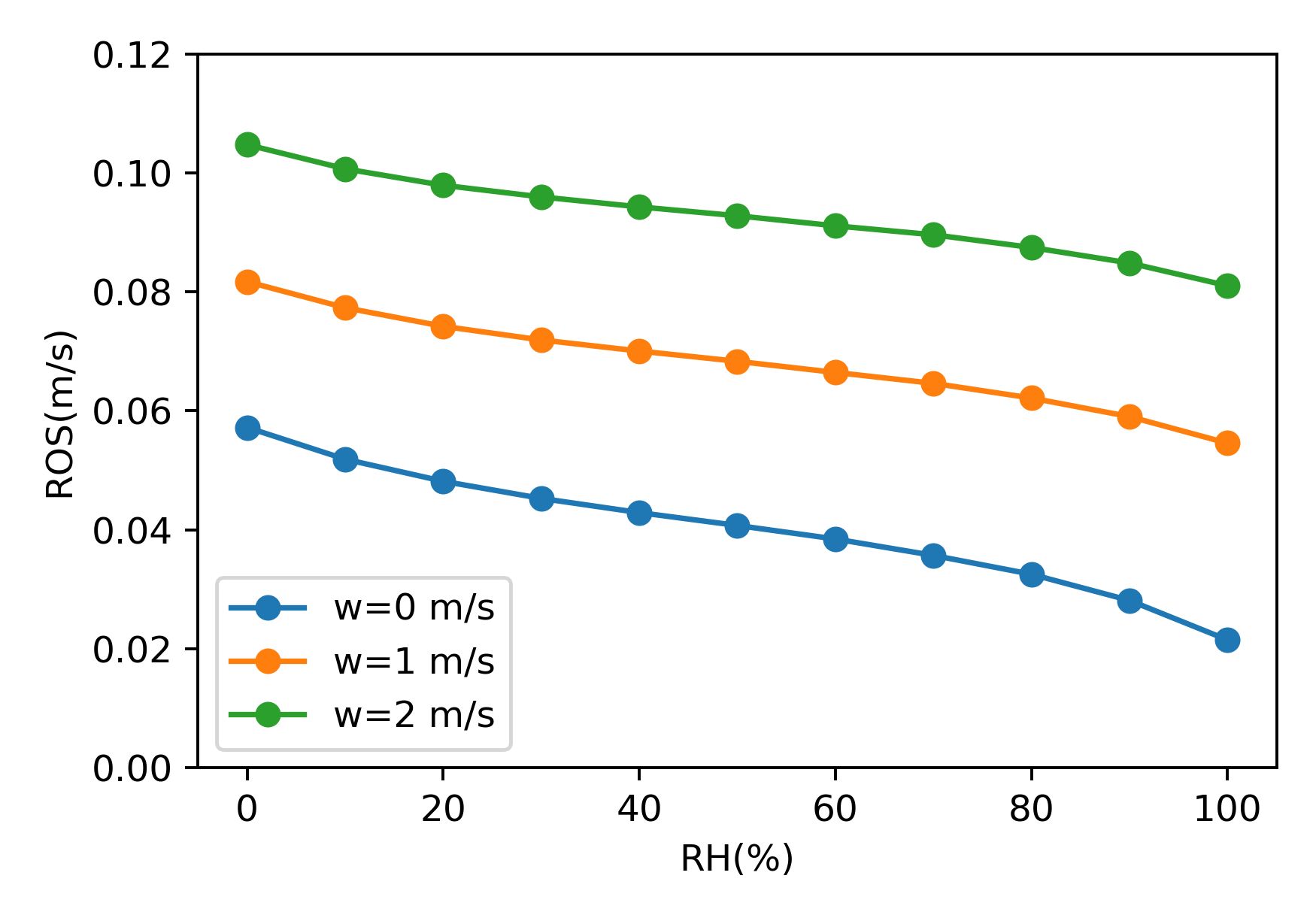} 
             \caption{Varying RH and wind velocity $w$ in only dead wood ($r=0.0$)}
             \label{fig:ROS_moistureC2M_a}
        \end{subfigure}
        \hfill
        \begin{subfigure}{0.49\textwidth}
            \includegraphics[width=\linewidth]{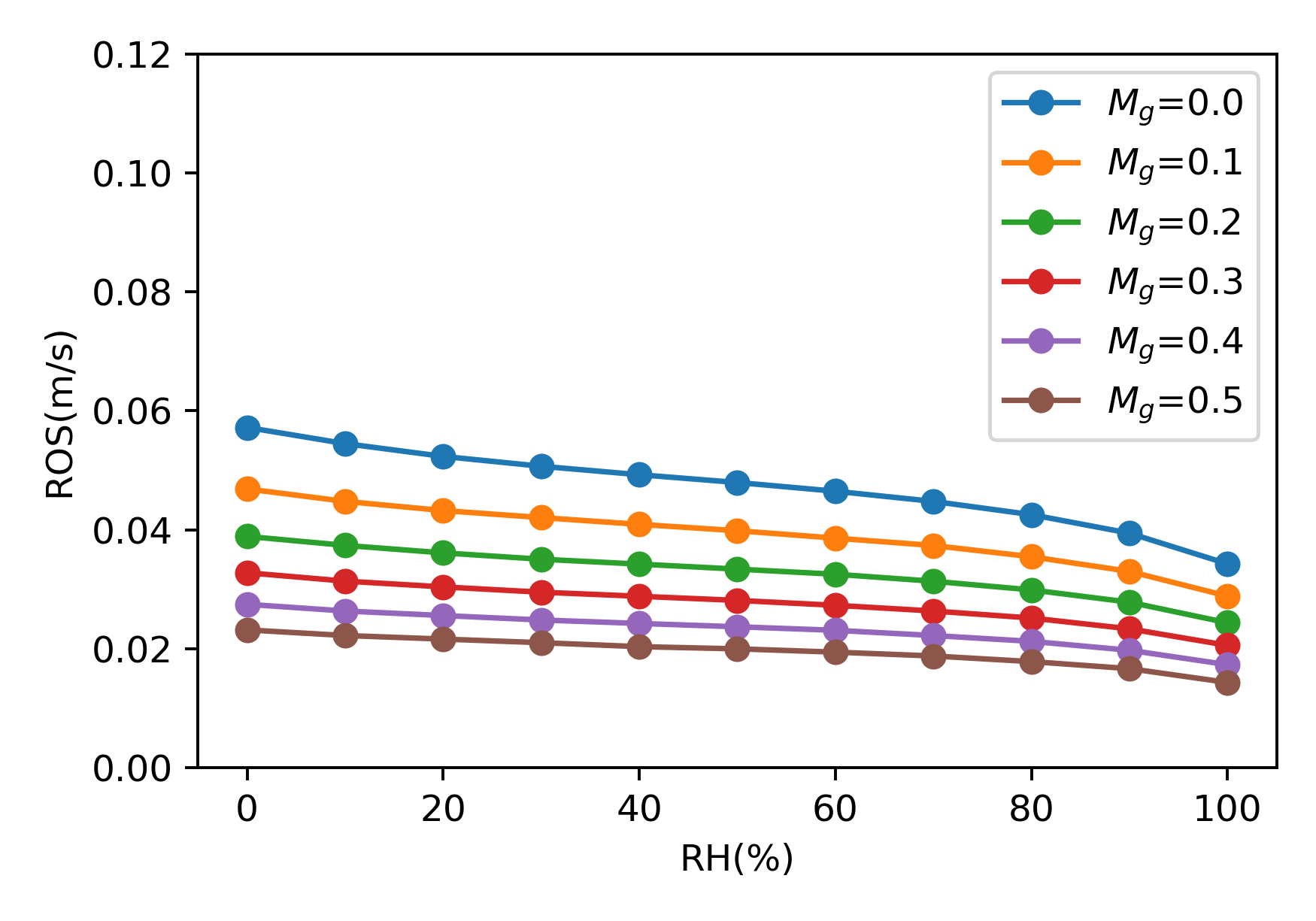}
            \caption{Varying RH and green wood moisture content $M_g$ in mixed wood ($r=0.5$)}
            \label{fig:ROS_moistureC2M_b}
        \end{subfigure}
    \caption{Dependencies of the rate of spread (ROS) with the C2M moisture model on the relative humidity RH ($\phi$) with varying wind velocity $w$ for $r=0.0$ (a), and varying green wood moisture content $M_g$ for $r=0.5$ (b).  An increase of the relative humidity leads to a slower rate of spread in both variations.}
    \label{fig:ROS_moistureC2M}
\end{figure}

The relative humidity, RH or $\phi$, influences the dead wood moisture, as described in Sec.~\ref{sec:C2M} and in particular in Eq.~\eqref{eq:ms} for ambient temperature.
The variation of relative humidity in Fig.~\ref{fig:ROS_moistureC2M_a} shows a decrease in the rate of spread for higher humidity values. 
In this case, only dead wood is regarded, $r=0$, because only this type of wood is influenced by the  humidity of the air. 
The different wind velocities affect, as already known for the basic model, the rate of spread and lead to a nearly constant offset between the different curves. 
The curve for $w=0$ shows for high relative humidity a decrease that is similar to the decrease in the rate of spread in Fig.~\ref{fig:ROS_moisture_a}. Due to the more complex influence of the relative humidity on the total moisture level in the C2M model, the drop in the rate of spread is not that drastically as in the S2M model when increasing the total moisture content. 

For mixed wood, Fig.~\ref{fig:ROS_moistureC2M_b} shows the dependency of the rate of spread on both, the relative humidity influencing the dead wood, and on the green wood moisture content $M_g$. 
Compared to Fig.~\ref{fig:ROS_moistureC2M_a}, the rate of spread is slower and the effect of relative humidity is less strong due to the wood mixture.

The two figures show a behavior that is expected from literature: In \cite[Fig. 6]{MOINUDDIN2021103422}, the comparison of different (empirical) models with simulation results is shown. 
All the models there show a decaying tendency. The different empirical models describe the decay either by a linear or an exponential decay function. 
The model results differ from this behavior in the case of very high relative humidity values, where  an additional drop occurs. 
As a relative humidity near 100\% is not the typical wildfire condition, the fit of the model to the literature values is still good. 

\paragraph{Influence of bulk density and packing ratio}

The dependence of the rate of spread on the bulk density of the fuel has been a widely studied topic for decades \cite{catchpole1998rate,cruz2018got}. This dependency is highly significant in heterogeneous environments, and recent ADR models have demonstrated the ability to accurately reproduce it \cite{vogiatzoglou2024interpretable}.
Here, the focus is on studying the influence of the fuel bulk density, $\bar{\rho}_f R_{f,0}$ in Eq.~\eqref{eq:rhocpApprx},  in the ROS. For this numerical experiment, a fixed fuel density is considered (see Tab.~\ref{table:general-parameters1}) and the fuel volume ratio, $R_{f,0}$,  also called packing ratio in this context, is changed.

Three different models for the dependency of the  bulk density on the fuel loadare compared. 
Eqs.~\eqref{eq:rhodefY} and~\eqref{eq:cpdefY} give the first approximation, where both, the gaseous and solid phase, are regarded. Additionally, their portion varies in time due to the combustion process. This modeling approach is referred to as ``mixture (time varying)". The next approximation is given by Eq.~\eqref{eq:rhocp0}, taking into account the mixture but not the time variation. As the simplest approximation, Eq.~\eqref{eq:rhocpApprx} does not depend on the gaseous phase but regards only the fuel (solid phase). 

\begin{figure}
    \centering
    \includegraphics[width=0.5\linewidth]{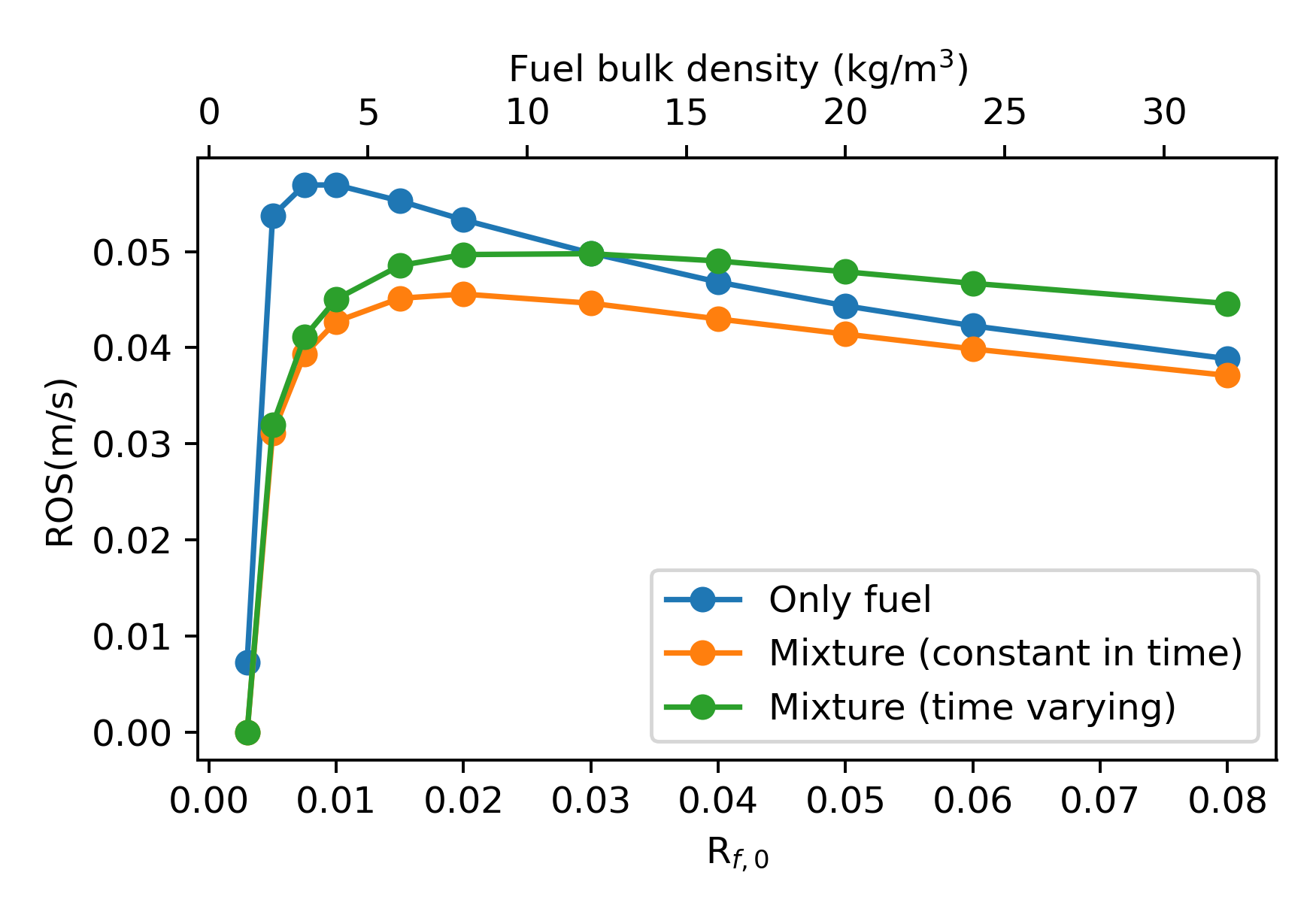}
    \caption{Rate of spread (ROS) depending on the bulk density and the packing ratio for the three different models. The simulations are without moisture $M=0$ and wind $w=0$.  For sufficiently large fuel bulk densities the rate of spread decreases with increasing fuel bulk densities.}
    \label{fig:ros_BD}
\end{figure}

Fig.~\ref{fig:ros_BD} shows the dependency of the rate of spread on the fuel bulk density for the three approaches. The results evidence that there are two well-defined regions. For lower values of the fuel bulk density (and packing ratio), there is a transition from no fire conditions to wave propagation, increasing the ROS up to a maximum value. For larger densities (and packing ratio), the ROS decreases slowly, in accordance with the results reported in the literature. We believe that the role of the $T^3$ term in the non-local radiation model is the key to reproduce this particular behavior. This pattern was reported in \cite{cruz2018got}, where a Ricker function was proposed as an approximation for this dependency relationship. The proposed Ricker function accounts for a constant contribution that dominates at low densities, transitioning to an exponentially decaying function at higher densities. The specific density value for this transition does not coincide with that in \cite{cruz2018got}  because no calibration has been performed in our model. Differences between the three models considered herein are more noticeable in the region near the maximum ROS. For the 'only fuel' model, the transition is very sharp, whereas the other models show a smoother transition. 
 
\paragraph{Influence of surface coverage}

The rate of spread depends as well on the surface coverage, which is relevant for heterogeneous environments. Here, the surface coverage is modeled as a factor multiplying the fuel packing ratio, $R_{f,0}$, when assuming a theoretical maximum value  $R_{f,max}$, as defined in in Eq.~\eqref{eq:SC}.  The previous study was carried out for a wide range of fuel packing ratios. Here, the maximum packing ratio is assumed to be $R_{f,0}=0.01$, and the propagation regime up to this value is analyzed to identify the transition from no fire to propagation conditions. Within this range, the fuel load can be considered a limiting factor for propagation.
Experiments show a non-linear dependency of the rate of spread on the surface coverage \cite{miller_upward_2015}. Again, the three different models described above for the dependency of the bulk density on the fuel load are compared.

\begin{figure}
    \centering
    \includegraphics[width=0.5\linewidth]{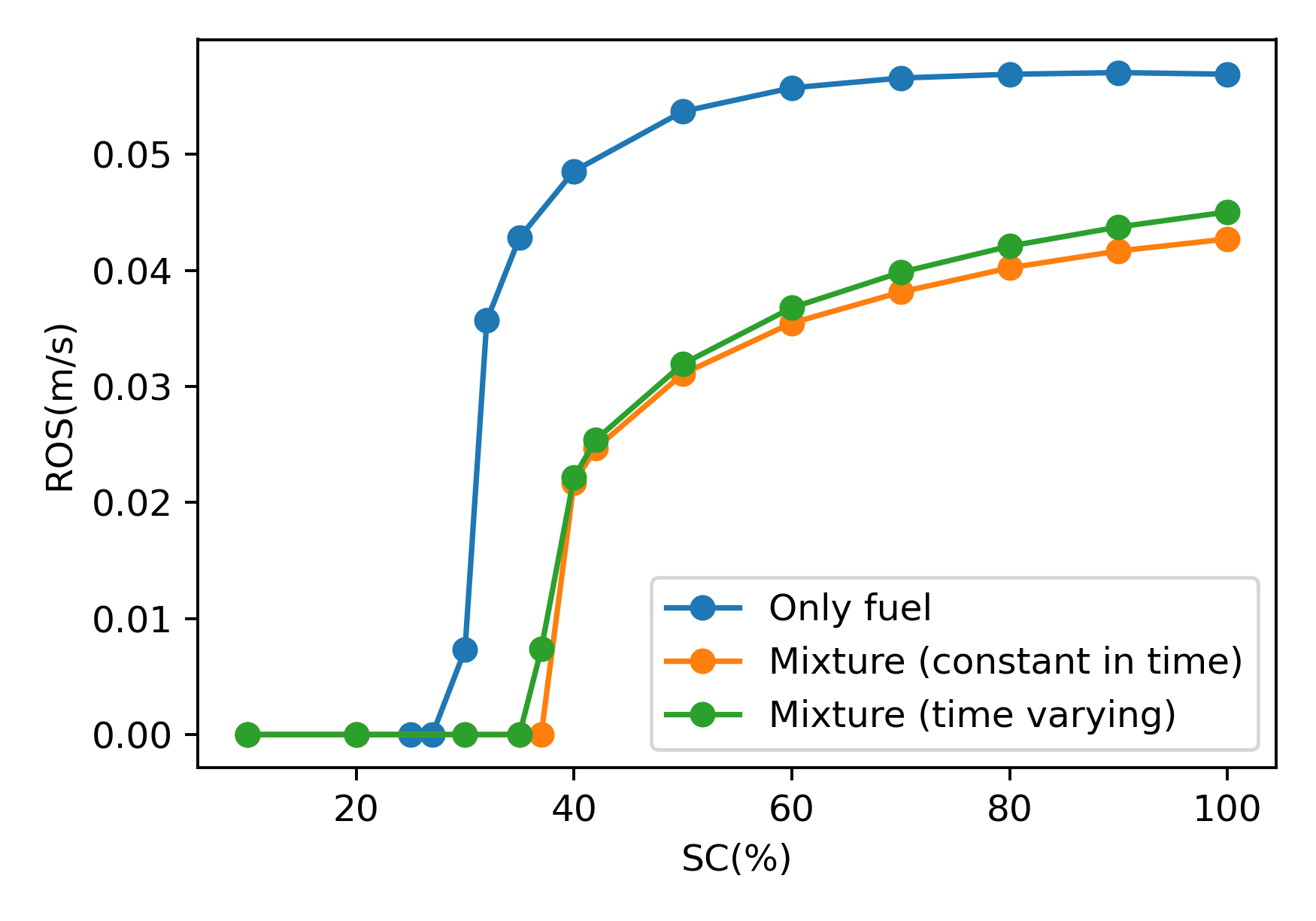}
    \caption{Rate of spread (ROS) depending on the surface coverage (SC) for three different fuel models. The simulations are without moisture $M=0$ and wind $w=0$. The fire stops propagating for very small surface coverage.}
    \label{fig:ros_SC}
\end{figure}

Fig.~\ref{fig:ros_SC} shows the dependency of the rate of spread on the surface coverage for all three approaches for describing the bulk density. 
The results of all three approaches have in common that there is a threshold value of surface coverage between 25 and 40\% such that the traveling wave solution is stopped. 
For higher surface coverage ratios, the temperature forms a traveling wave solution with a certain rate of spread. 

The two approaches representing the mixture of solid fuel and gases show only slight  differences, whereas the rate of spread for the most simplified model regarding only the fuel shows a higher rate of spread. From an applied point of view, the simplified 'only fuel' model for the bulk parameters provides a conservative approximation that estimates a higher rate of spread than the more complex models. 

\begin{remark}\label{rem:T3}
The key aspect when considering local radiation lies in the strong non-linearity of the diffusion term. In \cite{reisch_analytical_2024}, it was shown  that a variation of the diffusion coefficient produces a change in the ROS. In the present model, the dependency of the diffusion mechanism on $T^3$ allows to connect the ROS with the energy contained in the medium, which is proportional to its temperature. This dependency is required to produce physically consistent results, reproducing the trends observed in all the cases discussed above.
\end{remark}

\subsection{Two-dimensional domains}\label{sec:2dim}

Now the spatial domain is extended to two spatial dimensions (2D) and different scenarios with spatial heterogeneity are considered. Below, results are presented for four different cases and the observations made above for one spatial dimension are now extended to more realistic fire propagation patterns in two-dimensional landscapes.

\subsubsection{Propagation in a homogeneous environment: assessing the effect of fuel moisture}

The aim of this case is to assess the effect of fuel moisture in 2D fire propagation across homogeneous land, under a low wind  velocity $\mathbf{w}=(0.25,0.25)^T$ m/s. The computational domain is $[0,200]\times[0,200]$ m$^2$. The initial condition for the temperature is given by
\begin{equation}\label{eq:icreal}
   T(x,y,0) = 
		\left\{
		\begin{array}{lll}
      	670 \mbox{ K}  &\mbox{if} &  r(x,y)< 5 \mbox{ m}, \\
        300 \mbox{ K} & \multicolumn{2}{l}{\mbox{otherwise}}  \\  
		\end{array}
		\right. 
\end{equation}
with 
\begin{equation}\label{colliding3}
r(x,y)=\sqrt{(x-x_1)^2+(y-y_1)^2}
\end{equation}
with $(x_1,y_1)=(100,100)$ m.  Under these wind conditions, the predominant propagation direction points toward the top-right corner of the domain.

The computational mesh has 400 cells in each Cartesian direction. The simulation runs for 1400 seconds using $\mathsf{CFL}=0.4$. The complete model with radiation and constant reaction rate is used, assuming the simplification in Eq.~\eqref{eq:rhocpApprx}. The S2M model is used to represent the effect of moisture. The model parameters are taken from Tab.~\ref{table:general-parameters1}--\ref{table:general-parameters3}, unless otherwise specified.

\begin{figure}
    \centering
    \begin{subfigure}{0.49\textwidth}
              \includegraphics[width=\linewidth]{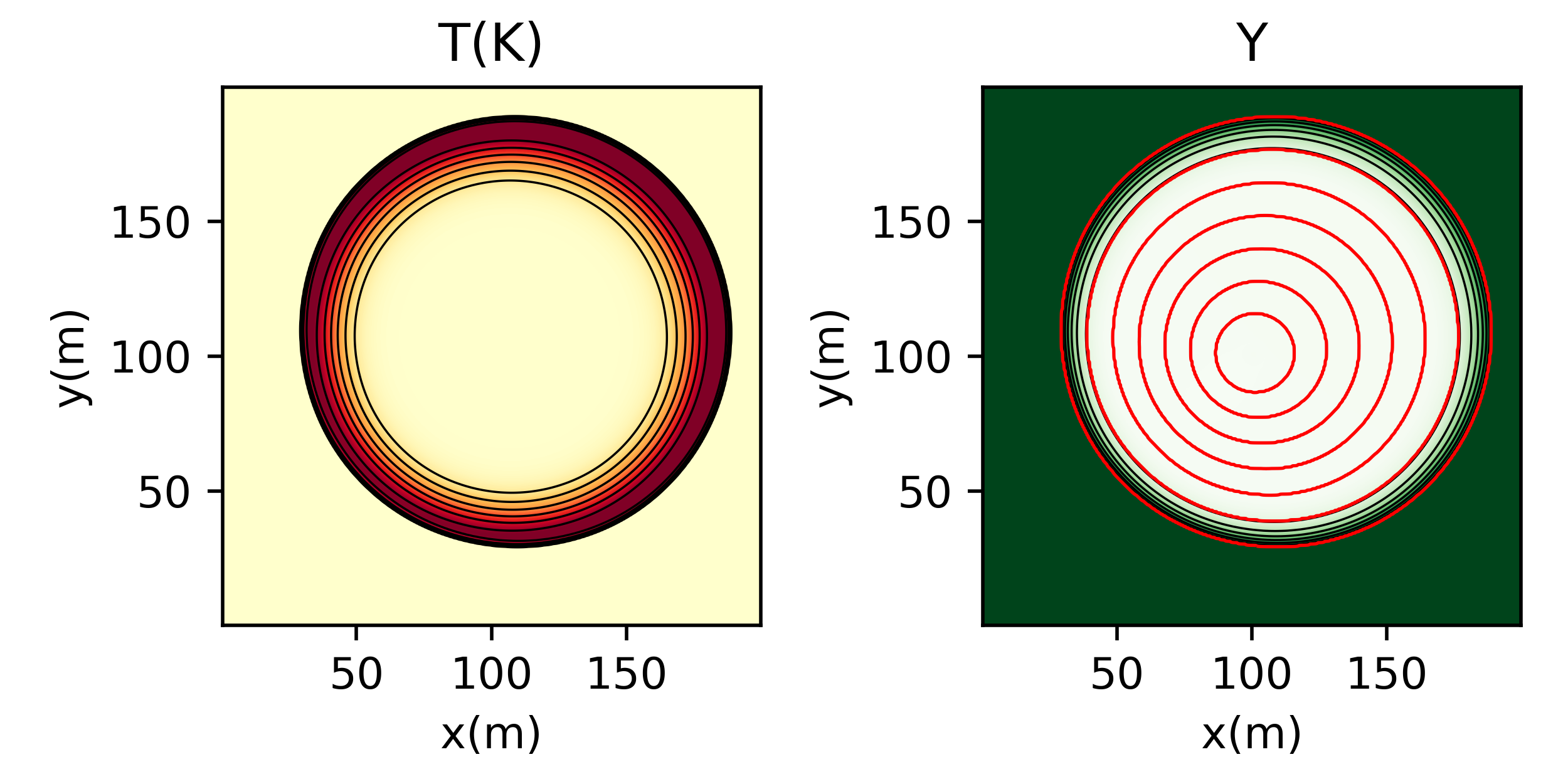} 
                  \caption{M=0.0}
    \end{subfigure}
  \begin{subfigure}{0.49\textwidth}
              \includegraphics[width=\linewidth]{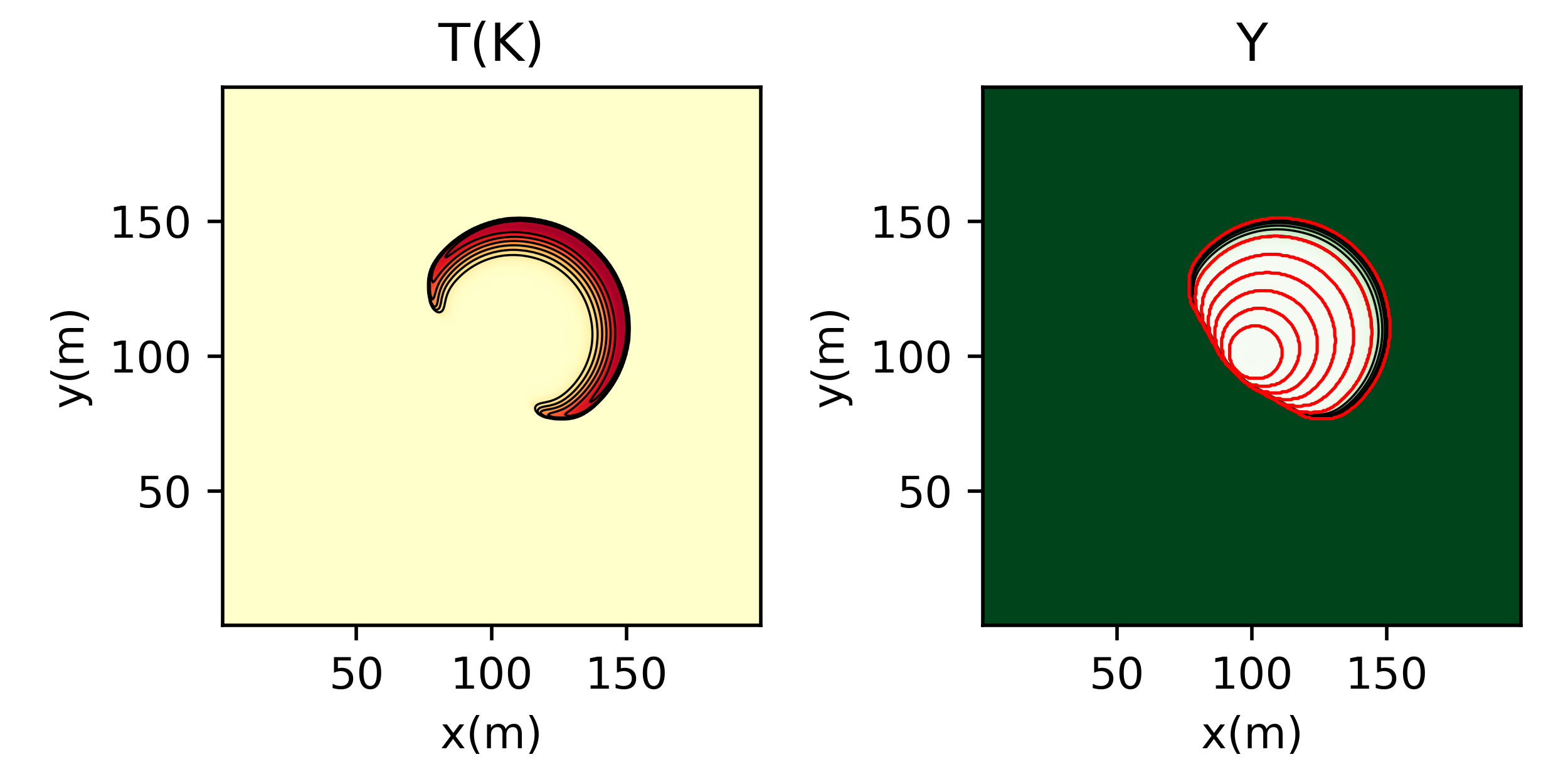}
               \caption{M=0.2}
    \end{subfigure}
  \begin{subfigure}{0.49\textwidth}
              \includegraphics[width=\linewidth]{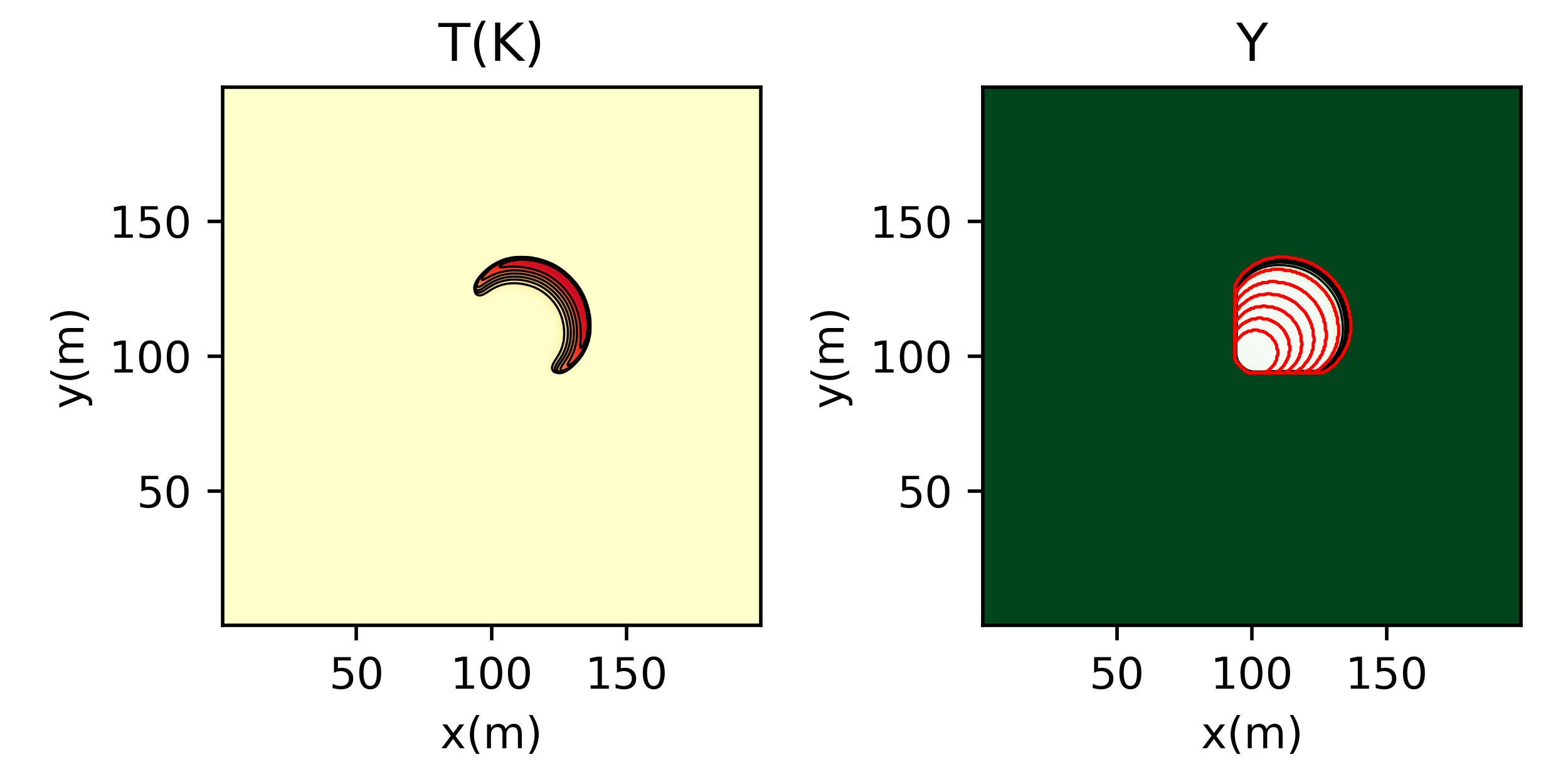} 
               \caption{M=0.4}
              \end{subfigure}
  \begin{subfigure}{0.49\textwidth}
              \includegraphics[width=\linewidth]{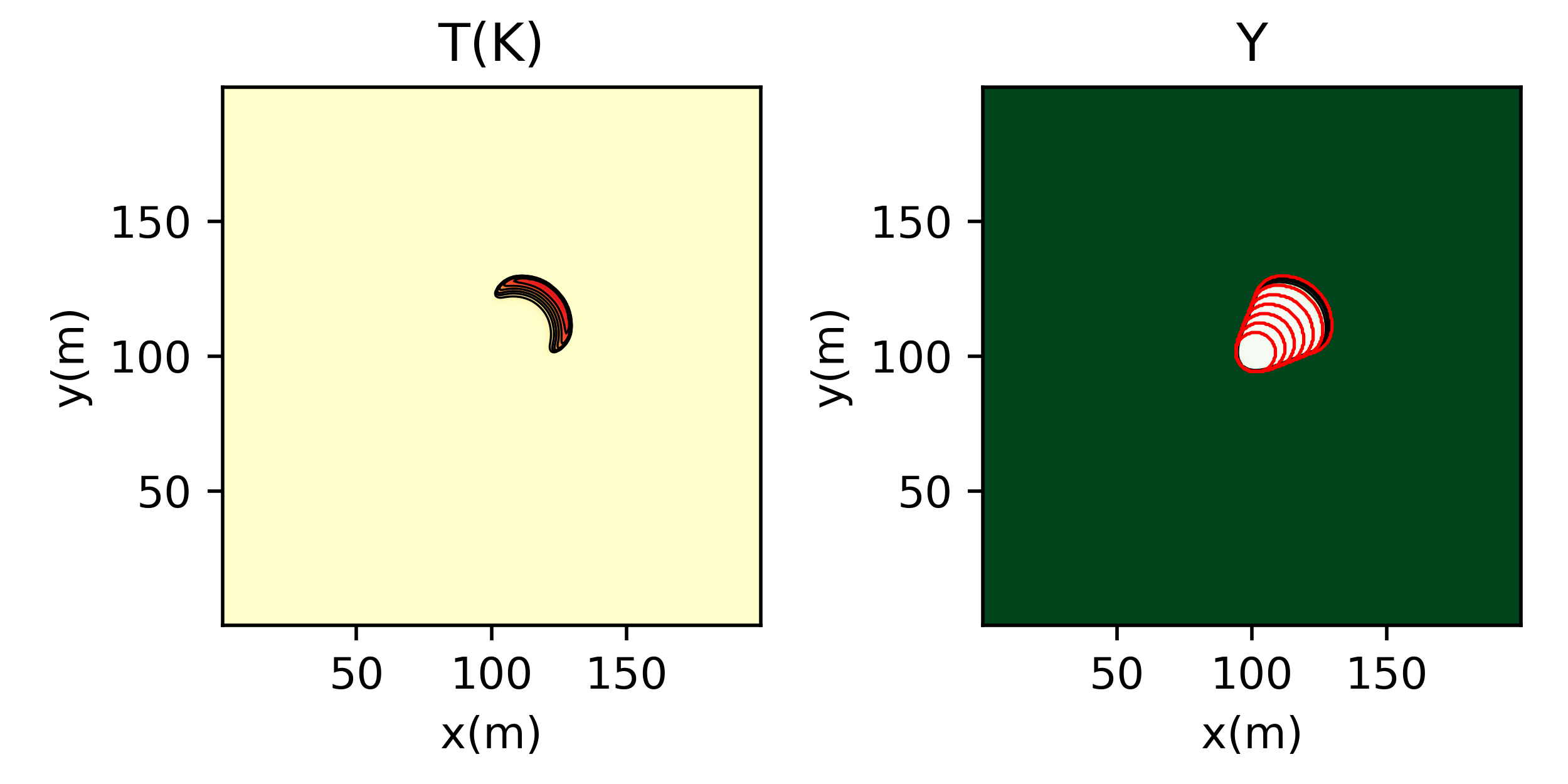} 
               \caption{M=0.6}
              \end{subfigure}
    \caption{Fire spread for different moisture content levels with a wind velocity $\mathbf{w}=(0.25,0.25)^T$ m/s, starting at $(x,y)=(100,100)$. Temperature and remaining fuel mass fraction are shown at the final time. The red isochrones provide information on the temporal evolution of the fire. Contour levels are depicted from 300~K to 1400~K.}
    \label{fig:ROS_moisture2D_exp1}
\end{figure}

Four different simulations are compared considering different moisture levels,  $M=0.0$, $M=0.2$, $M=0.4$, $M=0.6$. Fig. \ref{fig:ROS_moisture2D_exp1} shows the predicted temperature and biomass distribution at the final time. Besides, the fire isochrones are depicted in red every 200 seconds. These results evidence that the fuel moisture level has a strong impact on the shape of the burned area, as well as on the rate of spread. When considering zero moisture, the propagation takes place in all directions, being slightly faster in the direction of wind. Contrarily, when increasing the moisture level, propagation mainly takes place in the direction of wind. Besides, the width of the fire front and the rate of spread reduces as the moisture level increases. 
From a geometric perspective, the angle of the fire front decreases with increasing moisture level.

The reduction of the rate of spread in direction of the wind is in accordance with Fig.~\ref{fig:ROS_moisture} showing the decrease in spreading rate with increasing moisture level. 
More detailed, the spread orthogonal to the wind direction is reduced the most. This is in line with the zero wind case of Fig.~\ref{fig:ROS_moisture} where the rate of spread drops to zero for moisture levels above 80\%. 
In contrast, for non-zero wind speed, the decrease in the rate of spread is less significant and reaches a non-zero threshold. 
The two-dimensional simulation combines these two results when regarding either the direction of the wind or the orthogonal direction. As the moisture levels regarded are all below the critical threshold of $M=0.8$, there is still a fire spread orthogonal to the wind velocity, but the rate of spread in this direction is small. 

Tab.~\ref{table:burned_area} shows the total burned area at the final time of the simulation. 
The burned area is calculated considering the points with a biomass fraction below the threshold $Y=0.99$. The results show a strong reduction of burned area for higher moisture levels, e.g. a reduction of 80\% for a moisture level $M=0.2$ compared to the no moisture case.

\begin{table}
\centering
\caption{Burned area at the final time ($t=1400$ s) of the simulation, threshold for burning $Y<0.99$.}
\begin{tabular}{r|ccccc}
$M$ & 0.0  & 0.2  & 0.4 & 0.6   \\
\hline
Burned area (m$^2$)  & 19979.5  & 4052.0 & 1564.0 & 890.25 \\
\end{tabular}
\label{table:burned_area}
\end{table}

Including moisture in the model shows a significant difference in the system behavior, which is even more crucial in the two-dimensional setting. 
The burned area and the rate of spread differ strongly in the direction of the wind and orthogonal to it.

\subsubsection{Propagation in heterogeneous environments: assessing the effect of fuel moisture}

The aim of this case is to understand the model behavior when changing the fuel moisture in a heterogeneous landscape composed of four quadrants with different fuel moisture content. 
The computational domain is $[0,200]\times[0,200]$ m$^2$. The initial condition for the temperature is given by
\begin{equation}\label{eq:icreal2}
   T(x,y,0) = 
		\left\{
		\begin{array}{lll}
      	670 \mbox{ K}  &\mbox{if} &  r(x,y)< 5 \mbox{ m}, \\
        300 \mbox{ K} & \multicolumn{2}{l}{\mbox{otherwise}}  \\  
		\end{array}
		\right. 
\end{equation}
with $r$ in Eq.~\eqref{colliding3} and $(x_1,y_1)=(100,100)$ m. 
The moisture content differs in the four quadrants,
\begin{equation}\label{eq:fmic}
   M(x,y) = 
		\left\{
		\begin{array}{lll}
      	0.0   &\mbox{if} &  x<100 \mbox{ and } y>100, \\
        0.1   &\mbox{if} &  x>100 \mbox{ and } y>100, \\ 
        0.2   &\mbox{if} &  x<100 \mbox{ and } y<100 ,\\ 
        0.3   &\mbox{if} &  x>100 \mbox{ and } y<100 .
		\end{array}
		\right. 
\end{equation}

The computational mesh has 400 cells in each Cartesian direction. The simulation runs for 1400 seconds using $\mathsf{CFL}=0.4$. The complete model with radiation and constant reaction rate is used, assuming the simplification in Eq.~\eqref{eq:rhocpApprx}.
The S2M model is used to represent the effect of moisture. The model parameters are taken from Tabs.~\ref{table:general-parameters1}--\ref{table:general-parameters3}, unless otherwise specified, with zero wind conditions.

\begin{figure}
    \centering
        \includegraphics[width=0.9\linewidth]{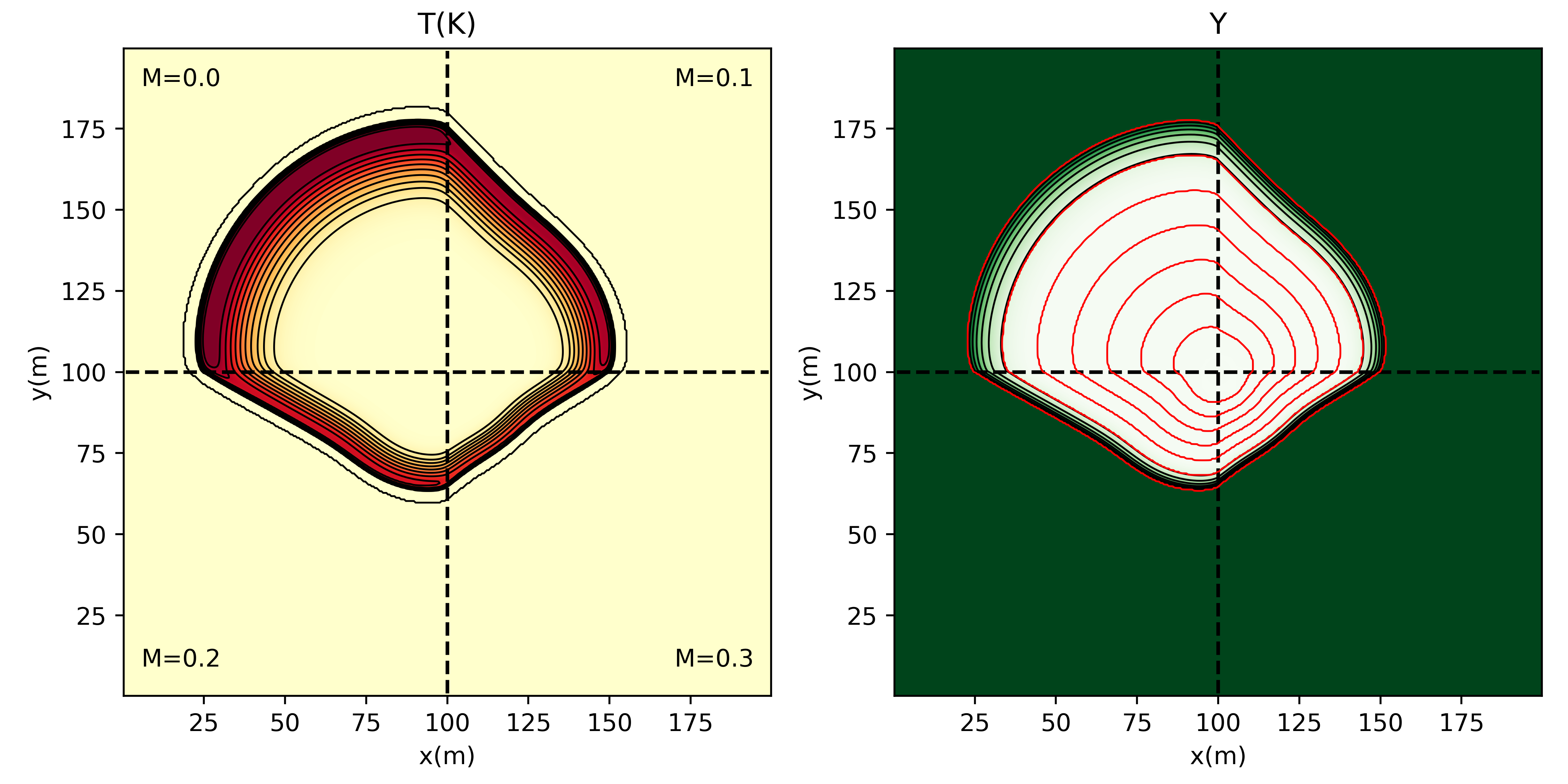} 
    \caption{Temperature (left) and biomass fraction (right) for fire propagation in a case with a different moisture content in each quadrant.  The isochrones show continuous connections of the solution at the intersections of different moisture content regions. Contour levels are depicted from 300 K to 1400 K.}
    \label{fig:ROS_FM2D_exp2}
\end{figure}

Figure \ref{fig:ROS_FM2D_exp2} shows the predicted temperature and biomass distribution at the final simulation time. Besides, the fire isochrones are depicted in red every 200 seconds. 
The results evidence that the ROS is higher in those quadrants with lower fuel moisture content, as observed in the one-dimensional setting of Fig.~\ref{fig:ROS_moisture_a} and the two-dimensional setting of Fig.~\ref{fig:ROS_moisture2D_exp1}.

The fuel moisture content has discontinuous jumps at the interface of two quadrants. However, the simulation results in Fig.~\ref{fig:ROS_FM2D_exp2} show that those sharp discontinuities can be handled by the model, without producing any numerical artifact at the interfaces. Based on these results, it is possible to have smaller patches of heterogeneous moisture content levels, modeling more realistic scenarios, without running into numerical problems due to discontinuous parameters.

\subsubsection{Propagation in heterogeneous environments: assessing the effect of surface coverage}

The aim of this case is to understand the model behavior when changing the fuel load by modifying the surface coverage, $SC$, assuming that the fuel load is a limiting factor  for the parameters chosen here --i.e., corresponding to the region of Fig \ref{fig:ros_BD} where the ROS increases with fuel load--.  Those areas with a lower value of  $SC$ are expected to act as a firebreak. The computational domain is $[0,200]\times[0,200]$ m$^2$. The initial condition for the temperature is given by
\begin{equation}\label{eq:icreal33}
   T(x,y,0) = 
		\left\{
		\begin{array}{lll}
      	670 \mbox{ K}  &\mbox{if} &  r(x,y)< 5 \mbox{ m}, \\
        300 \mbox{ K} & \multicolumn{2}{l}{\mbox{otherwise}}  \\  
		\end{array}
		\right. 
\end{equation}
with $r$ in Eq.~\eqref{colliding3} and $(x_1,y_1)=(50,50)$ m.   
The surface coverage is set as a piecewise constant distribution in space
\begin{equation}\label{eq:icreal3}
   SC(x,y) = 
		\left\{
		\begin{array}{lll}
      	1.0  &\mbox{if} &  x\leq 100 \mbox{ m}, \\
        SC_r  &\mbox{if} &  x> 100 \mbox{ m}  \\  
		\end{array}
		\right. 
\end{equation}

The computational mesh has 400 cells in each Cartesian direction. The simulation runs for 2200 seconds using $\mathsf{CFL}=0.4$. The complete model with radiation and constant reaction rate is used, assuming the simplification in Eq.~\eqref{eq:rhocpApprx}. The S2M model is used to represent the effect of moisture. The model parameters are taken from Tabs.~\ref{table:general-parameters1}--\ref{table:general-parameters3}, unless otherwise specified above.

\begin{figure}
    \centering
    \begin{subfigure}{0.49\textwidth}
                \includegraphics[width=\linewidth]{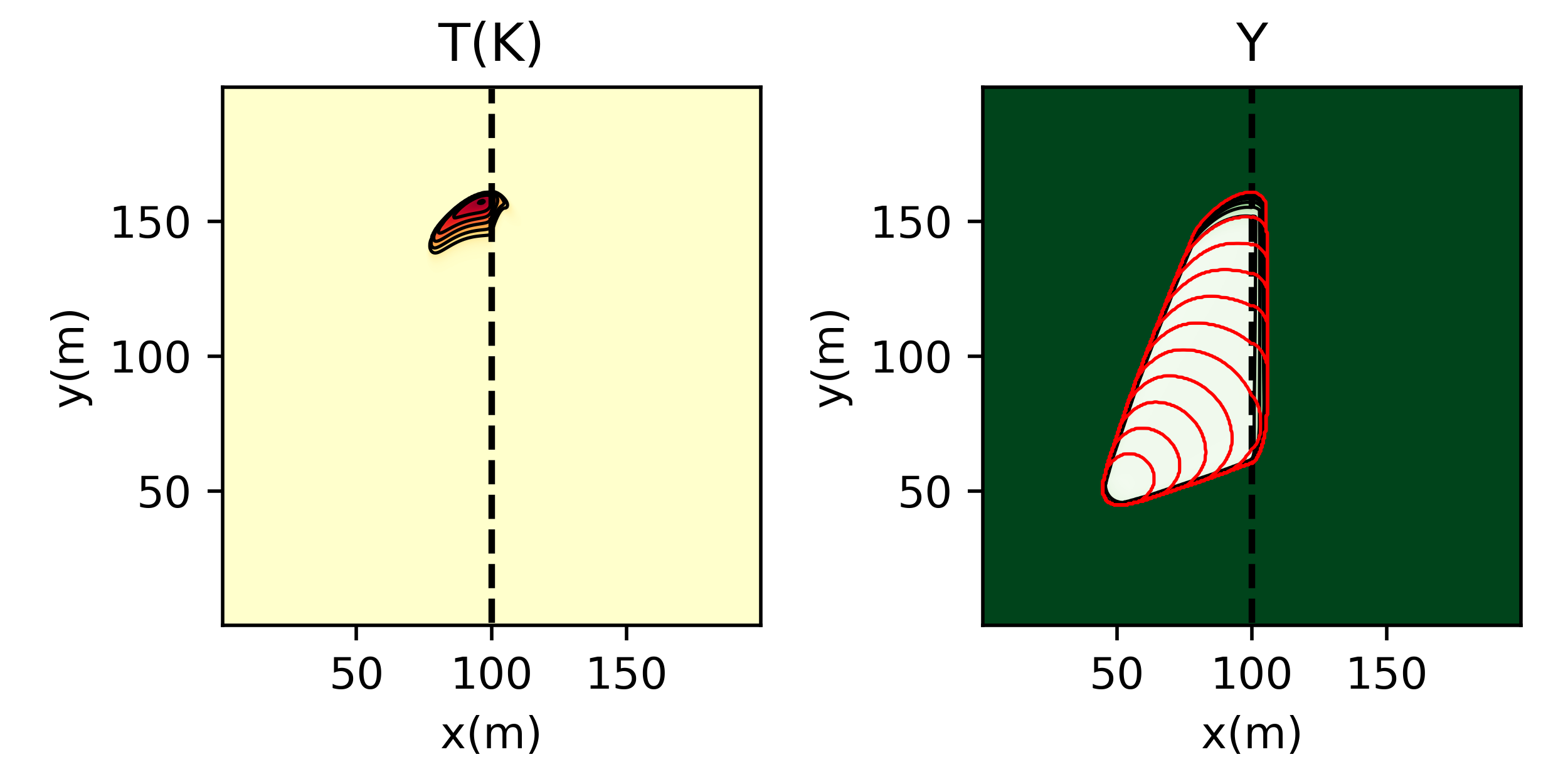} 
                \caption{$SC_r=0.2$}
                \label{fig:ROS_SC2D_exp3_a}
    \end{subfigure}
        \begin{subfigure}{0.49\textwidth}
                 \includegraphics[width=\linewidth]{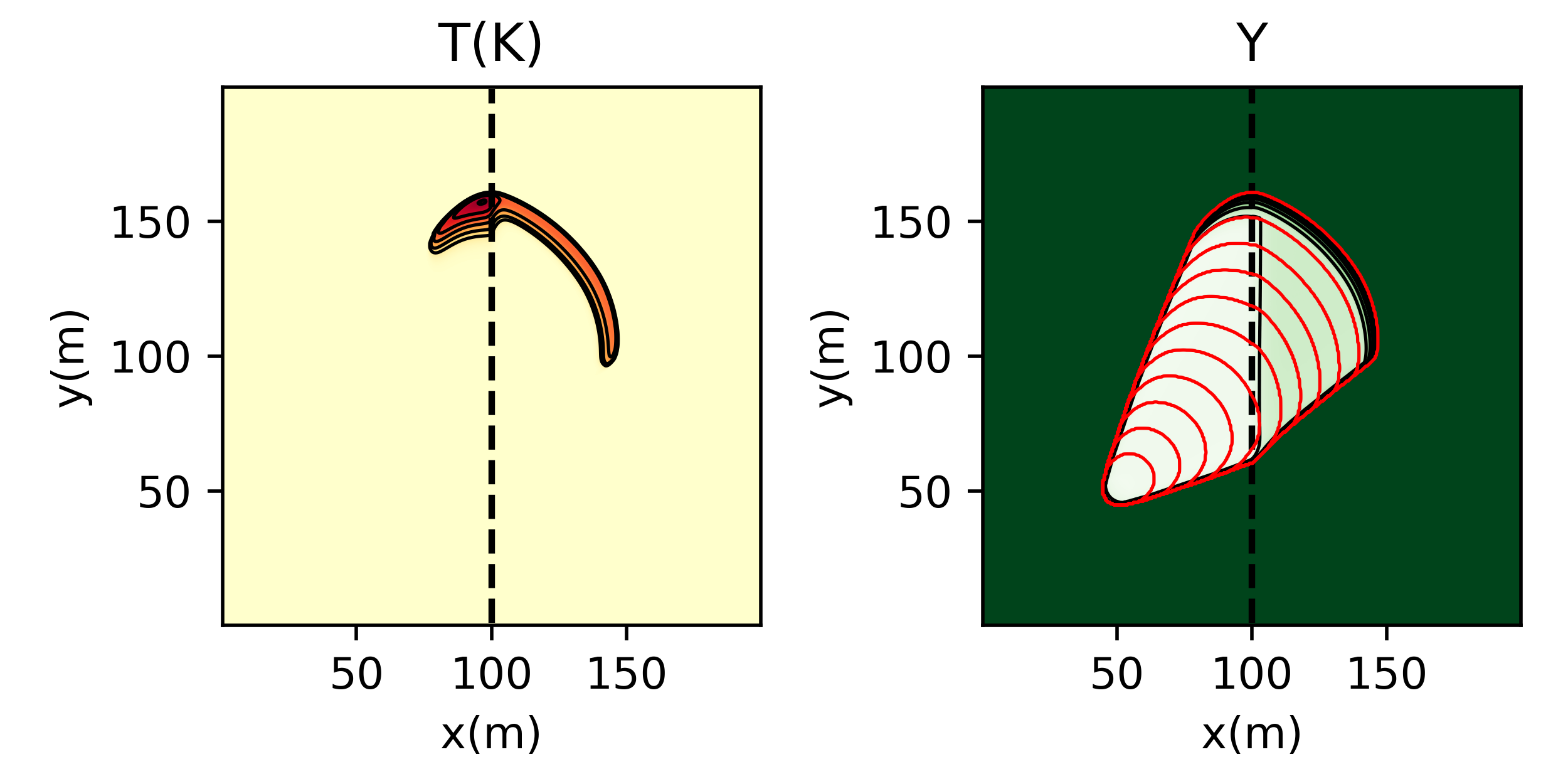} 
                    \caption{$SC_r=0.4$}
    \end{subfigure}
        \begin{subfigure}{0.49\textwidth}
                \includegraphics[width=\linewidth]{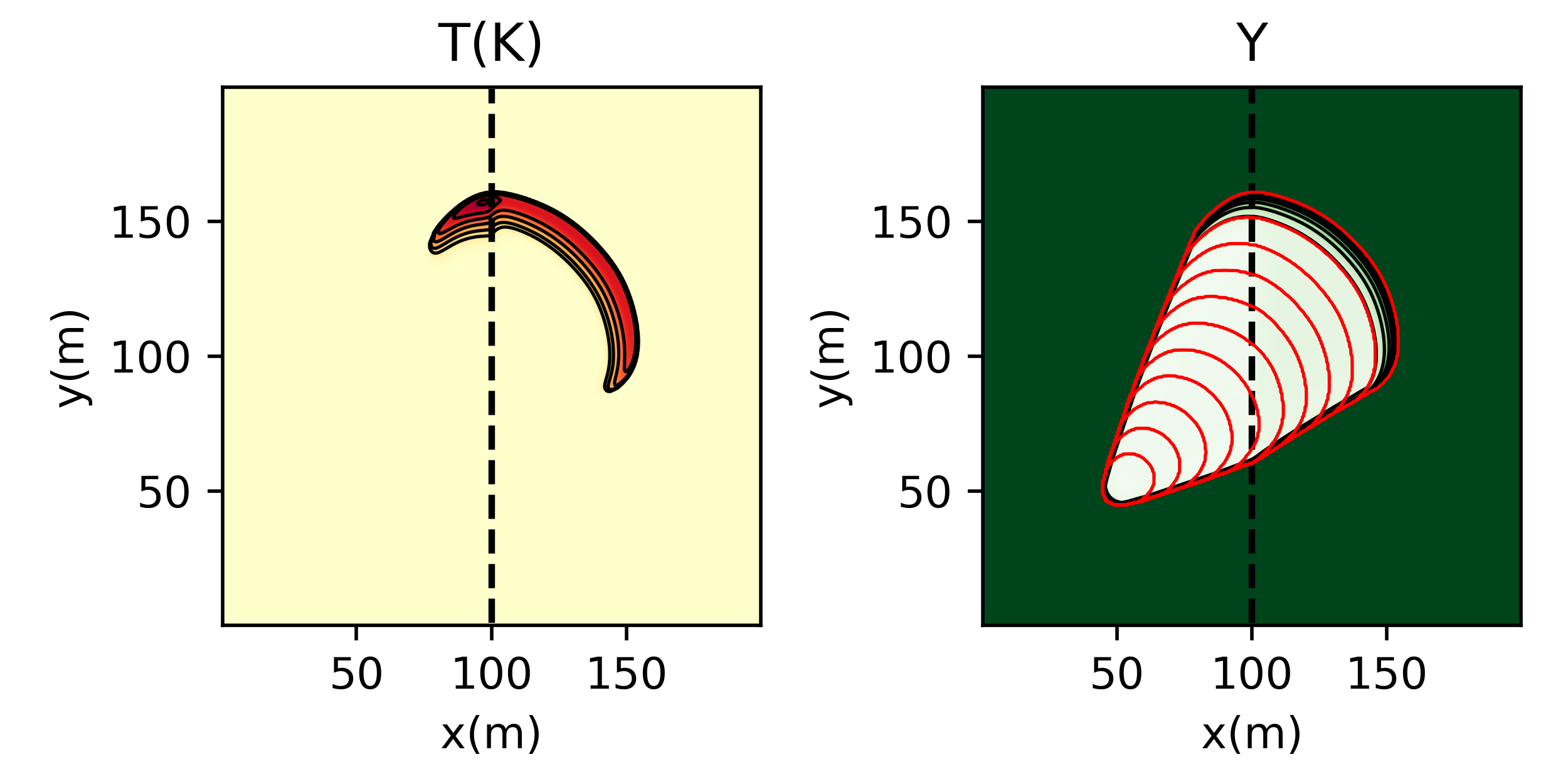} 
                   \caption{$SC_r=0.6$}
    \end{subfigure}
        \begin{subfigure}{0.49\textwidth}
                   \includegraphics[width=\linewidth]{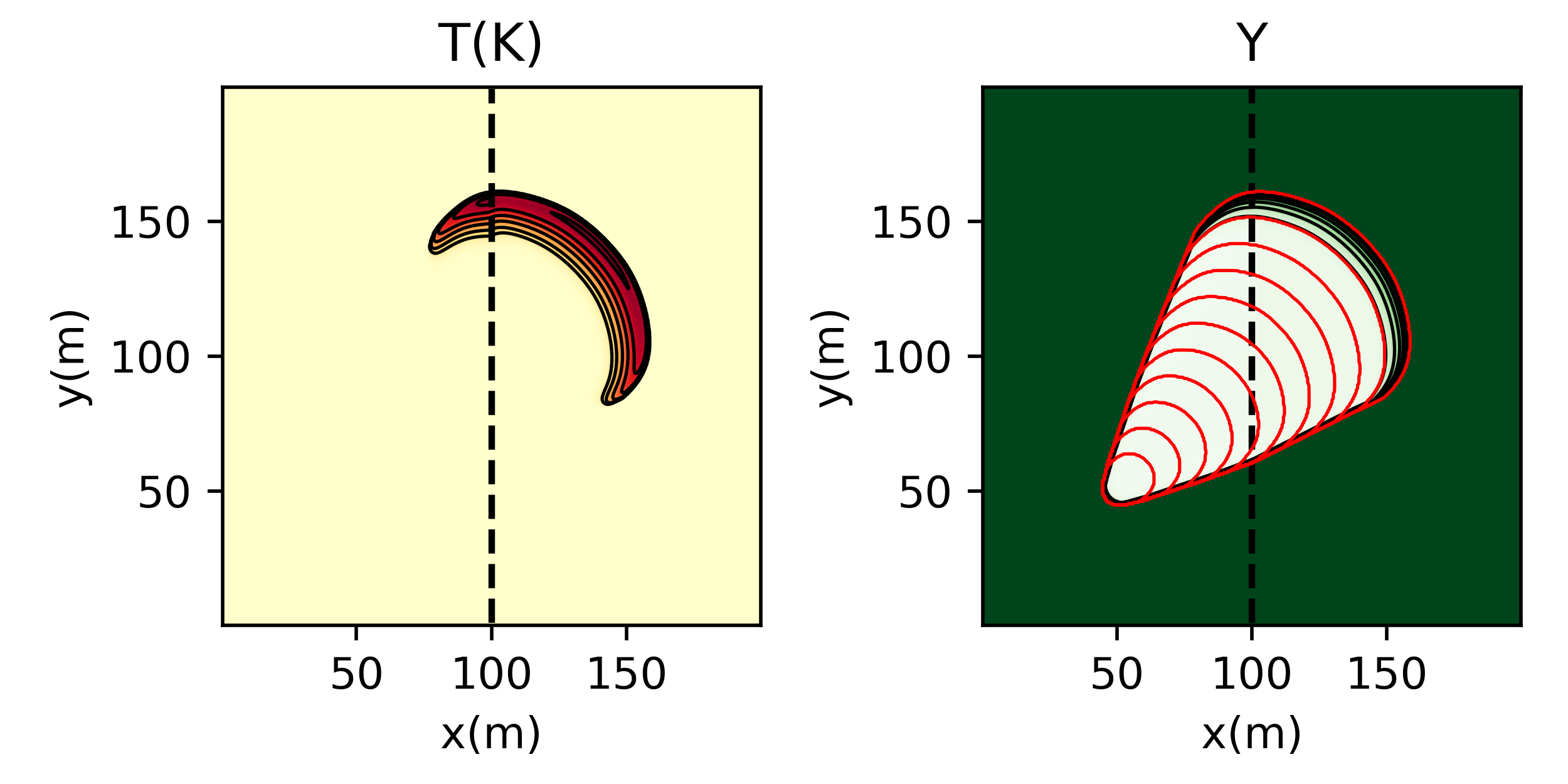} 
                      \caption{$SC_r=0.8$}
    \end{subfigure}
    \caption{Fire spread at the interface of a changing surface coverage ($SC$), starting from $(50,50)$ m with a wind velocity $\mathbf{w}=(0.25,0.25)^T$ m/s. The temperature and remaining fuel mass fraction are shown at final time, the isochrones give information on the temporal evolution. The area with $SC=0.2$ acts as a fire break, areas with low $SC$ slow down the rate of spread. Contour levels are depicted from 300 K to 1400 K.  }
    \label{fig:ROS_SC2D_exp3}
\end{figure}

Four different simulations consider different fuel surface coverages, $SC_r$, namely $SC_r=0.2$, $SC_r=0.4$, $SC_r=0.6$, $SC_r=0.8$, setting a wind velocity $\mathbf{w}=(1,1)^T$ m/s. Fig. \ref{fig:ROS_SC2D_exp3} shows the predicted temperature and biomass distribution at the final time. Besides, the fire isochrones are depicted in red every 200 seconds. 
These results confirm that higher fuel loads, regarded as higher $SC$, enhance fire propagation. Besides, there is a threshold at which the fire extinguishes and propagation stops, acting as a firebreak. In this example, the threshold appears for a value of $SC$ between 0.2 and 0.3.  
Fig.~\ref{fig:ROS_SC2D_exp3} shows a strong dependency on stopping the fire on the surface coverage but, in oppose to the influence of the moisture content, the rate of spread is less strongly affected by the reduced surface coverage. 
This tendency is supported by the one-dimensional simulation in Fig.~\ref{fig:ros_SC}, where the drop from a higher rate of spread for high surface coverage values to zero for small surface coverage values is very abrupt. 

As in the previous case, these results also evidence that sharp discontinuities in the fuel load can be adequately handled by the model, without producing numerical artifacts at the interface.

\subsubsection{Propagation in heterogeneous environments with topography }

This case examines fire propagation across uneven vegetation, both with and without the influence of topography. The computational domain is $[0,200]\times[0,200]$ m$^2$. The initial condition for the temperature is given by
\begin{equation}\label{eq:icreal34}
   T(x,y,0) = 
		\left\{
		\begin{array}{lll}
      	670 \mbox{ K}  &\mbox{if} &  r(x,y)< 5 \mbox{ m}, \\
        300 \mbox{ K} & \multicolumn{2}{l}{\mbox{otherwise}}  \\  
		\end{array}
		\right. 
\end{equation}
with $r$ in Eq.~\eqref{colliding3} and $(x_1,y_1)=(100,50)$ m.   
A heterogeneous biomass distribution  is modeled by the function
\begin{equation}\label{eq:Rfcase}
W_{f,0}(x,y)=\sin\left(\frac{x}{20.0}\right) \cos\left(\frac{y}{10.0} - 20.0\right) + \sin\left(\frac{x}{20.0} - 40.0\right)^2 \sin\left(\frac{y}{10.0}\right) + 2.0 .
\end{equation}
setting $\bar{\rho}_{f}=400$~kg/m$^3$ and $R_{f,0}=0.01$.

Two different scenarios are considered, the first subcase considers flat topography whereas the second subcase considers a smoothly varying topography given by
\begin{equation}\label{eq:Zcase}
Z(x,y)=50  \exp\left(-\frac{(x - 200)^2 + (y - 200)^2}{20000}\right).
\end{equation}

The computational mesh has 400 cells in each Cartesian direction. The simulation runs for 2600 seconds using $\mathsf{CFL}=0.4$. The complete model with radiation and constant reaction rate  is used, assuming the simplification in Eq.~\eqref{eq:rhocpApprx}. The bulk density and heat capacity are computed using the more realistic mixture approximation in Eq.~\eqref{eq:rhocp0}.  The moisture is zero and the wind is given by $\mathbf{w}=(0.0,0.5)^T$~m/s. The model parameters are taken from Tables~\ref{table:general-parameters1}--\ref{table:general-parameters3}, unless otherwise specified above.

\begin{figure}
    \centering
        \begin{subfigure}{0.99\textwidth}
          \includegraphics[width=\linewidth]{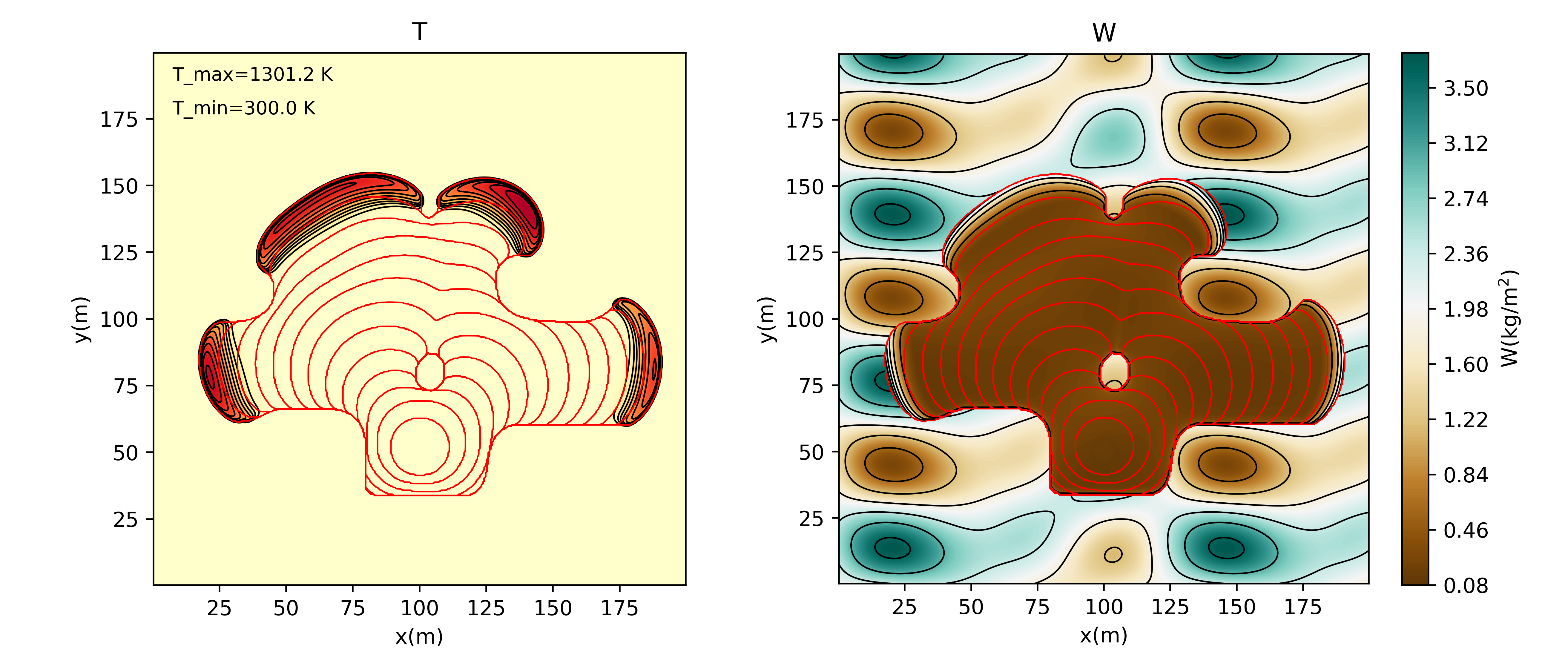}   
        \caption{Flat terrain}
        \end{subfigure}
        \begin{subfigure}{0.99\textwidth}\label{fig:ROS_FM2D_exp4_2_b}
        \includegraphics[width=\linewidth]{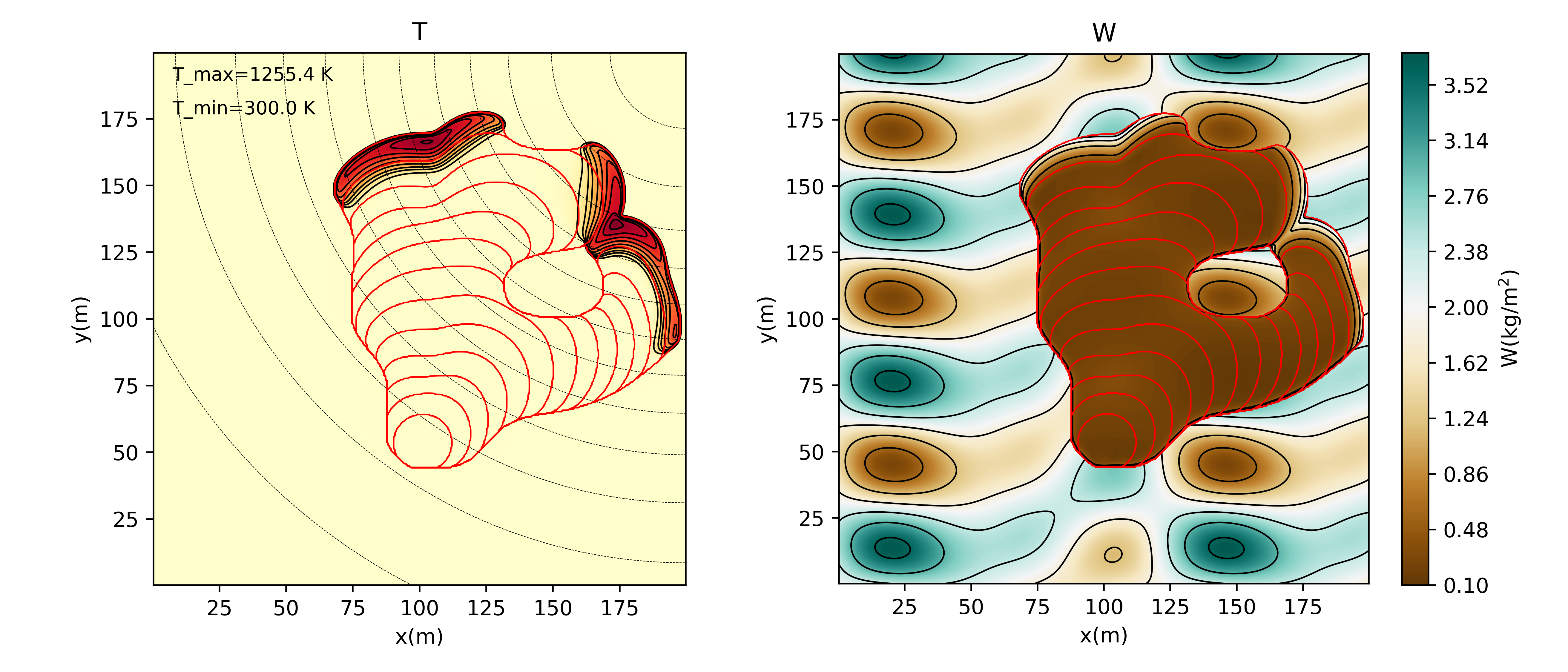} 
        \caption{Topography with gradient pointing towards $(x,y)=(200,200)$ }
        \end{subfigure}
    \caption{Temperature (left) and fuel load (right) at the final time for the simulation of fire with (b) and without topography (a), considering spatially heterogeneous fuel with wind $\mathbf{w}=(0.0,0.5)^T$ m/s, $M=0.0$, and the mixture approximation in Eq.~\eqref{eq:rhocp0}.  
    (a) Flat terrain is considered: The fire spreads faster in areas with high fuel load, influenced by the wind in $y-$direction. (b) Gaussian topography is considered: The slope towards $(x,y)=(200,200)$ enhances the fire spread in this direction. }
    \label{fig:ROS_FM2D_exp4_2}
\end{figure}

Fig.~\ref{fig:ROS_FM2D_exp4_2} shows the evolution of the fire (isocrhones) and the final state for the temperature and fuel load. A tendency to spread to areas with higher fuel load is observed. 
Additionally, there are fire breaks for small fuel loads, which is comparable to the situation in Fig.~\ref{fig:ROS_SC2D_exp3_a}. 
The wind direction is parallel to the $y$-axis, but there is a strong propagation as well in the areas with high fuel load orthogonal to the wind direction. 

Fig.~\ref{fig:ROS_FM2D_exp4_2} shows the spread of fire in the same biomass configuration but now with additional uneven topography. 
The topography acts as an additional advective influence, compare Sec.~\ref{subsec:modelling_wind}.
The topography in Eq.~\eqref{eq:Rfcase} has a gradient pointing towards $(x,y)=(200, 200)$ m, therefore there is an additional advective influence towards the upper right-hand corner. 
Fig.~\ref{fig:ROS_FM2D_exp4_2} shows this influence by having a stronger tendency to spread towards this corner.

\subsubsection{Validation using an experimental field study: case F19 }

The experimental case F19 \cite{mell2007physics} belongs to a set of experiments conducted in the Northern Territory of Australia, during July and August 1986 \cite{cheney1995fire,cheney1993influence}. For all the experiments, fuels were open grasslands and grasses were continuous and normally fully cured; besides, the topography was flat. For case F19, the fuel was \textit{Themeda} grass. The grass-land plot was 200 × 200 m and the ignition line fire was 175 m long. This line fire was created with drip torches carried by two field workers walking for 56 s in opposite directions from the center point to the ends of the line fire. The average wind speed was  4.8 m/s. The resulting average rate of spread is around 1.45 m/s (87 m/min).

The experiment F19 has been extensively used in the literature for model validation  \cite{mell2007physics} and is herein reproduced using the proposed model for the same purpose. A spatial domain $[0,250]\times[0,250]$ m$^2$ is considered, setting the remaining biomass fraction at zero along a 25 m wide strip along the edges of the domain. The  resolution of the mesh is 1 m, which leads to an average time step of 0.05 s. The initial condition is set as a constant temperature --ambient temperature-- in the whole domain. To reproduce the ignition process, a moving heat source term is added on the left side of the domain, mimicking the operation carried out by the field workers.

The parameter values for this case are shown in Table \ref{table:F19param}. It must be noted that there are different choices for the parameters reproducing the observed ROS, with differences in the width of the fire front, maximal temperature, and other variables. Furthermore, note that the ROS in this case is unusually high, which amplifies the difference between the parameters in Table \ref{table:F19param} and the default parameters. A domain height of $l=1$ m is considered, but other values will be explored in the future to obtain more suitable parameter values, e.g. for the optical path length. 

\begin{figure}[!ht]
    \centering
        \includegraphics[width=0.6\linewidth]{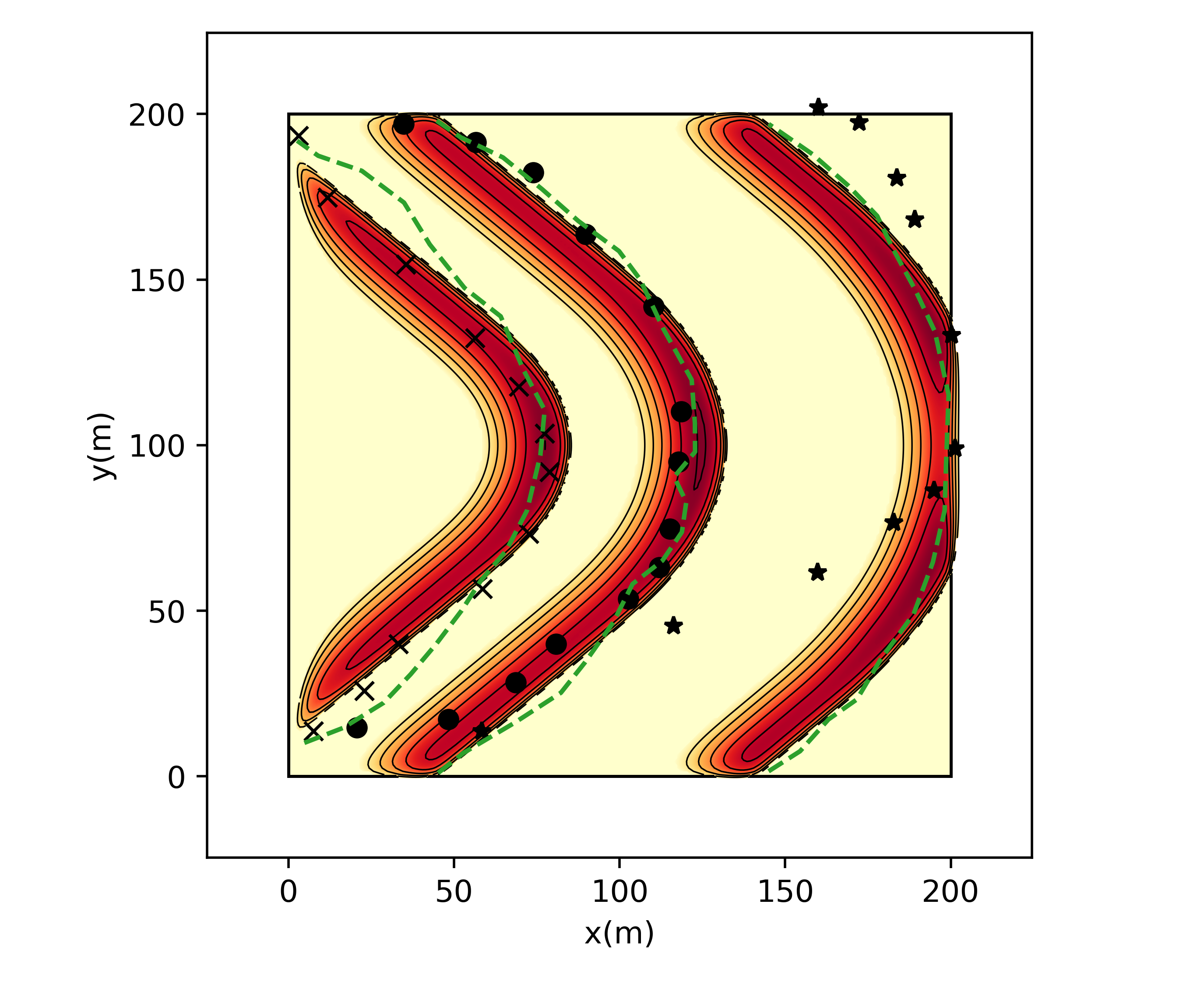} 
    \caption{Fire fronts at $t=56$ s, $t=86$ s and $t=138$ s reproduced by the model (shading and contour lines), compared with the experimental measurements from \cite{cheney1995fire,cheney1993influence}  (black markers) and the numerical results provided by the WFDS model from  \cite{mell2007physics} (green dashed line). Contour lines are shown for the temperature every 100 K, from 900 K to 1400 K. Temperatures below 900 K are represented in light yellow (background color) to better visualize the fire front.}
    \label{fig:caseF19}
\end{figure}

Figure \ref{fig:caseF19} shows the fire front at $t=56$ s, $t=86$ s and $t=138$ s reproduced by the model and it is compared with the experimental measurements from \cite{cheney1993influence,cheney1995fire}  and the numerical results provided by the WFDS model from  \cite{mell2007physics}. The proposed model reproduces the reported ROS and the shape of the fire front, which evolves from a triangular shape to a more rounded shape. The computed results closely match those of the WFDS model  \cite{mell2007physics}.

\begin{table}
\centering
\caption{Parameter values for the case F19, adapted from  \cite{mell2007physics}. }
\label{table:F19param}
\begin{tabular}{lccc}
Parameter & Symbol  & Value & Units \\
\hline
Ground elevation  & $Z$ & 0 &\hbox{m} \\
Wind velocity  & $\mathbf{w}$ & 4.8 &\hbox{m·s$^{-1}$} \\
\hline
Initial fuel volume ratio & $R_{f,0}$  & 0.00061 &\hbox{-}  \\
Fuel density  & $\bar{\rho}_{f}$ & 512 &\hbox{kg·m$^{-3}$}  \\
Initial fuel load & $W_{f,0}$ & 0.31 &\hbox{kg·m$^{-2}$}  \\
Specific heat of dry fuel  & $c_{p,f0}$ & 1.5 &\hbox{kJ·kg$^{-1}$·K$^{-1}$} \\
Fuel pyrolysis temperature  & $T_{pc}$ & 400 &\hbox{K} \\
Fuel combustion heat  & $\mathcal{H}$ &15000 &\hbox{kJ·kg$^{-1}$} \\
Reaction rate  & $A_c$ & 0.1 &\hbox{s$^{-1}$} \\
Fuel moisture content& $M$ & 0.06 &\hbox{-} \\
\hline
Thermal conductivity  & $k_c$ & 0.0001&\hbox{kW·m$^{-1}$· K$^{-1}$}  \\
Optical path length & $\delta$ & 10.0 & \hbox{m} \\
Ambient cooling coefficient  & $\hat{\alpha} $ & 0.2&\hbox{kW·m$^{-2}$·K$^{-1}$}  \\
Wind correction coefficient  & $\beta$ & 0.15 &\hbox{-}  \\
\end{tabular}
\end{table}

\section{Conclusions}\label{sec:conclusion}

We presented an ADR wildfire propagation model that accounts for the effect of fuel moisture, and also considers wind, local radiation, natural convection, and topography. Special attention has been given to the derivation of the governing equations from the fundamental conservation laws within an idealized vegetation stratum which is modelled as a two-phase porous flow. The proposed model includes different mechanisms for live and dead fuel by using the apparent calorific capacity method, which models the effect of thermal phase changes through an apparent heat capacity.
Three different moisture models--the S2M, C2M, and S3M--have been presented  to represent the evaporation process and the constituents of fuel with different levels of detail, being the S2M the preferred option when total fuel  moisture data are available and the C2M model used when willing to distinguish between live and dead fuels. The use of the C2M model will benefit wildfire propagation predictions, because dead fuel moisture is often considered to be the main control on fire characteristics and fire impact on fuel and vegetation, see, e.g., \cite{ rakhmatulina2021soil,wagtendonk1977refined}. A limitation of this model is that the relative humidity  does not vary over time. A more complete approach, see, e.g., \cite{perello2024adaptable}, could be easily adapted. Also note that the utilization of dead fuel moisture data in the model, if available,  is straightforward by substituting this approach --i.e., $M_d$ would be considered as an input parameter--.

Our work provides new insights on ADR-based wildfire propagation modeling that complement previous studies, e.g., \cite{babak_effect_2009, grasso_two-dimensional_2018,mandel2008wildland,mitra_studying_2024,nieding_impact_2024,reisch_analytical_2024,weber_toward_1991}. Key modeling decisions have been carefully analyzed and justified through both theoretical analysis and simulation results. One of the fundamental findings of this analysis is related to the degree of simplification that can be applied to the terms in the advection-diffusion-reaction equation. Whereas the combustion heat and atmospheric cooling source terms can be significantly simplified without excessively compromising the physical realism of the solution, the diffusion term requires, at least, a non-linear model for the diffusion parameter to obtain physically consistent results.  This is addressed through Rosseland's approximation of local radiation. With regard to the combustion heat source term, it is shown that the commonly used Arrhenius combustion function can be approximated by a constant as long as the switching function for activating and deactivating the combustion process is still included. This approximation is in line with the macroscopic scale of the wildfire model: The Arrhenius law is based on chemical processes on a much smaller length scale, and the resulting non-linear combustion term contrasts the model accuracy on the macroscopic scale. For the combustion model, the key parameters are the combustion heat of the fuel, $\mathcal{H}$, and the kinetic constant, $A_c$, the latter  depending on the physio-chemical structure of the fuel. With respect to radiation, the fundamental aspect when considering local radiation through Rosseland's approximation lies in the strong non-linearity of the diffusion term.  The dependence of the diffusion mechanism on $T^3$ links the ROS to the energy of the medium, which is proportional to its temperature, ensuring physically consistent results and reproducing the observed trends. The optical path length, $\delta$, is the key parameter for the Rosseland approximation and can be retained as a calibration parameter.

The parametric analyses presented herein provide evidence of the model's ability to reproduce the expected fire propagation behavior and the dependence of the ROS on the main drivers of fire spread. The model reproduces both the exponentially decaying trend of the ROS as fuel moisture increases (live fuels) and the nearly linearly  decaying trend of the ROS as relative humidity increases (dead fuels). Moreover, the model reproduces an increase in ROS with the fuel load when the latter is a limiting factor, showing a threshold for no-fire conditions. When the fuel load is not a limiting factor, a Ricker-like function behavior is reproduced, with an exponentially decaying ROS for large fuel load values.

A limitation of this model is that the effect of radiation is only considered locally. The model uses the Rosseland approximation to model the local radiative effects as a diffusion mechanism. Therefore, heat is transferred between adjacent areas in a manner similar to conduction and the long-distance effect of radiation is not represented, which results in short pre-heating times. Consequently, the influence of topography must be approximated as an advective phenomenon. The use of non-local radiation models would avoid this unphysical representation of topography and would improve the effect of wind, which should not be only represented through the advective term. In any case, the present approach based on local radiation offers a good balance between accuracy and complexity in the model.

So far, the optical path length $\delta$ was treated as a constant. A more detailed modeling approach may include a connection between the optical path length and the vegetation structure, for example, through the packing ratio $R_f$, the fuel fraction $Y$, or the radiation parameter $k$. Something else to explore in the future is the dependence of the wind correction coefficient, $\beta$, on the moisture content, which may improve the model's predictions in presence of strong wind conditions. Additionally, a more complete definition of the chemical reaction processes through the consideration of oxygen, water vapor, and combustion products would also improve the physical realism of the model.

Accounting for fuel moisture effects on wildfire dynamics opens possibilities to consider hydrological and plant physiological controls on the wildfire propagation.  This is already done in the context of wildfire risk, see, for example, \cite{ruffault2020increased, Ruffault:2022b}, but is not common for wildfire propagation.  Another interesting topic that could be explored with this model is the link between vegetation pattern dynamics in water-limited systems and wildfire propagation in these systems.  A spatially aggregated modeling approach has been applied in \cite{crompton2022fire} to study wildfire regime--vegetation feedback. In this context,  spatially distributed modeling of wildfire propagation could further provide insights into wildfire--ecosystem structure relations.

As a next step of this research, we plan to use the presented model to reproduce real-world fire events by calibrating the parameters in Table~\ref{table:general-parameters2}.  While we are confident that the model will reproduce the fire dynamics sufficiently well, a systematic study on real-world fire events is necessary to build confidence in the model's predictive capability. 

\section*{Acknowledgments}

This work has been funded by the Spanish Ministry of Science, Innovation and Universities - Agencia Estatal de Investigación (10.13039/501100011033) and FEDER-EU under project-nr. PID2022-141051NA-I00. This work was partially funded by the Aragón Government, DGA (Spain), through the Fondo Europeo de Desarrollo Regional (FEDER) under project-nr. T32-23R.  I. \"Ozgen-Xian's work is funded by the \textit{Tenure-Track-Programm} by the German Federal Ministry of Education and Research. A. Navas-Montilla thanks Prof. César Dopazo for the insightful discussions on wildfire modelling, which offered valuable perspectives and inspired potential future developments.

\appendix

\section{Energy conservation equations in the model}\label{appendix_eqs}

In this section, the conservation of energy and the assumptions made to develop the model herein are presented. Refer to  \cite{bird2002transport} for a complete description of these equations. For the sake of clarity, notation has been simplified and subscripts for the material properties for the fluid and solid cases have been omitted when analyzed independently.\vspace{6pt}

\noindent\textbf{Conservation of energy in a fluid:}\vspace{6pt}

The partial differential equation describing the conservation of internal energy, $e$, in a gas is expressed as follows
\begin{equation}\label{eq:energyeq1} 
     \frac{\partial \rho e}{\partial t}  + \nabla \cdot ( \rho e\mathbf{v}
     )   =  -\nabla\cdot  \mathbf{q}     -p\nabla\cdot \mathbf{v} + \phi_v +q_s\, ,
\end{equation} 
where $\rho$ is density, $p$ is pressure, $\mathbf{v}$ is the flow velocity, $\mathbf{q}$ is a diffusion heat flux --e.g., Fourier law and Rosseland approximation of radiation--, $p\nabla\cdot \mathbf{v}$ is the volume expansion term, $\phi_v$ is the viscous dissipation and $q_s$ represents other heat sources (expressed per unit volume). It must be noted that the volume expansion term, $p\nabla\cdot \mathbf{v}$, cannot be neglected even for low Mach number flows near the incompressibility condition, such as those considered in this paper.
 
 Making use of the definition of enthalpy, that is
\begin{equation}
     h=e+\frac{p}{\rho},
\end{equation} 
one can rewrite Eq.~\eqref{eq:energyeq1} as
\begin{equation}\label{eq:energyeq2} 
     \frac{\partial \rho h}{\partial t}  + \nabla \cdot ( \rho h\mathbf{v}
     )   = \frac{Dp}{Dt} -\nabla\cdot  \mathbf{q}    + \phi_v +q_s\, ,
\end{equation} 
where the term $\frac{D}{Dt}=\frac{\partial }{\partial t}  + \mathbf{v} \cdot \nabla $ represents the material derivative operator.

Now, by expanding the left-hand side of Eq.~\eqref{eq:energyeq2} and making use of the mass conservation equation---i.e.  $\frac{\partial \rho}{\partial t}  + \nabla \cdot ( \rho \mathbf{v}
     ) =0$, Eq.~\eqref{eq:energyeq2} is rewritten in its non-conservative form
\begin{equation}\label{eq:energyeq3} 
      \rho \frac{Dh}{Dt}   = \frac{Dp}{Dt} -\nabla\cdot  \mathbf{q}    + \phi_v +q_s\,.
\end{equation} 

By expressing enthalpy differential in terms of pressure and temperature \cite{bird2002transport},  Eq.~\eqref{eq:energyeq3} becomes
\begin{equation}\label{eq:energyeq4} 
      \rho c_p \frac{DT}{Dt}   = -\left(\frac{\partial \ln \rho}{\partial \ln T} \right)_p \frac{Dp}{Dt} -\nabla\cdot  \mathbf{q}    + \phi_v +q_s\, ,
\end{equation} 
where $c_p$ is the specific heat at constant pressure. Considering an ideal gas, the relation $\left(\frac{\partial \ln \rho}{\partial \ln T} \right)_p=-1$ holds and the previous equation becomes
\begin{equation}\label{eq:energyeq5} 
      \rho c_p \frac{DT}{Dt}   =  \frac{Dp}{Dt} -\nabla\cdot  \mathbf{q}    + \phi_v +q_s\,.
\end{equation}

When considering the application to wildfire modeling in this paper, a simple analysis of order of magnitude of the terms in Eq.~\eqref{eq:energyeq5} shows that the total derivative of pressure as well as the viscous dissipation term can be neglected. Comparing the magnitude of these terms with the left-hand side of Eq.~\eqref{eq:energyeq5}, we obtain
\begin{equation}
\left\{
\begin{aligned}
   \frac{\phi_v}{\rho c_p \frac{DT}{Dt} } \sim \frac{\nu U^2 \Delta t}{c_pL^2 \Delta T } \ll 1 \, , \qquad  \frac{\phi_v}{\rho c_p u\frac{\partial T}{\partial x} } \sim \frac{\nu U }{c_pL \Delta T } \ll 1 \, , \\
      \frac{\frac{Dp}{Dt}}{\rho c_p \frac{DT}{Dt} } \sim \frac{\Delta p }{\rho c_p \Delta T } \ll 1 \, ,
\end{aligned}
\right.      
\end{equation}
for the characteristic scales of our problem, being the time scale related to the heating time from ambient temperature to the maximal temperature ($\Delta T \sim 600$ K). The term $\Delta p$ is negligible for a constant atmospheric pressure.

Considering the previous assumptions, the equation for the conservation of energy can be written as follows
\begin{equation}\label{eq:energyeq5v2} 
      \rho c_p \frac{DT}{Dt}   =  -\nabla\cdot  \mathbf{q} +q_s\,.
\end{equation} 
Analogously,  the conservative form of the equation becomes
\begin{equation}\label{eq:energyeq2v2} 
     \frac{\partial \rho h}{\partial t}  + \nabla \cdot ( \rho h\mathbf{v}
     )   = -\nabla\cdot  \mathbf{q}    +q_s\, .
\end{equation} \vspace{6pt}

\noindent\textbf{Conservation of energy in a solid:}\vspace{6pt}

The partial differential equation for the conservation of energy, $e$, in a solid, also known as the heat equation, is expressed as 
\begin{equation}\label{eq:solideq1} 
     \frac{\partial \rho e}{\partial t}    =  -\nabla\cdot  \mathbf{q}   +q_s\, ,
\end{equation} 
where $\rho$ is the density of the solid, $\mathbf{q}$ is a diffusion heat flux and $q_s$ represents other heat sources. 

For a solid assumed incompressible, the specific heats at constant volume and pressure coincide, $c=c_v=c_p$. Therefore, internal energy and enthalpy can be considered equal, $e=h$.  If neglecting the variation in time of the material properties, we can rewrite equation Eq.~\eqref{eq:solideq1}  as 
\begin{equation}\label{eq:solideq2} 
      \rho c_p\frac{\partial T}{\partial t}    =  -\nabla\cdot  \mathbf{q}   +q_s\, .
\end{equation} 

\noindent\textbf{Conservation of energy in a two-phase porous medium:}\vspace{6pt}

Following  Sec. \ref{sec:mathmodel}, the medium can be represented by a two-phase porous flow, where the solid phase or fuel is denoted by $f$ and the gaseous phase, or air, is denoted by $a$. The volume fraction of fuel in the domain is given by $R_f$. Assume having a perfect mixture and all variables are homogenized. All the material properties --e.g., density, specific heat-- are defined in  Sec. \ref{sec:mathmodel}. By combining the conservation equations for enthalpy in a fluid and a solid, see \cite{vafai2015handbook}, under the assumption of thermal equilibrium ($T_f = T_a = T$), we obtain a single-equation model as described in \cite{sero2002modelling}:
\begin{equation}\label{eq:twophase1} 
     \frac{\partial \rho h}{\partial t}  +  \nabla \cdot (  \bar{\rho}_a h_a (1-R_{f})\mathbf{v}_a
     )   = -\nabla\cdot  \mathbf{q}    +q_s\, ,
\end{equation}
where  $\rho h = \bar{\rho}_fh_f R_{f} +  \bar{\rho}_a h_a (1-R_{f})$ is the enthalpy per unit volume of the mixture, $\mathbf{q}$ is a the total diffusion heat flux and $q_s$ represents all the heat sources (in both phases). Analogously, 
\begin{equation}\label{eq:twophase2} 
      \rho c_p   \frac{\partial T}{\partial t} +  \bar{\rho}_a \bar{c}_{p,a} (1-R_f) \mathbf{v}_a \cdot \nabla T    =  -\nabla\cdot  \mathbf{q} +q_s\, ,
\end{equation}
where $ \rho c_p=\bar{\rho}_f \bar{c}_{p,f} R_{f} +  \bar{\rho}_a \bar{c}_{p,a} (1-R_f)  $ is the specific heat per unit volume of the mixture and $\mathbf{q}$ and $q_s$ represent heat fluxes and sources in both phases.

\section{Verification of the numerical solvers -- Mesh convergence analysis}\label{appendix}

\begin{figure}
    \centering
    \begin{subfigure}{0.49\textwidth}
           \includegraphics[width=\linewidth]{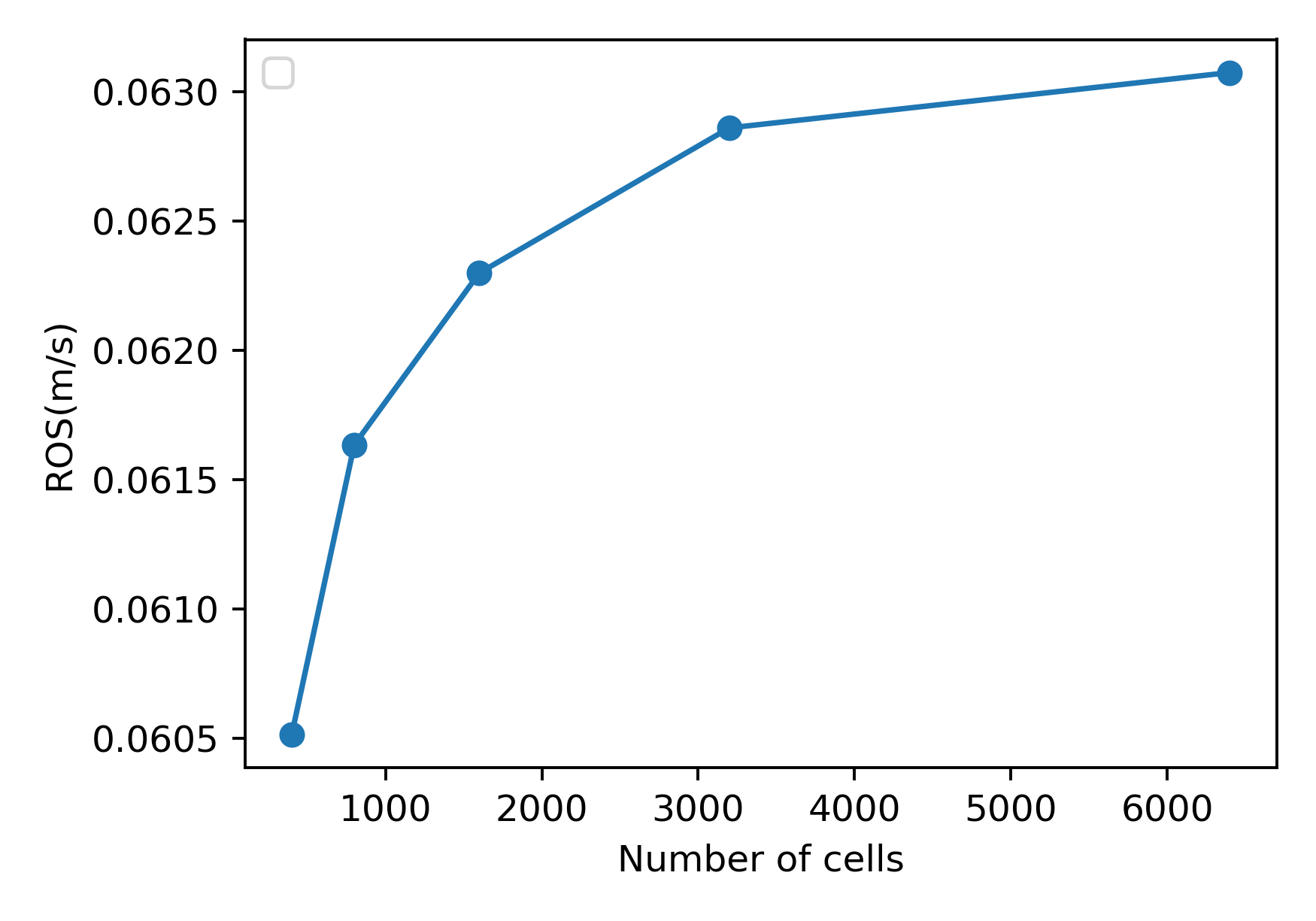}
           \caption{$M=0.0$, $w=0.2$}
    \end{subfigure}
        \begin{subfigure}{0.49\textwidth}\includegraphics[width=\linewidth]{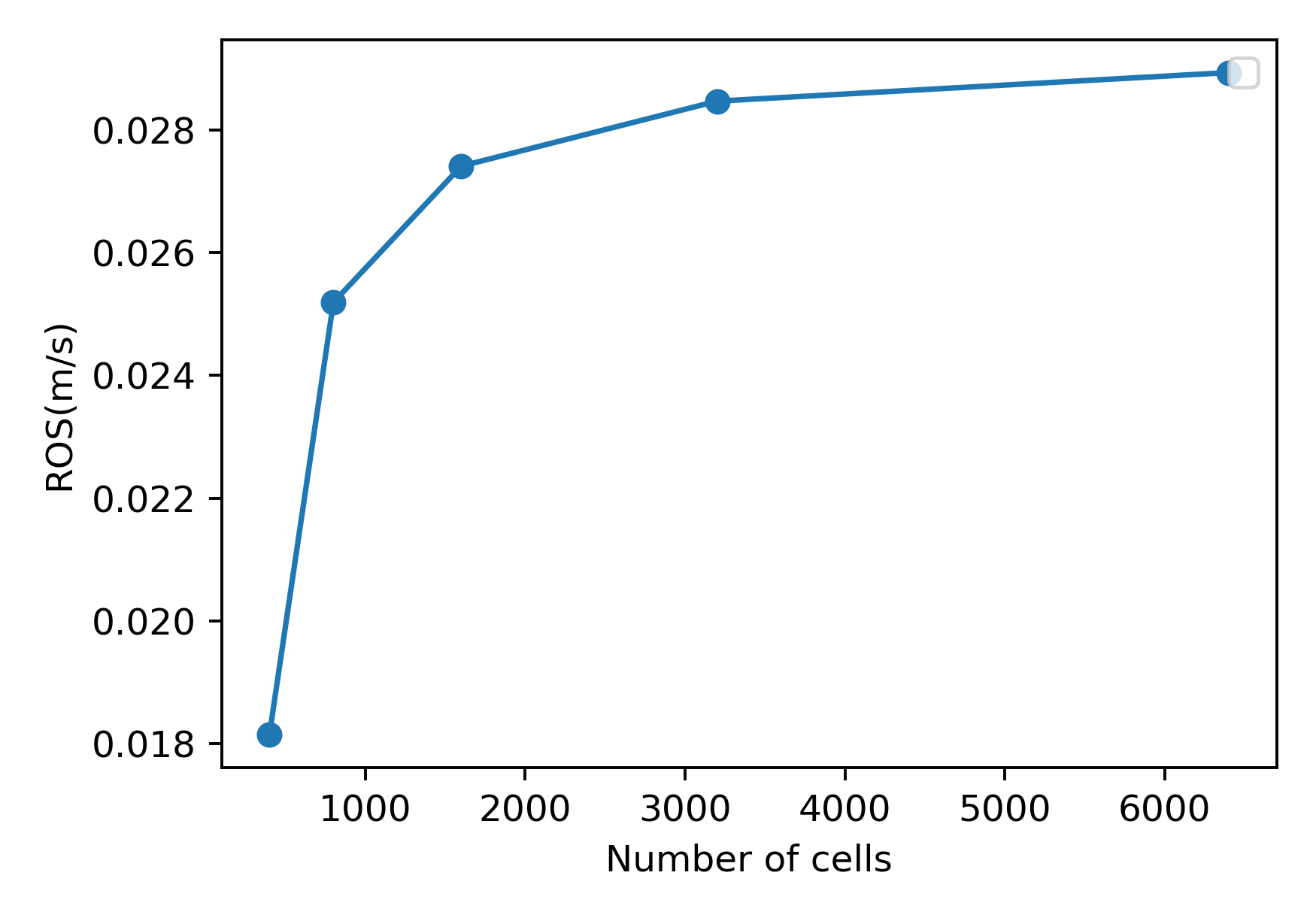}
        \caption{ $M=0.2$, $w=0.0$}
        \end{subfigure}
            \begin{subfigure}{0.49\textwidth}
    \includegraphics[width=\linewidth]{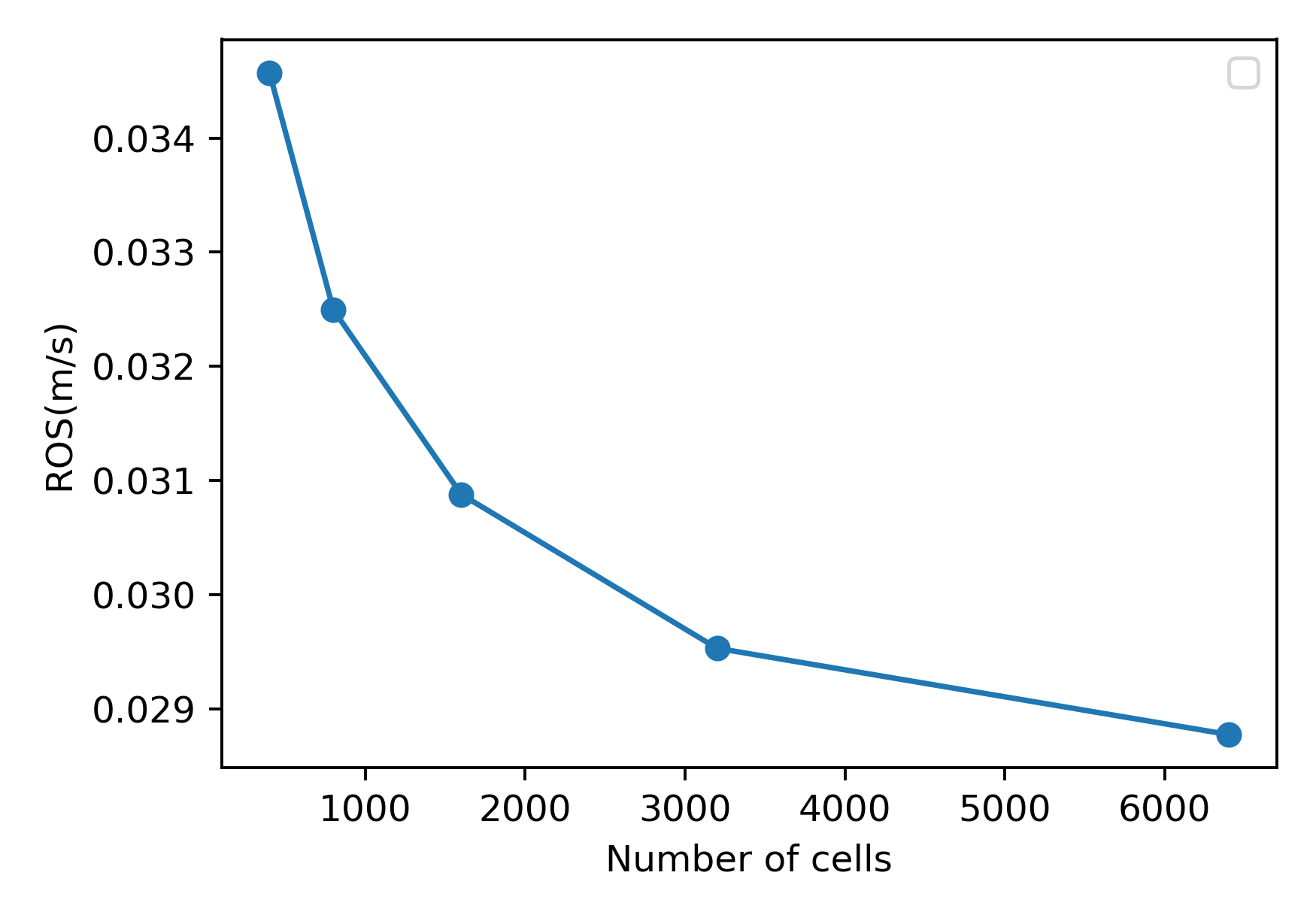}
    \caption{$M=0.5$, $w=0.5$}
    \end{subfigure}
    \caption{Rate of spread depending on the number of cells  for different moisture and wind scenarios at $t_f=2000$ s. All scenarios show a convergence tendency for high number of cells. 
    }
    \label{fig:convergence}
\end{figure}

This section provides a mesh convergence analysis of the model (using S2M), with and without moisture and wind. Fig.~\ref{fig:convergence} shows the rate of spread depending on the number of mesh cells $N_x$ in the spatial discretization, compare Sec.~\ref{subsec:fvm}. Low wind conditions, i.e. $w=0.2$ m/s and the default parameters from Tabs.~\ref{table:general-parameters1}--\ref{table:general-parameters3} are used. The initial condition in Eq.~\eqref{eq:ic} is used. The domain length is 500 m. 

\begin{figure}
    \centering
    \includegraphics[width=0.49\linewidth]{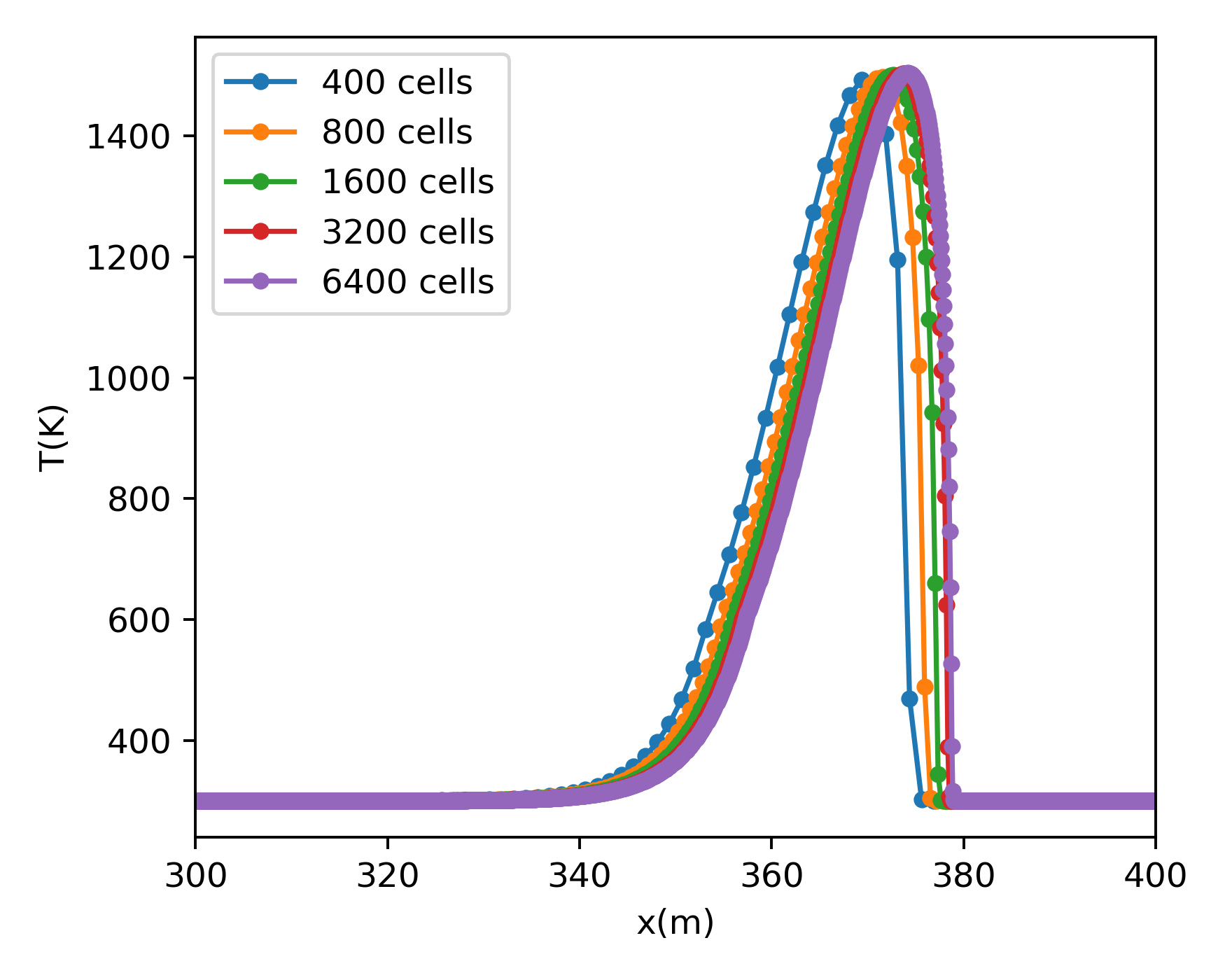}
    \includegraphics[width=0.49\linewidth]{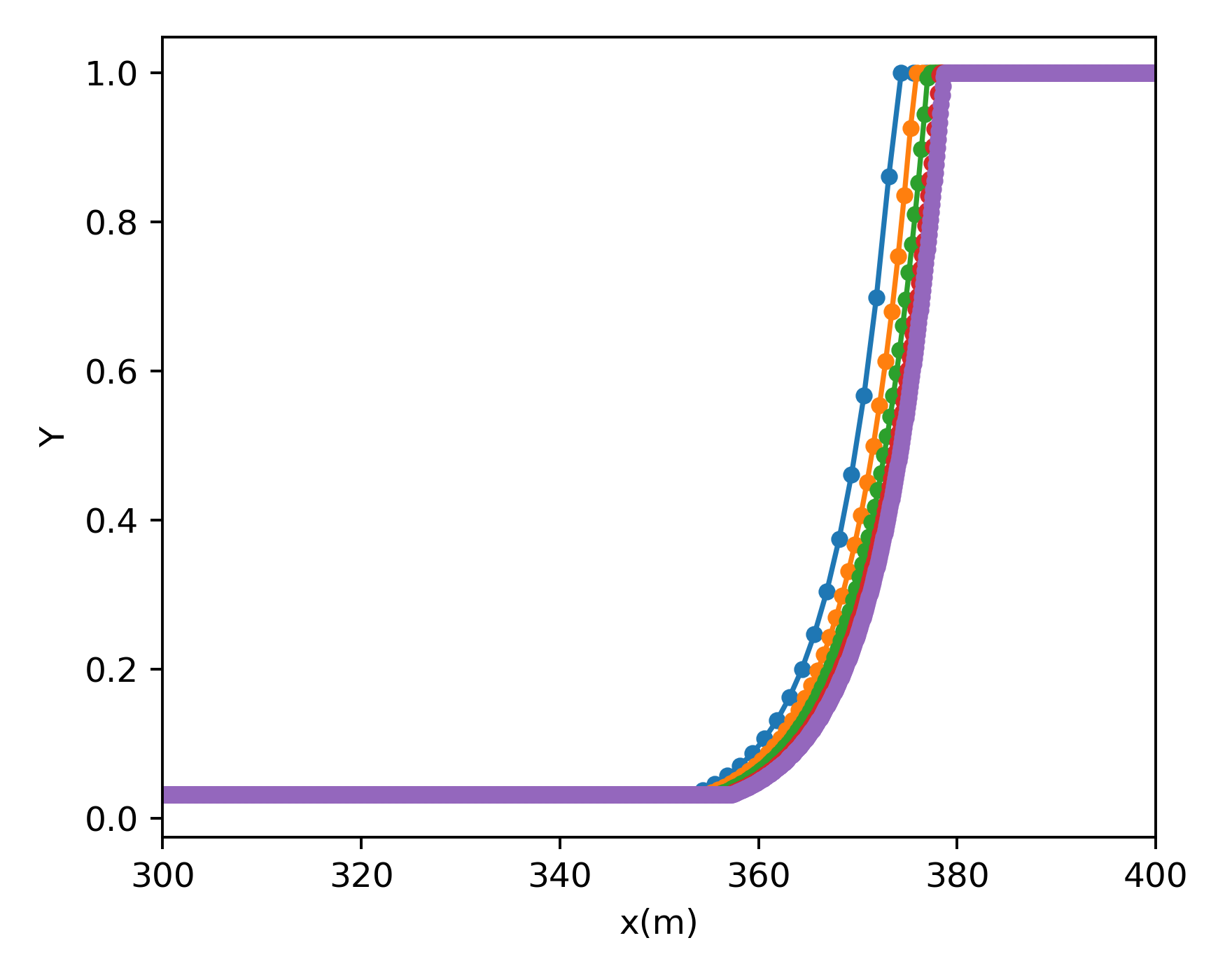}\\    
     \includegraphics[width=0.49\linewidth]{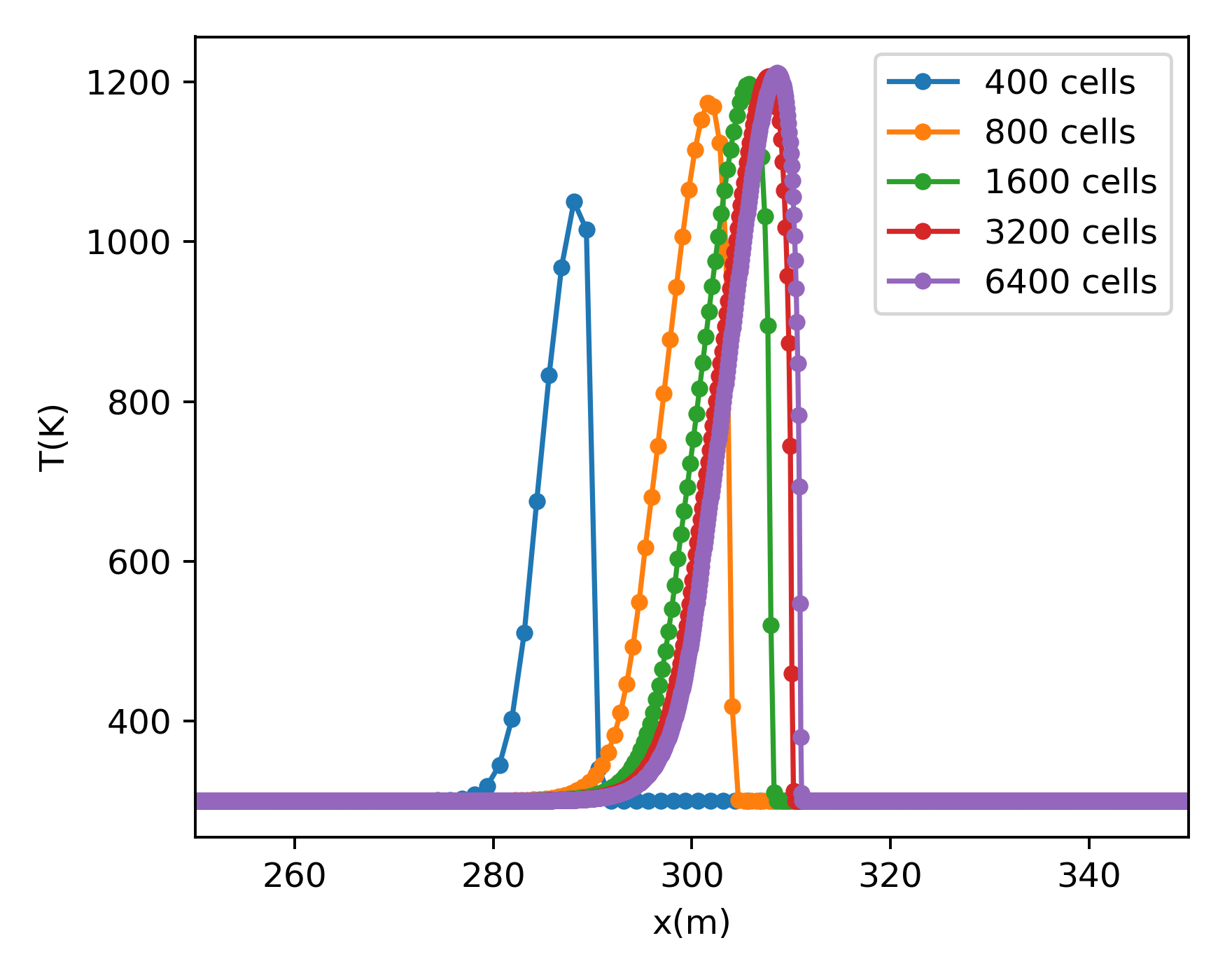}
    \includegraphics[width=0.49\linewidth]{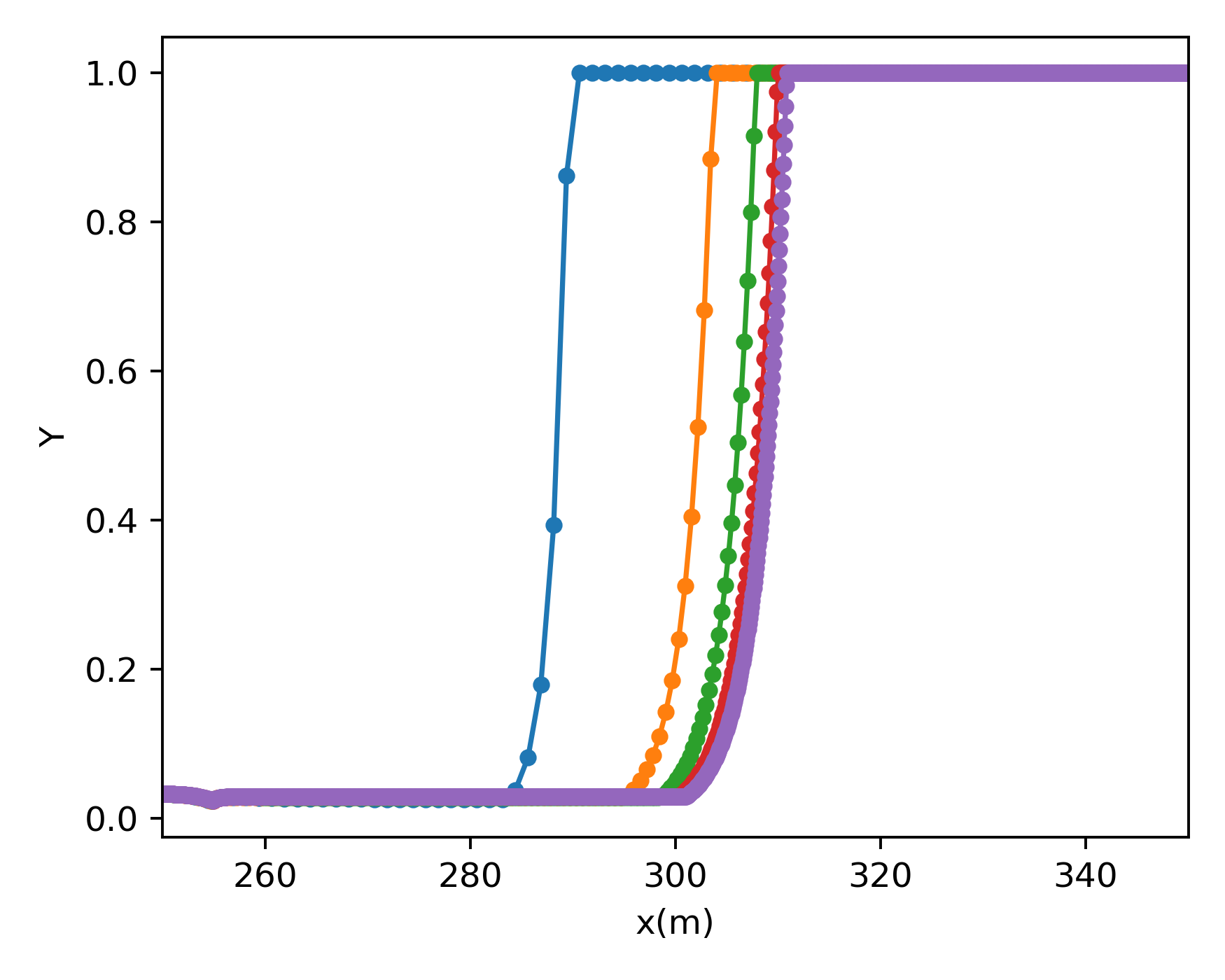}\\ 
      \includegraphics[width=0.49\linewidth]{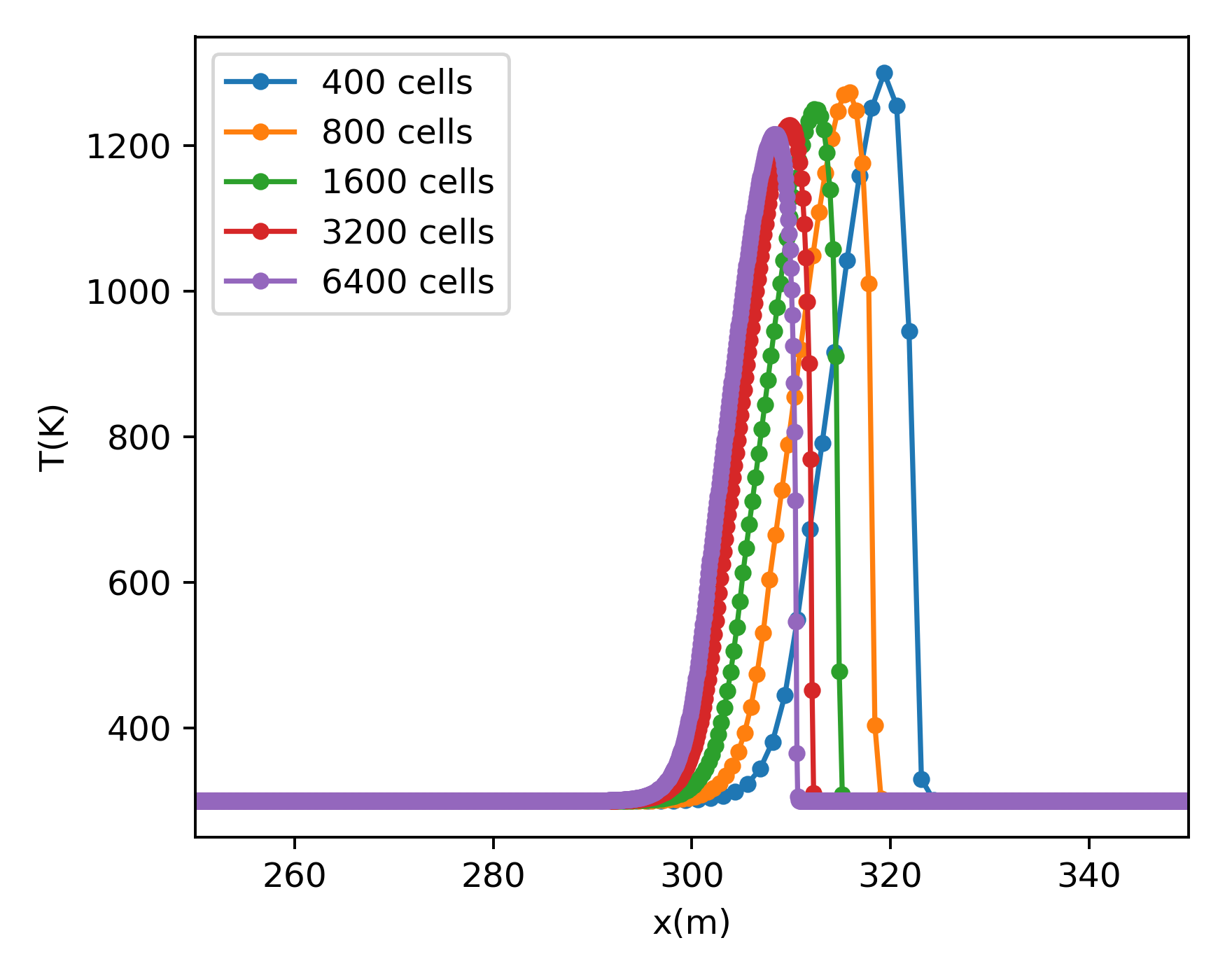}
    \includegraphics[width=0.49\linewidth]{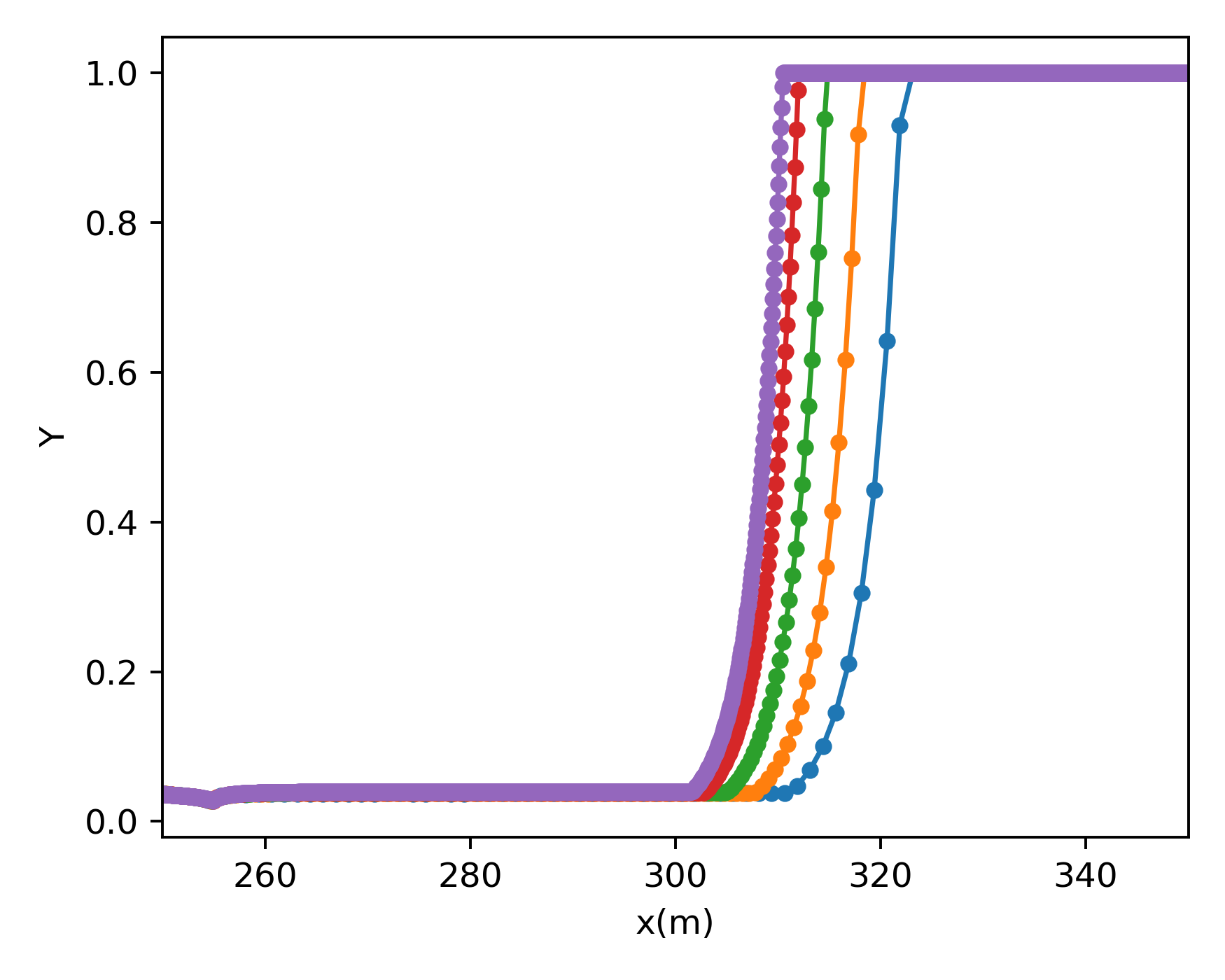}\\    
    \caption{Profiles of the solution depending on the number of cells  for different wind and moisture scenarios at a final time $t_f=2000$~s.  Top: $M=0.0$, $w=0.2$; middle:   $M=0.2$, $w=0.0$; bottom:  $M=0.5$, $w=0.5$.
    The simulations show a convergence tendency for the traveling wave front and the maximal temperature.}
    \label{fig:profile_grid}
\end{figure}

Additionally, Fig.~\ref{fig:profile_grid} shows the profiles of the solutions. The traveling wave fronts are approaching a limit and the shape of the fronts is almost identical for $N_x=1600$ and more. In the case of having both moisture and wind, convergence is slightly slower than in the case without the combined effect. Nevertheless, smooth convergence is achieved for a sufficient number of cells and the error in ROS is very small for meshes of $N_x=1600$ cells and above, e.g. the position of the front varies less than 6 m. These results motivate our choice of $N_x=1600$ throughout the paper (for 1D cases in a domain with length $L=500$ m). 

\bibliographystyle{plainnat}
\bibliography{references}

\end{document}